\documentclass[conf]{new-aiaa}
\usepackage[utf8]{inputenc}
\usepackage{iepc2022}
\usepackage{lineno}

\usepackage{amsmath}
\usepackage{graphicx}
\usepackage[version=4]{mhchem}
\usepackage{siunitx}
\usepackage{longtable,tabularx}
\usepackage[export]{adjustbox}
\newcommand{\diff}{\mathrm{d}}

\allowdisplaybreaks[1]

\IEPCsubmissionnumber{486}

\title{Enabling direct kinetic simulation of dense plasma plume expansion for laser ablation plasma thrusters}

\author{Wai Hong Ronald Chan\footnote{Postdoctoral Associate, Department of Aerospace Engineering Sciences, WaiHongRonald.Chan@colorado.edu} and Iain D. Boyd\footnote{H.T. Sears Memorial Professor, Department of Aerospace Engineering Sciences, Iain.Boyd@colorado.edu\\\indent \enskip \small Copyright \copyright~2022 by W.H.R. Chan and I.D. Boyd. Published by the Electric Rocket Propulsion Society with permission.}}
\affil{University of Colorado, Boulder, CO, 80309, USA}

\begin{document}

    \maketitle

    \begin{abstract}
        Laser ablation plasma thrusters are an emerging space propulsion concept that provides promise for lightweight payload delivery. Predicting the lifetime and performance of these thrusters hinges on a comprehensive characterization of the expansion dynamics of the ablated plasma plume. While state-of-the-art techniques for simulating plasmas are often particle-based, a grid-based direct kinetic solver confers advantages in such a transient and inhomogeneous problem by eliminating statistical noise. A direct kinetic solver including interparticle collisions is employed on a plume expansion model problem spanning one dimension each in configuration and velocity space. The high degree of thermodynamic nonequilibrium inherent in plume expansion is characterized, justifying the need for a kinetic rather than a hybrid or fluid solver. Thruster-relevant metrics such as the momentum flux are also computed. The plume dynamics are observed to be highly inhomogeneous in space with insufficient time for thermalization in the region preceding the expansion front, and the theoretical possibility of reducing the local grid resolution by up to two orders of magnitude at the far end of the domain is established. These grid-point requirements are verified via the employment of nonuniform grids of various expansion ratios, several of which also employ coarsening in velocity space. Longer domain lengths are explored to characterize thruster-scale phenomena and larger ambient pressures are simulated as a testbed to probe facility effects due to collisions with background particles.
    \end{abstract}

    \section{Nomenclature}

    {\renewcommand\arraystretch{1.0}
    \noindent\begin{longtable*}{@{}l @{\quad=\quad} l@{}}
        $*$ & particle type ($n=\text{neutral}$, $i=\text{ion}$, $e=\text{electron}$)\\
        $d_\text{atom}$ & characteristic atomic diameter ($10^{-10}\text{ m}$)\\
        $E$ & axial electric field strength\\
        $\mathbf{E}$ & electric field vector\\
        $e$ & elementary charge\\
        $f_*$ & probability density function of particle type $*$\\
        $f_*^\text{eq}$ & equilibrium (Maxwellian) probability density function of particle type $*$\\
        $k_B$ & Boltzmann constant\\
        $L$ & domain length\\
        $m_*$ & mass of particle type $*$\\
        $n_*$ & number density of particle type $*$\\
        $n_{*,0}$ & inlet number density of particle type $*$\\
        $n_{*,j}$ & reference number density of particle type $*$ at domain edge $j$\\
        $n_*(x=0^+;t)$ & number density of particle type $*$ in the first (leftmost) computational cell at time $t$\\
        $q_*$ & charge of particle type $*$\\
        $S$ & collision term\\
        $t$ & time coordinate\\
        $t_\text{mfp}$ & neutral--neutral collisional mean free time\\
        $t_\text{sim}$ & total simulation time\\
        $T_*$ & temperature of particle type $*$\\
        $T_j$ & reference temperature at domain edge $j$\\
        $v$ & axial velocity coordinate\\
        $v_{\text{th},*,j}$ & reference thermal velocity of particle type $*$ at domain edge $j$\\
        $\mathbf{v}$ & velocity vector\\
        $x$ & axial spatial coordinate\\
        $x_c$ & transition location between two reference distributions ($x_c=0.2L$)\\
        $\mathbf{x}$ & spatial displacement vector\\
        $\cdot_e$ & electron quantities\\
        $\cdot_i$ & ion quantities\\
        $\cdot_n$ & neutral quantities\\
        $\cdot_j$ & reference quantities at left ($j=1$) and right ($j=2$) of computational domain\\
        $\Delta t$ & time step\\
        $\Delta t_\text{sub}$ & time step used for substepping\\
        $\Delta v$ & axial velocity grid size\\
        $\Delta x$ & axial spatial grid size\\
        $\varepsilon_0$ & vacuum permittivity\\
        $\lambda_{D,j}$ & reference Debye length at domain edge $j$\\
        $\lambda_\text{mfp}$ & neutral--neutral collisional mean free path\\
        $\tau$ & characteristic collision time\\
        $\phi$ & electric potential\\
        $\phi_{\text{th},j}$ & reference thermal potential at domain edge $j$

    \end{longtable*}}

    \section{Introduction}\label{sec:intro}
    
    The laser ablation plasma thruster is a novel advanced space propulsion concept that has emerged as a promising candidate at, but not limited to, microthrust levels~\citep{kantrowitz1972propulsion,phipps2010review,keidar2015electric,levchenko2018prospects,zhang2019review,levchenko2020perspectives}. These thrusters can be associated with high specific impulses and have the potential to be accelerated by laser sources external to the spacecraft, thus decreasing payload sizes and masses. Thruster design requires a detailed understanding of the physics of ablation and the ensuing plasma plume expansion. In particular, the dynamics of plume expansion directly influence thruster efficiency and lifetime. The plume expansion process is sensitive to ambient conditions and inhomogeneities in the ablation process, both at the laser beam and propellant substrate levels~\cite{keidar2015electric,keidar2004plasma,keidar2004propellant,zhang2018investigation}. A robust computational tool that predicts thruster performance in a variety of plume expansion scenarios is key to the development and characterization of reliable and durable thrusters. We lay the foundations to this end by developing a direct kinetic (DK) solver that computes plume expansion in an accurate manner with increased efficiency and the ability to prototype a variety of component models for the underlying physics, such as interparticle collisions and ablated particle distributions, in a broad range of ambient operating conditions.
    
    The setup of plume expansion is deceptively simple, for the challenge of simulating ablation plasma plume expansion lies in the broad range of physical regimes that it spans. Near the substrate, the plume is dense, often approaching atmospheric number densities at typical laser powers, and the corresponding plasma plume is strongly coupled. The plasma parameter is large and collisions dominate long-range interactions. At the expansion front, the plume expands into a sparsely populated or evacuated ambient, and the corresponding plasma plume is weakly coupled. The plasma parameter is small and the background electromagnetic fields become important for particle dynamics. Any instabilities that emerge in the domain can propagate quickly between the two regions at the electron thermal speed, far exceeding the expansion front speed that characterizes the baseline plume dynamics~\citep{cui2021grid}.
    
    The state-of-the-art technique for simulating plasmas is the particle-in-cell method, where ions and/or electrons are represented by superparticles interacting in configuration space. Particle methods are susceptible to statistical noise since a sufficiently large number of superparticles is required in each computational cell to adequately sample the particle distribution function in the configuration and velocity spaces. This is exacerbated in plume expansion problems that span a broad range of baseline number densities due to the spatial inhomogeneity of the setup, as well as mean axial speeds due to ambipolar acceleration by fast electrons. Such statistical noise is eliminated in DK methods, where the distribution function is directly simulated in these spaces without discretization into superparticles. This work focuses on justification and setup of efficient application of the DK method to a simplified plume expansion problem.
    
    Grid-based kinetic solvers, also known as DK, continuum kinetic, discrete velocity, and Vlasov solvers, are an increasingly relevant method of high-fidelity plasma simulation. Since they eliminate the statistical noise associated with particle-in-cell methods, they are inherently more suitable for transient, multiscale, and spatially inhomogeneous problems with low-noise-floor requirements, which constitute many complex natural systems and modern engineering challenges. These solvers involve the direct Eulerian simulation of the particle distribution function according to the Boltzmann or Vlasov equations (depending on whether interparticle collisions and the self-consistent electric field are included), which describe the kinetic transport of this probability density function in the configuration (physical) and velocity spaces, as reviewed in detail by~\citet{filbet2003comparison} and~\citet{hara2018test}. In particular, the DK method differs from typical Vlasov solvers in that interparticle collisions are included. To date, DK solvers have been implemented both in a hybrid sense with fluid schemes~\citep{kolobov2007unified}, as well as in a fully kinetic setting~\citep{dimarco2014numerical}, with applications to nuclear fusion~\citep{thomas2012review}, space physics and astrophysics~\citep{palmroth2018vlasov}, and space propulsion and materials processing~\citep{hara2018test}. The DK method is suitable for unsteady problems where time averaging cannot be used, as well as steady problems with nascent instabilities where the amount of time averaging required to resolve the instabilities can become prohibitive. This grid-based method is also superior to particle methods in regions with lower particle densities, such as the expansion front in ablation plumes. The tradeoff is the higher dimensionality associated with direct solution of the Vlasov and Boltzmann equations. This work highlights the nonequilibrium nature of plume expansion in order to justify the use of a kinetic solver, and explores avenues to reduce the computational cost of the DK method through the employment of a suitable nonuniform grid, building on the grid-point requirements identified in our earlier work~\citep{chan2022grid}. A nonuniform grid enables the employment of a DK solver to thruster-scale domains. In addition, thruster-relevant investigations such as the computation of momentum flux and the nascent characterization of facility effects are carried out.
    
    With these objectives in mind, the paper is structured as follows. In \S~\ref{sec:method}, the methodology adopted in this work is introduced, including details of the employed solver, numerical setup of the expanding plume, and a brief description of the key plume dynamics. Results are discussed in \S~\ref{sec:results}, including characterization of thermodynamic nonequilibrium and spatial inhomogeneity, computation of the momentum flux, and discussion of nonuniform grid design accounting for previously identified grid-point requirements, which are used as reference points in a comparison of uniform and nonuniform grid results. Preliminary investigations into thruster-scale domain lengths and the characterization of facility effects through variations in the ambient pressure are also discussed. Conclusions are provided in \S~\ref{sec:conc}.

    \section{Methodology}\label{sec:method}
    
    \subsection{Solver description: Vlasov-Poisson-BGK equations}\label{sec:solver}
    
    The grid-based DK solver employed in this work was originally developed at the University of Michigan with extensive verification and validation against canonical and complex plasma problems~\citep{hara2018test, hara2012one, hara2014mode, hara2015quantitative, hara2017kinetic, raisanen2019two, vazsonyi2020non}. Under the electrostatic approximation, the solver computes the time evolution of $f_*$ according to the following transport equation
    \begin{linenomath*}
    \begin{equation}
    \frac{\partial f_* (\mathbf{x},\mathbf{v};t)}{\partial t} + \mathbf{v}\cdot\nabla_\mathbf{x} f_*(\mathbf{x},\mathbf{v};t) + \frac{q_* \mathbf{E}}{m_*}\cdot\nabla_\mathbf{v} f_*(\mathbf{x},\mathbf{v};t) = S(f_*).
    \end{equation}
    \end{linenomath*}
    Note that $S$ is set to zero in a collisionless simulation, yielding the Vlasov equation, and is otherwise modeled. Multiple particle types and species may be simultaneously simulated with appropriate cross-collision terms. The computational domain is discretized in $\mathbf{x}$--$\mathbf{v}$ space with a finite-volume method, and advection is performed to second-order accuracy using the monotonic upstream-centered scheme for conservation laws (MUSCL) with Strang splitting and a modified Arora-Roe limiter to preserve positivity. Gauss' law is expressed using $\mathbf{E}=-\nabla_\mathbf{x}\phi$ as a Poisson equation
    \begin{linenomath*}
    \begin{equation}
    \nabla_\mathbf{x}^2 \phi = -\frac{e(n_i-n_e)}{\varepsilon_0},\label{eqn:poisson}
    \end{equation}
    \end{linenomath*}
    where
    \begin{linenomath*}
    \begin{gather}
    n_i(\mathbf{x};t) = \int_\mathbf{v} f_i(\mathbf{x},\mathbf{v'};t) \, \diff \mathbf{v'},\label{eqn:ni}\\
    n_e(\mathbf{x};t) = \int_\mathbf{v} f_e(\mathbf{x},\mathbf{v'};t) \, \diff \mathbf{v'}.\label{eqn:ne}
    \end{gather}
    \end{linenomath*}
    All collisions in this work are modeled using the Bhatnagar-Gross-Krook (BGK) operator
    \begin{linenomath*}
    \begin{equation}
    S(f_*) = \frac{f_*^\text{eq} - f_*}{\tau}.
    \end{equation}
    \end{linenomath*}
    The simulations are parallelized using the Message Passing Interface, with domain decomposition performed solely in configuration space to avoid communication during the computation of Eqs.~\eqref{eqn:ni} and \eqref{eqn:ne}, and the SuperLU\_DIST library implemented with the help of the PETSc toolkit for direct solution of Eq.~\eqref{eqn:poisson}. Electron advection and collisions are substepped to speed up the simulation runtime. The current version of the solver has been tested with the physical plume setup used in this work on up to 432 computational cores with at least 80\% weak scaling efficiency over an order-of-magnitude variation in core count. For these weak scaling tests, the spatial-cell-per-core ratio and the number of velocity cells per spatial location both exceed 100. The maximum baseline simulation size tested is about 40 million cells per particle type with a wall time of about 16 hours on 288 Intel Skylake cores~\cite{chan2022grid}, and the chief scaling bottleneck is the parallel Poisson solver. While the longer-domain simulations in \S~\ref{sec:longdomain} have fewer cells, their wall time can approach 2 weeks on up to 432 Skylake cores due to correspondingly longer simulation durations $t_\text{sim}$. Our DK solver has the capability to handle a nonuniform grid in all dimensions to alleviate costs as discussed in \S~\ref{sec:nonuniform}.

    \subsection{Plume setup}\label{sec:setup}
    
\begin{figure}
      \centerline{
    \includegraphics[width=0.9\linewidth]{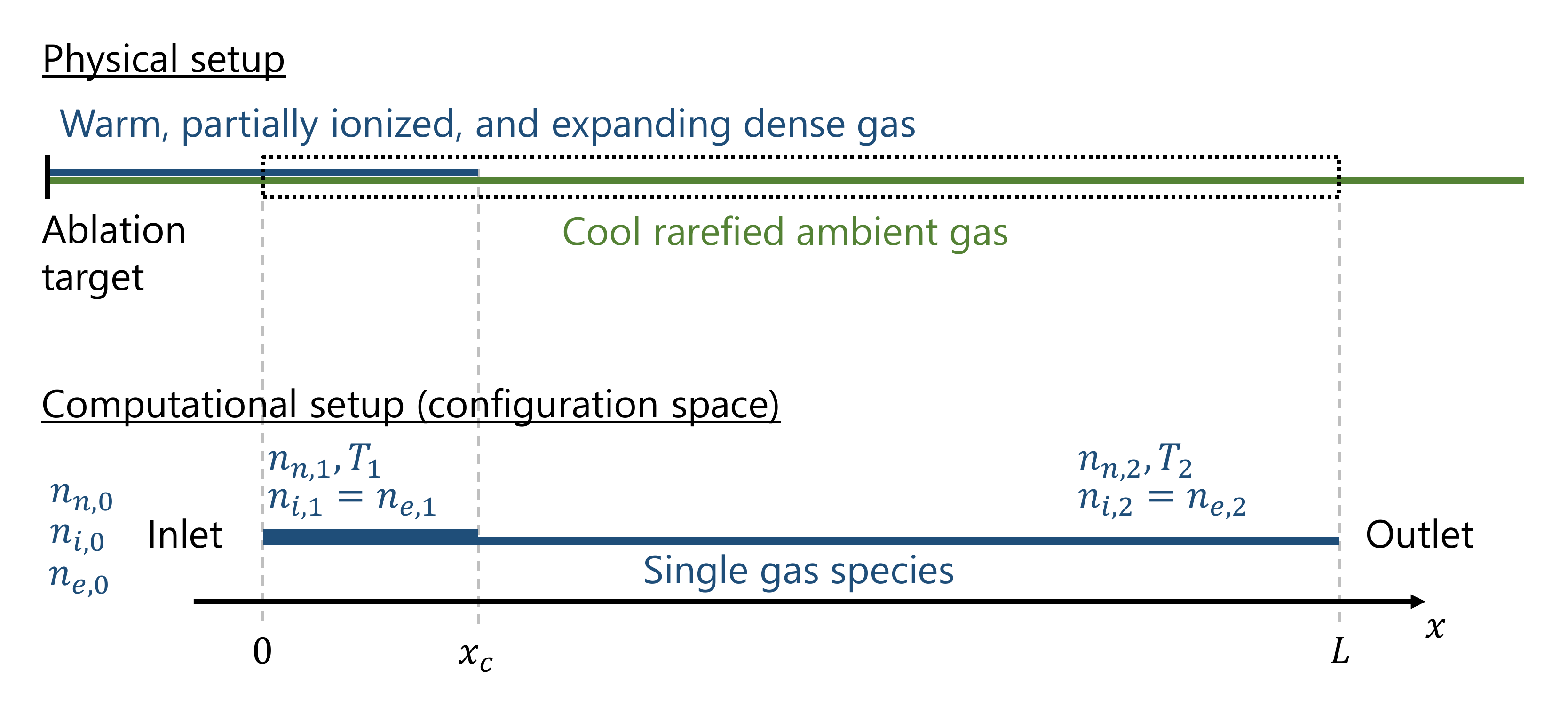}
    }
      \caption{Schematic of the physical and computational setups for the plume expansion problem in configuration (physical, $x$) space.}
    \label{fig:plumeschematic}
    \end{figure}

    This work primarily involves an unsteady, spatially nonperiodic, and weakly collisional plume expansion model problem involving one dimension each in space and velocity (1D1V) for ease of numerical prototyping. A schematic of the physical and computational setups corresponding to the model problem is depicted in Fig.~\ref{fig:plumeschematic}. In a physical setting, an impinging laser ablates material from a target surface, forming a warm dense plume that is often almost as dense as the standard atmosphere, which then expands away from the surface. The plume expansion often occurs in ambient surroundings with low particle density, such as rarefied or evacuated outer space. The left boundary of the computational domain is set in the middle of this plume near the target surface, such that the left end of the domain contains a high-density partially ionized quasineutral gas at local thermodynamic equilibrium, and inlet boundary conditions are imposed on the left with gas of the same composition entering the domain. The computational boundary is offset some distance from the physical wall to obviate the need to resolve the wall sheath. The right boundary of the computational domain is set sufficiently far away from this plume, such that the right end of the domain contains a low-density gas of a cooler temperature, and outlet boundary conditions are imposed on the right with no backflow permitted. These boundaries are modeled by homogeneous Neumann conditions at both domain boundaries for the electric potential with a fixed value on the right boundary. While the expanding and ambient gases are typically of different species in a physical setting, they are modeled by the same gas species (aluminum) in this work for simplicity.

    The quantitative initial and boundary conditions of the simulation are as follows. The domain is initialized with a partially ionized aluminum gas that has a higher particle density on the left near the inlet. The initial ($t=0$) spatial variation of the particle distribution function for each of the considered particle types (neutral, ion, electron) is given by
    \begin{linenomath*}
    \begin{equation}
    f_*(x,v;t=0) = \frac{f_{*,1}+f_{*,2}}{2} - \frac{f_{*,1}-f_{*,2}}{2} \tanh\left(200\frac{x-x_c}{L}\right).
    \end{equation}
    \end{linenomath*}
    A hyperbolic tangent smoothing profile is introduced to allow the underlying grid to resolve the transitions in particle density and temperature, and prevent the generation of numerical instabilities at the transition location. All reference particle distributions are Maxwellian, i.e.,
    \begin{linenomath*}
    \begin{gather}
    f_{*,1}(v) = \frac{n_{*,1}}{\sqrt{2\pi}v_{\text{th},*,1}}\exp\left(-\frac{v^2}{2 v_{\text{th},*,1}^2}\right),\\
    f_{*,2}(v) = \frac{n_{*,2}}{\sqrt{2\pi}v_{\text{th},*,2}}\exp\left(-\frac{v^2}{2 v_{\text{th},*,2}^2}\right),
    \end{gather}
    \end{linenomath*}
    where $v_{\text{th},*,j} = \sqrt{k_B T_j/m_*}$. All considered cases are initialized with the following baseline reference number density and temperature ratios
    \begin{linenomath*}
    \begin{gather}
    \frac{n_{i,1}}{n_{i,2}} \sim \frac{n_{e,1}}{n_{e,2}} \sim \frac{n_{n,1}}{n_{n,2}} = 10^{10},\label{eqn:nnratio}\\
    \frac{T_1}{T_2} = 20,\label{eqn:Tratio}\\
    \frac{n_{i,1}}{n_{n,1}} = \frac{n_{e,1}}{n_{n,1}} = 3.24437\times10^{-2},\label{eqn:saha}
    \end{gather}
    \end{linenomath*}
    where the left reference temperature is $T_1 = 6{,}000\text{ K}$, which is commonly associated with nanosecond laser pulses traditionally used in ablation experiments. The ratio~\eqref{eqn:saha} is chosen to match the ionization state for a gas with neutral number density $n_{n,1} = 10^{25}\text{ m}^{-3}$ at temperature $T_1$, and the neutral--neutral ratio in Eq.~\eqref{eqn:nnratio} corresponds in turn to a characteristic ambient density $n_{n,2}$ for low Earth orbit. For the densities, baseline domain length, and baseline simulation duration considered in this work, $\lambda_\text{mfp}/L \sim t_\text{mfp}/t_\text{sim} \sim 1$, where $\lambda_\text{mfp}\sim \left(n_{n,1}d_\text{atom}^2\right)^{-1}$ and $t_\text{mfp} \sim \lambda_\text{mfp}/v_{\text{th},n,1}$.
    
    \begin{figure}
      \centerline{
    \includegraphics[trim=50 50 50 50, clip, width=0.65\linewidth]{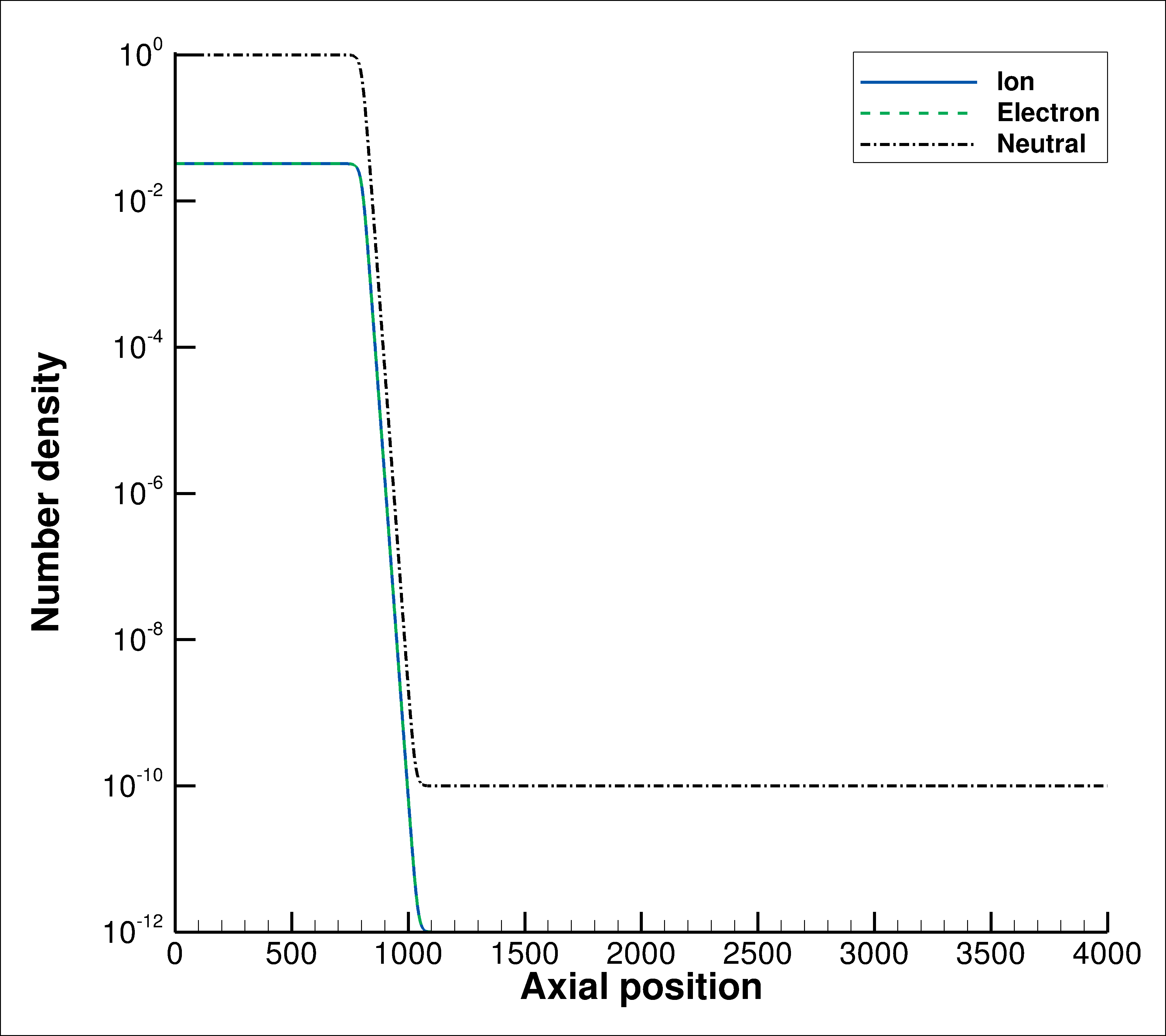}
    }
      \caption{Initial neutral (dash-dotted), ion (solid), and electron (dashed) number density profiles, normalized by $n_{n,1}$.}
    \label{fig:initprofile}
    \end{figure}
    
    The initial number density profiles of the various particle types, normalized by $n_{n,1}$, are plotted in Fig.~\ref{fig:initprofile}. Number densities and lengths are hereinafter nondimensionalized by $n_{*,1}$ and $\lambda_{D,1}=\sqrt{\varepsilon_0 k_B T_1/(2 n_{i,1} e^2)}$, respectively. A list of these and other reference quantities used for nondimensionalization is provided in Table~\ref{tab:nondim}. All variables and plot axes are nondimensional from this point on unless otherwise stated. 
    
    \begin{table}
    \begin{center}
    \caption{Summary of reference quantities used for nondimensionalization.}\label{tab:nondim}
    \begin{tabular} { c || >{\centering\arraybackslash} m{0.10\textwidth} | >{\centering\arraybackslash} m{0.35\textwidth}}
    Quantity & Symbol & Characteristic scale \\
    &&\\
    Number density		& $n$ 		& $n_{*,1}$\\
    Velocity			& $v$		& $v_{\text{th},*,1} = \sqrt{k_B T_1/m_*}$\\
    Position, length		& $x$, $L$		& $\lambda_{D,1} = \sqrt{\varepsilon_0 k_B T_1/(2 n_{i,1} e^2)}$\\
    Time				& $t$			& $\lambda_{D,1}/v_{\text{th},i,1} = \sqrt{m_i \varepsilon_0 / (2 n_{i,1} e^2)}$\\
    Electric potential		& $\phi$		& $\phi_{\text{th},1} = k_B T_1/e$\\
    Electric field		& $E$		& $\phi_{\text{th},1}/\lambda_{D,1}$
    \end{tabular}
    \end{center}
    \end{table}
    
    For the baseline simulations of this work, the domain length is set to be $L = 4{,}000$, and thus 99\% of the dip in the hyperbolic tangent function occurs across about $O(100)$ Debye lengths. We discuss the motivation behind this initial condition in more detail in Ref.~\citep{chan2022grid}. The baseline total simulation duration is $t_\text{sim} = 288$. At the inlet, Maxwellian particle distributions are injected at $T_1$ matching the interior temperature. However, the following inlet particle number densities $n_{*,0}$ are used instead of $n_{*,1}$ to maintain quasineutrality near the inlet
    \begin{linenomath*}
    \begin{gather}
    n_{n,0} = n_{n,1},\\
    n_{i,0} = n_{i,1},\\
    n_{e,0}(t) = n_{e,1}\frac{n_i(x=0^+;t)}{n_e(x=0^+;t)} + \frac{0.1}{\Delta t_\text{sub}} \int_0^t \left[n_i(x=0^+;t')-n_e(x=0^+;t')\right] \diff t'.\label{eqn:PIcontroller}
    \end{gather}
    \end{linenomath*}
    The expression \eqref{eqn:PIcontroller} mimics the action of a proportional-integral controller with integral gain $K_i = 0.1$ and is further discussed in Ref.~\citep{chan2022grid}. The extent of the velocity domain is $[-6,20]$ for neutrals and ions, and $[-6,6]$ for electrons. The extended positive range for ions is to account for ion acceleration. Otherwise, six thermal velocities in each direction can capture the bulk of the distribution of a particle population with zero mean velocity. Neutral--neutral, ion--neutral, and electron--neutral momentum-exchange collisions are included using a BGK collision model. Charge-exchange collisions are not considered here since the model problem is meant to mimic a physical setup where the ablated and background gases are different. The convective Courant number based on the maximum resolved ion speed is set at $0.9$. All mass ratios employed in this work take their actual physical values. Simulation parameters are summarized in Table~\ref{tab:param}.
    
    \begin{table}
    \caption{Summary of simulation parameters.}\label{tab:param}
    \begin{center}
    \begin{tabular} { c || >{\centering\arraybackslash} m{0.15\textwidth}}
    Quantity & Value \\
    &\\
    Baseline simulation length $L$					& $4{,}000$\\
    Baseline simulation duration $t_\text{sim}$		& $288$\\
    Velocity domain for ions and neutrals		& $[-6,20]$\\
    Velocity domain for electrons			& $[-6,6]$\\
    Courant number						& $0.9$\\
    Left reference temperature $T_1$, K		& $6{,}000$\\
    Ion mass, amu						& $26.98$\\
    Electron mass, amu					& $5.486\times10^{-4}$
    \end{tabular}
    \end{center}
    \end{table}

    \subsection{Plume dynamics}\label{sec:dynamics}

    \begin{figure}
      \centerline{
    (a)
    \includegraphics[trim=50 50 50 50, clip, width=0.42\linewidth,valign=t]{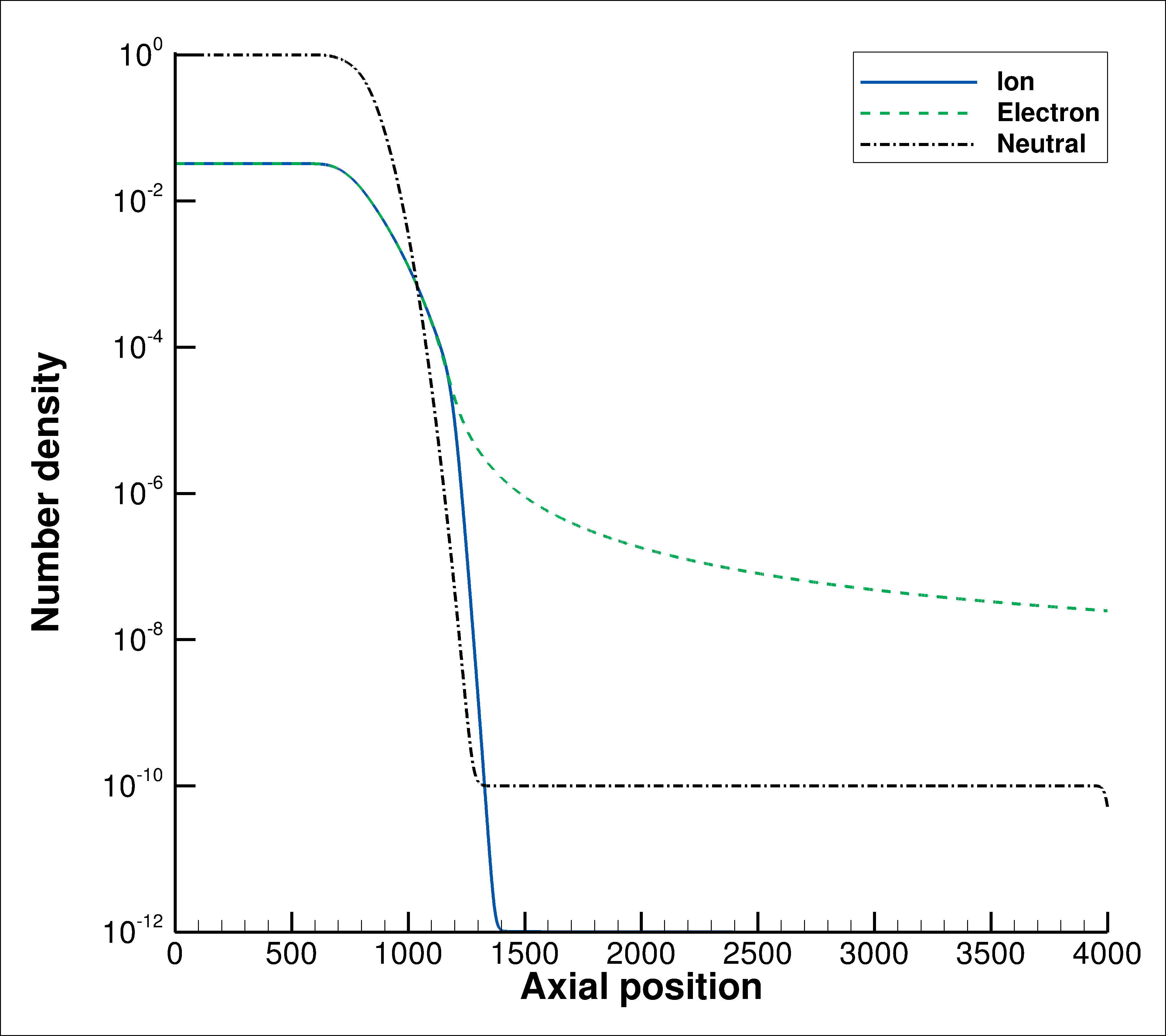}
    \quad
    (b)
    \includegraphics[trim=50 50 50 50, clip, width=0.42\linewidth,valign=t]{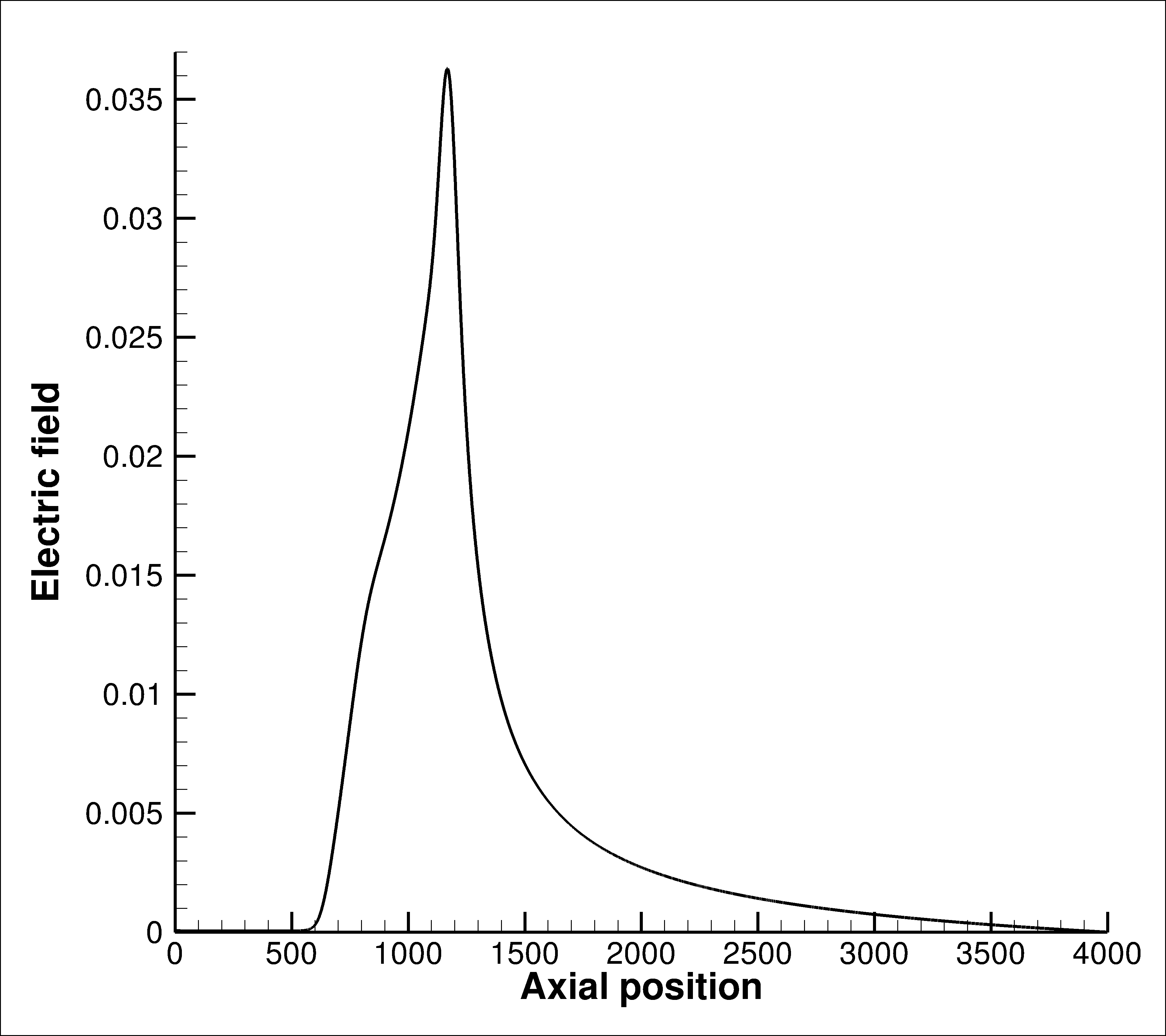}
    }
      \caption{Number density (a) and electric field (b) profiles at $t=72$. Number densities are normalized by $n_{n,1}$ similar to Fig.~\ref{fig:initprofile}.}
    \label{fig:nE_early}
    \end{figure}
    
\begin{figure}
      \centerline{
    (a)
    \includegraphics[trim=50 50 50 50, clip, width=0.42\linewidth,valign=t]{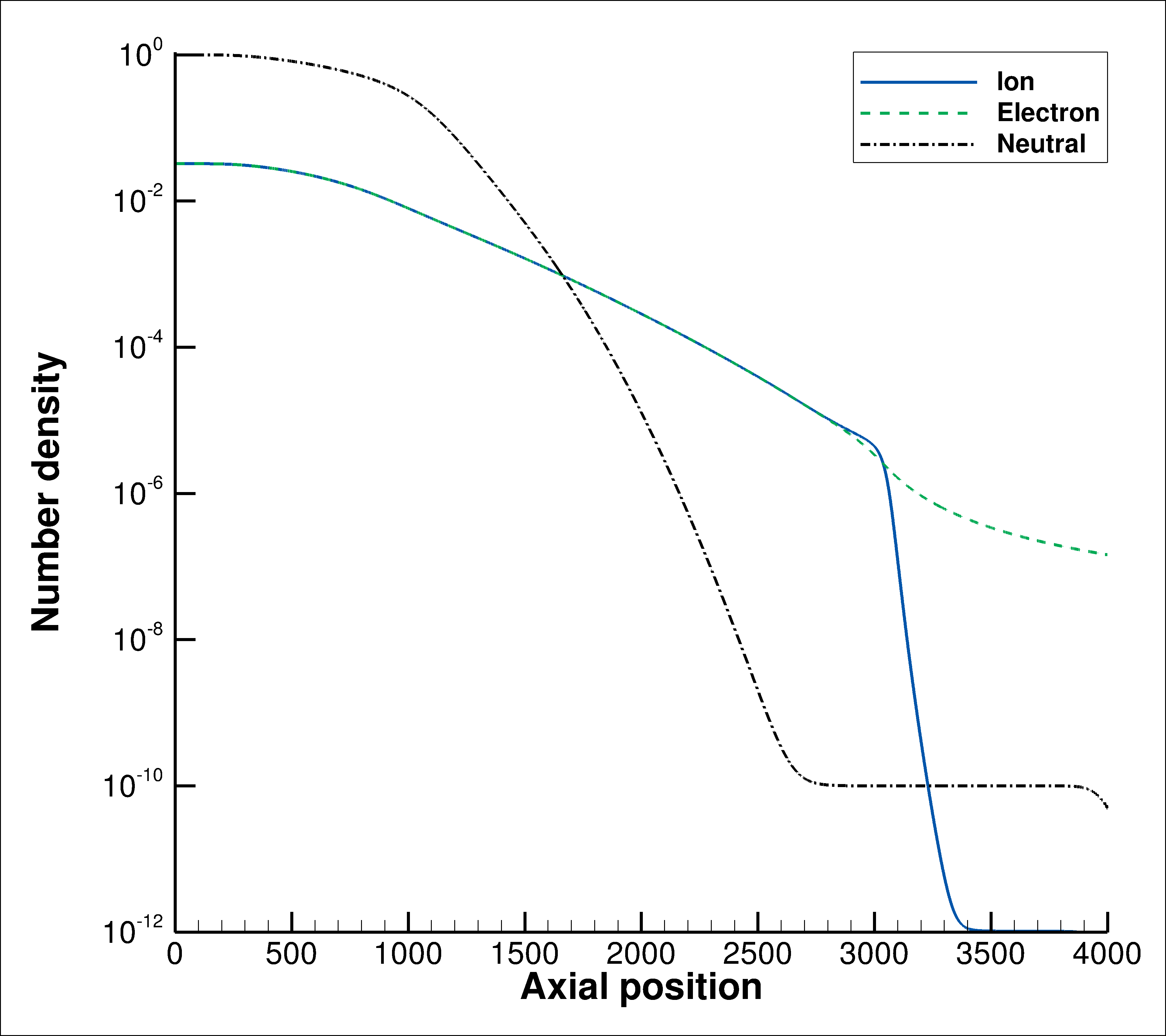}
    \quad
    (b)
    \includegraphics[trim=50 50 50 50, clip, width=0.42\linewidth,valign=t]{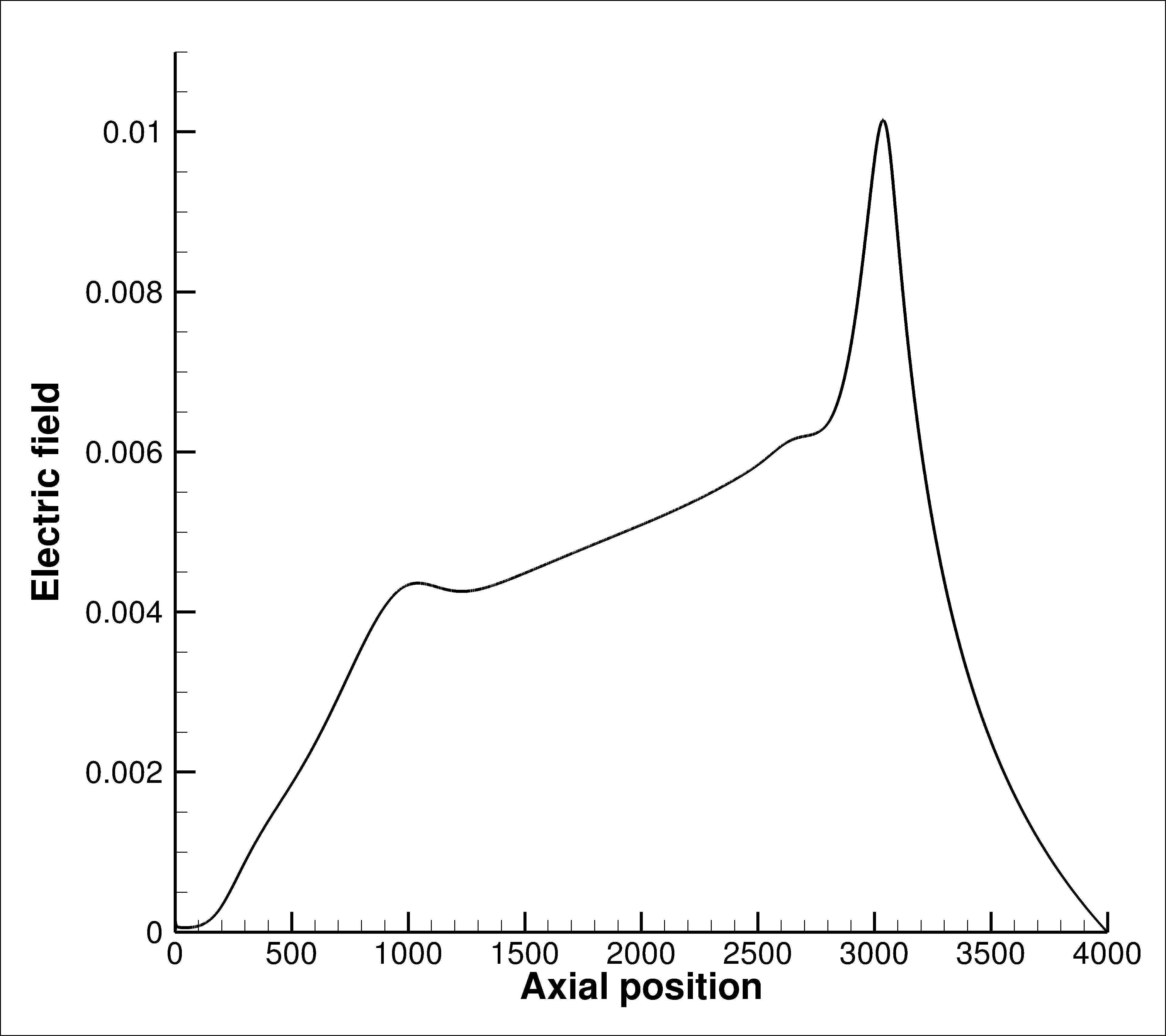}
    }
      \caption{Number density (a) and electric field (b) profiles at $t=288$. Number densities are normalized by $n_{n,1}$ similar to Fig.~\ref{fig:initprofile}.}
    \label{fig:nE_late}
    \end{figure}

    Figures~\ref{fig:nE_early} and \ref{fig:nE_late} plot the number density and electric field profiles at two time instances ($t=72 \text{ and } 288$). In the simulations corresponding to these plots, $\Delta x = 1/2$ and $\Delta v = 1/72$ for each particle type. At this grid resolution, both low-order macroscopic quantities and particle distribution functions are converged to $O(0.1\%)$~\citep{chan2022grid}. Right after initialization, electrons expand more quickly due to their higher thermal velocity and accelerate ions forward as a result. A quasineutral region of ions and electrons forms behind the expansion front. After some time has elapsed, a double layer forms at the expansion front as evidenced in Fig.~\ref{fig:nE_late}. The secondary peak in the electric field profile to the left of the primary peak at late times is a direct consequence of interparticle collisions and is not present when collisions are reduced (not shown here)~\citep{chan2022grid}. The dip in the density profiles at the right edge of the domain is due to the lack of backflow and does not significantly impact the results in the interior of the domain at the presented times.
    
\begin{figure}
      \centerline{
    (a)
    \includegraphics[trim=50 50 50 50, clip, width=0.42\linewidth,valign=t]{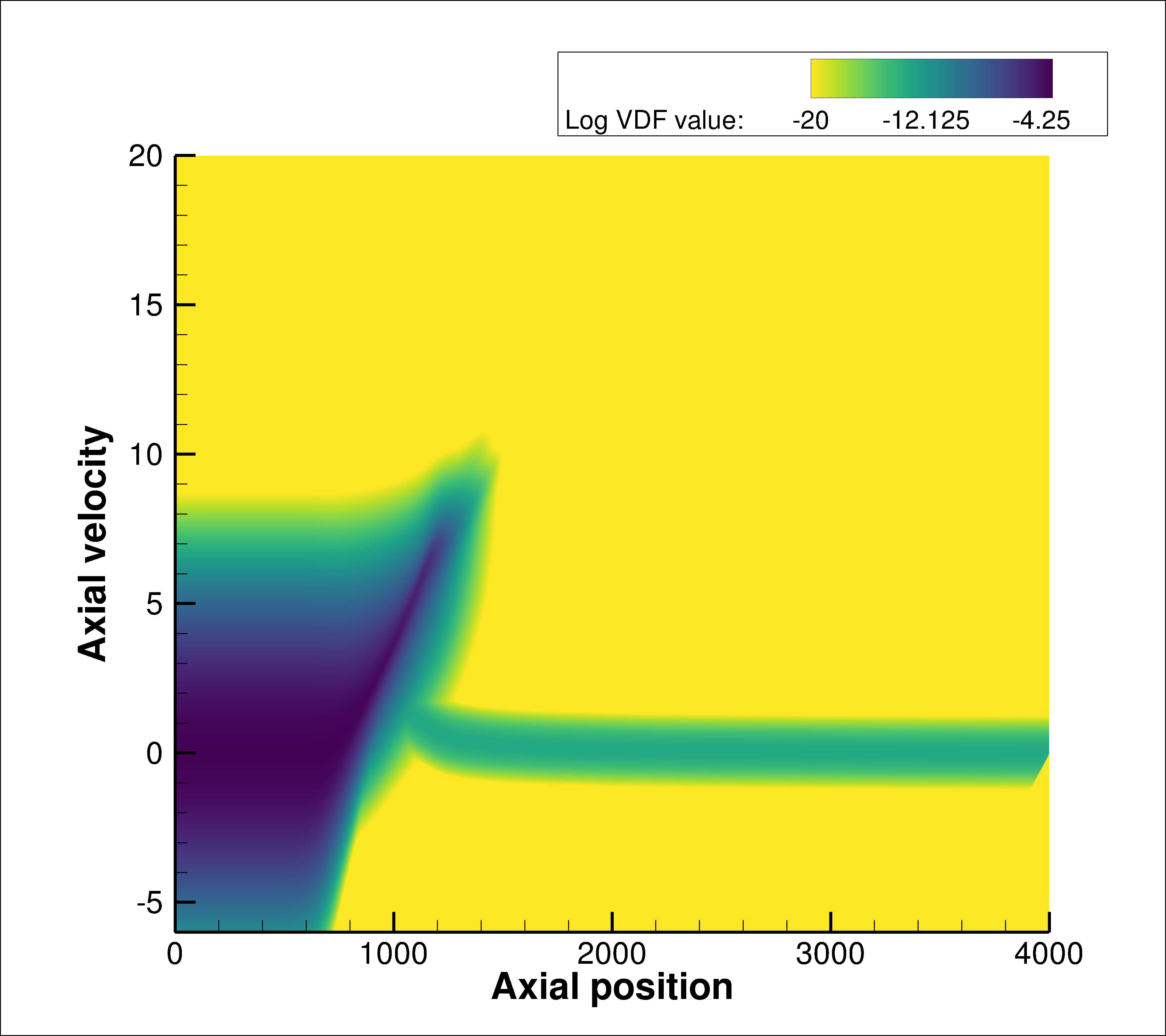}
    \quad
    (b)
    \includegraphics[trim=50 50 50 50, clip, width=0.42\linewidth,valign=t]{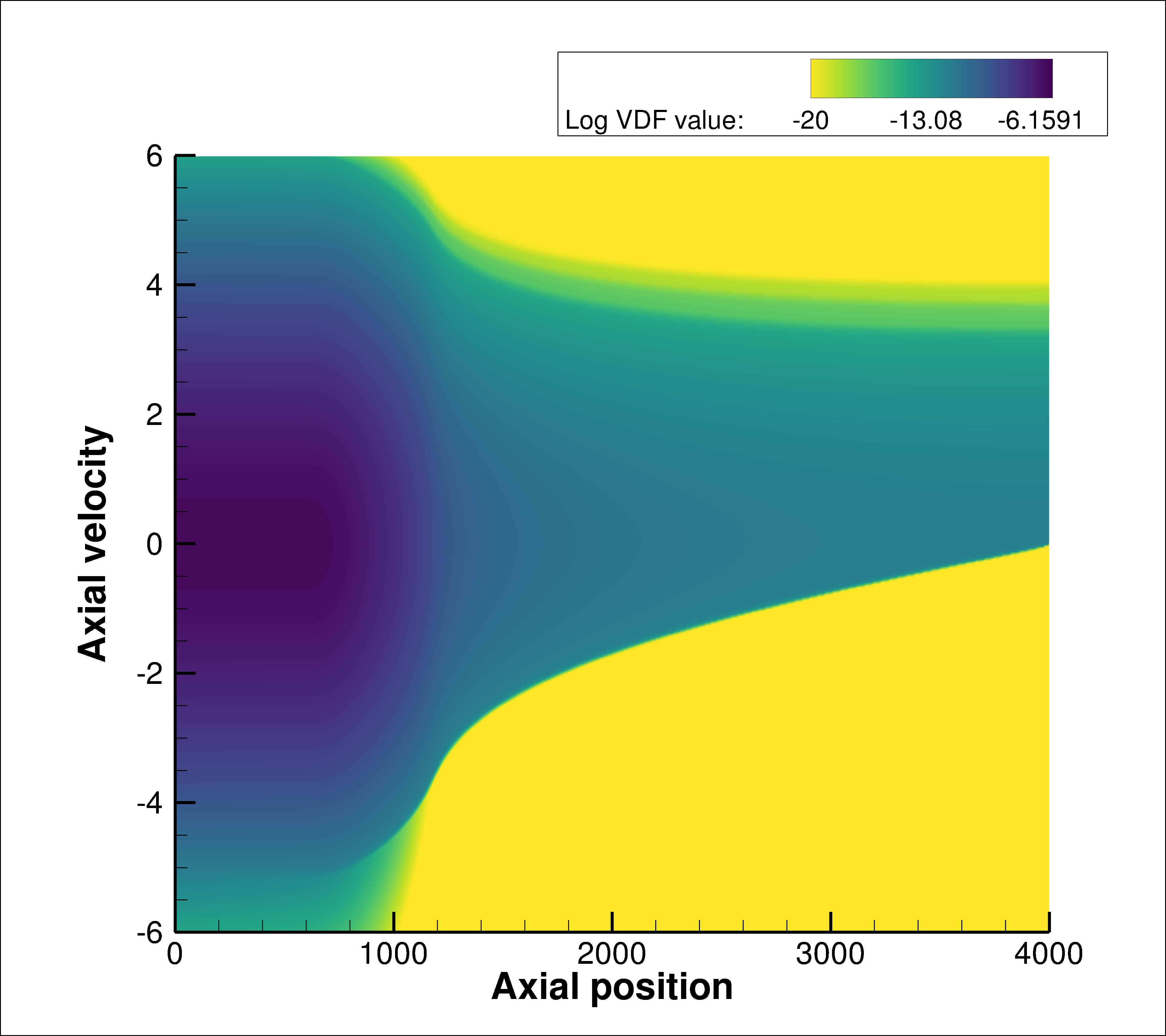}
    }
      \caption{Ion (a) and electron (b) velocity distribution functions at $t=72$. All logarithms in this and subsequent figures are taken to base 10.}
    \label{fig:vdf_early}
    \end{figure}

\begin{figure}
      \centerline{
    (a)
    \includegraphics[trim=50 50 50 50, clip, width=0.42\linewidth,valign=t]{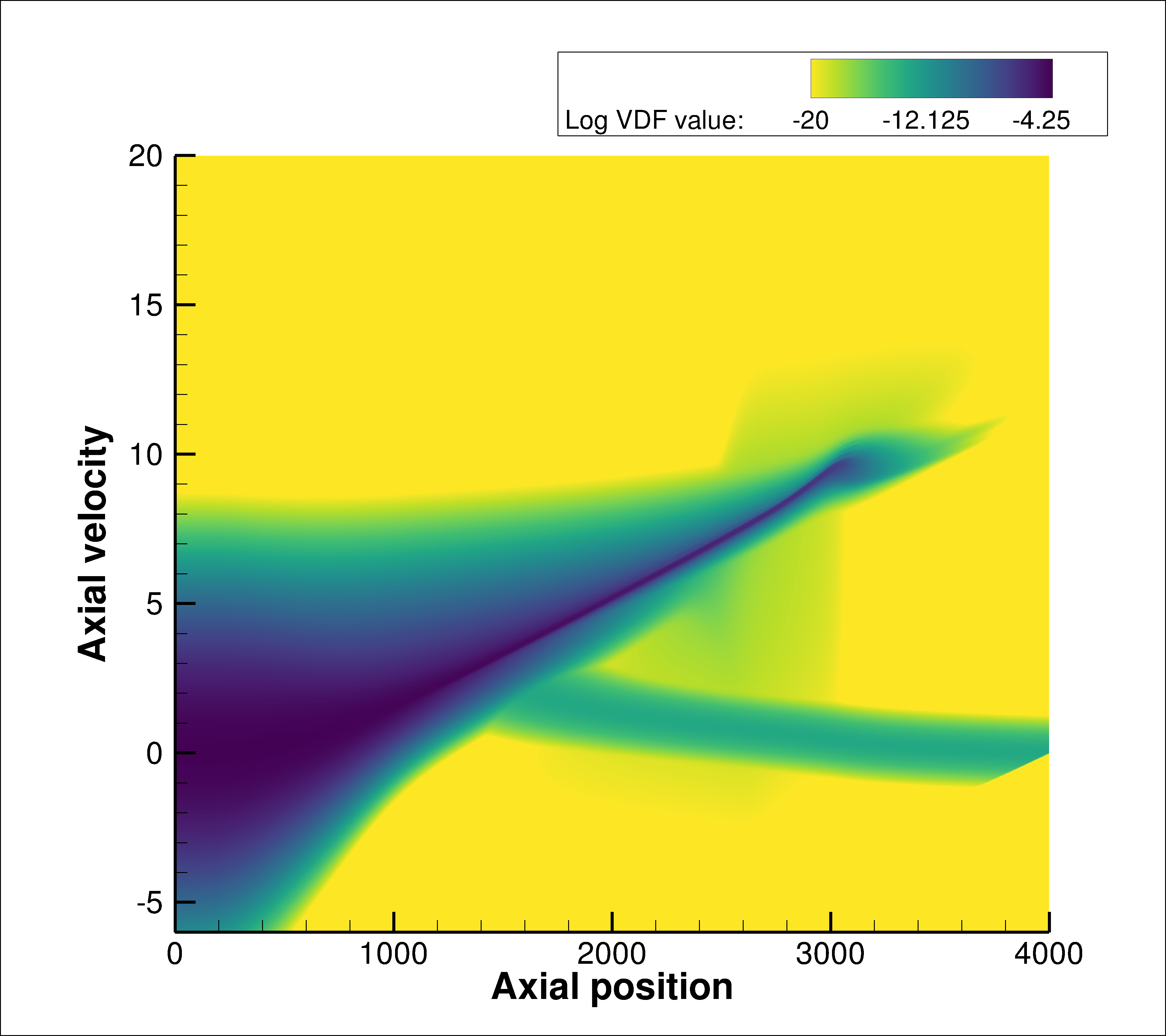}
    \quad
    (b)
    \includegraphics[trim=50 50 50 50, clip, width=0.42\linewidth,valign=t]{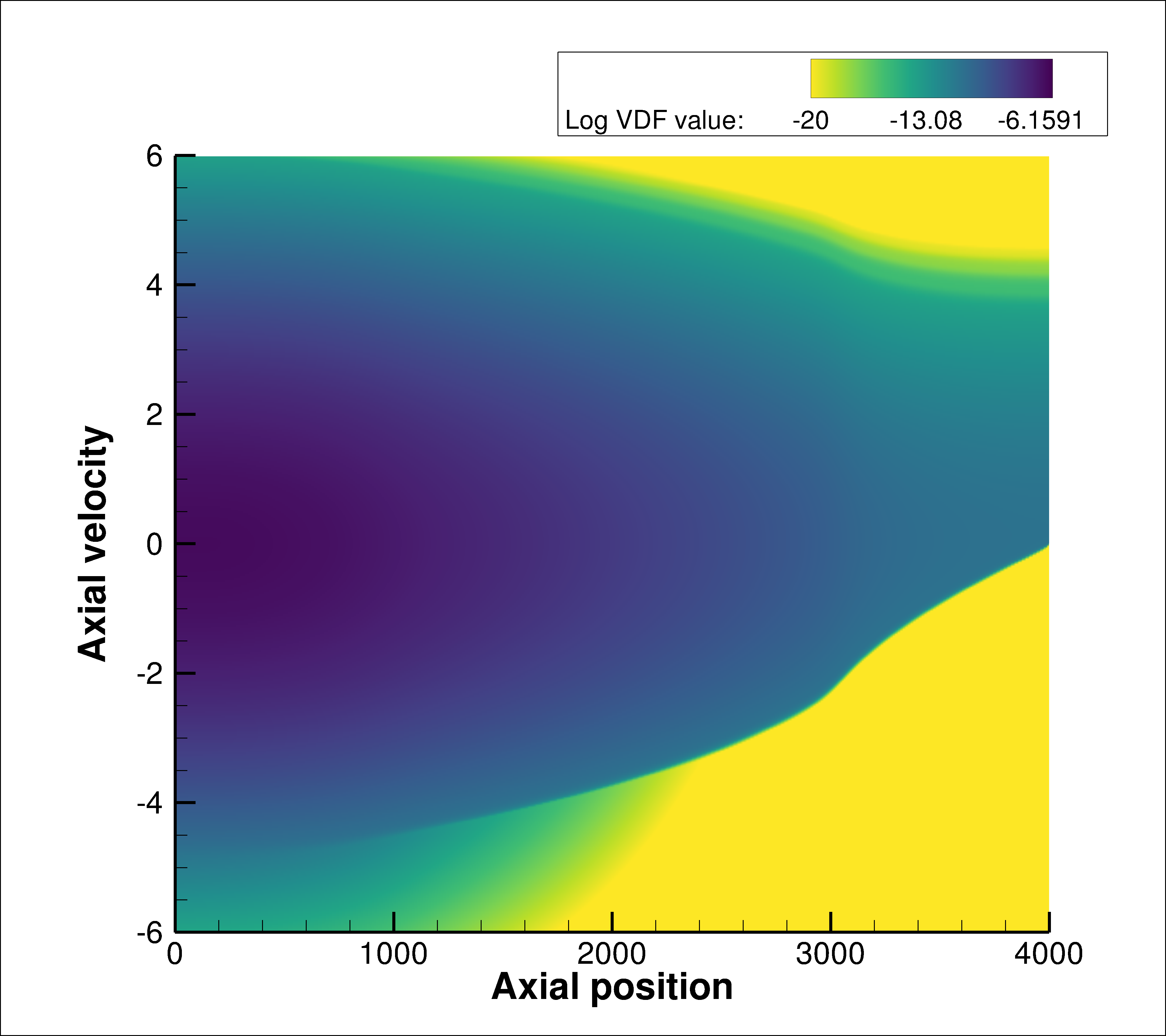}
    }
      \caption{Ion (a) and electron (b) velocity distribution functions at $t=288$.}
    \label{fig:vdf_late}
    \end{figure}

    Figures~\ref{fig:vdf_early} and \ref{fig:vdf_late} plot the ion and electron velocity distribution functions (VDFs) at the same two time instances. Hereinafter, all particle distributions are divided by the magnitude of their left reference charged-particle number density, $n_{i,1}=n_{e,1}$. The ion distributions are highly non-Maxwellian, particularly close to the expansion front, whereas the electron distributions seem to largely retain their Maxwellian behavior throughout the simulation upon a first inspection. A more detailed analysis of the distributions will be carried out in \S~\ref{sec:noneq}.
    
    \section{Results \& Discussion}\label{sec:results}
    
    \subsection{Thermodynamic nonequilibrium}\label{sec:noneq}
    
    \begin{figure}
      \centerline{
    (a)
    \includegraphics[trim=50 50 50 50, clip, width=0.42\linewidth,valign=t]{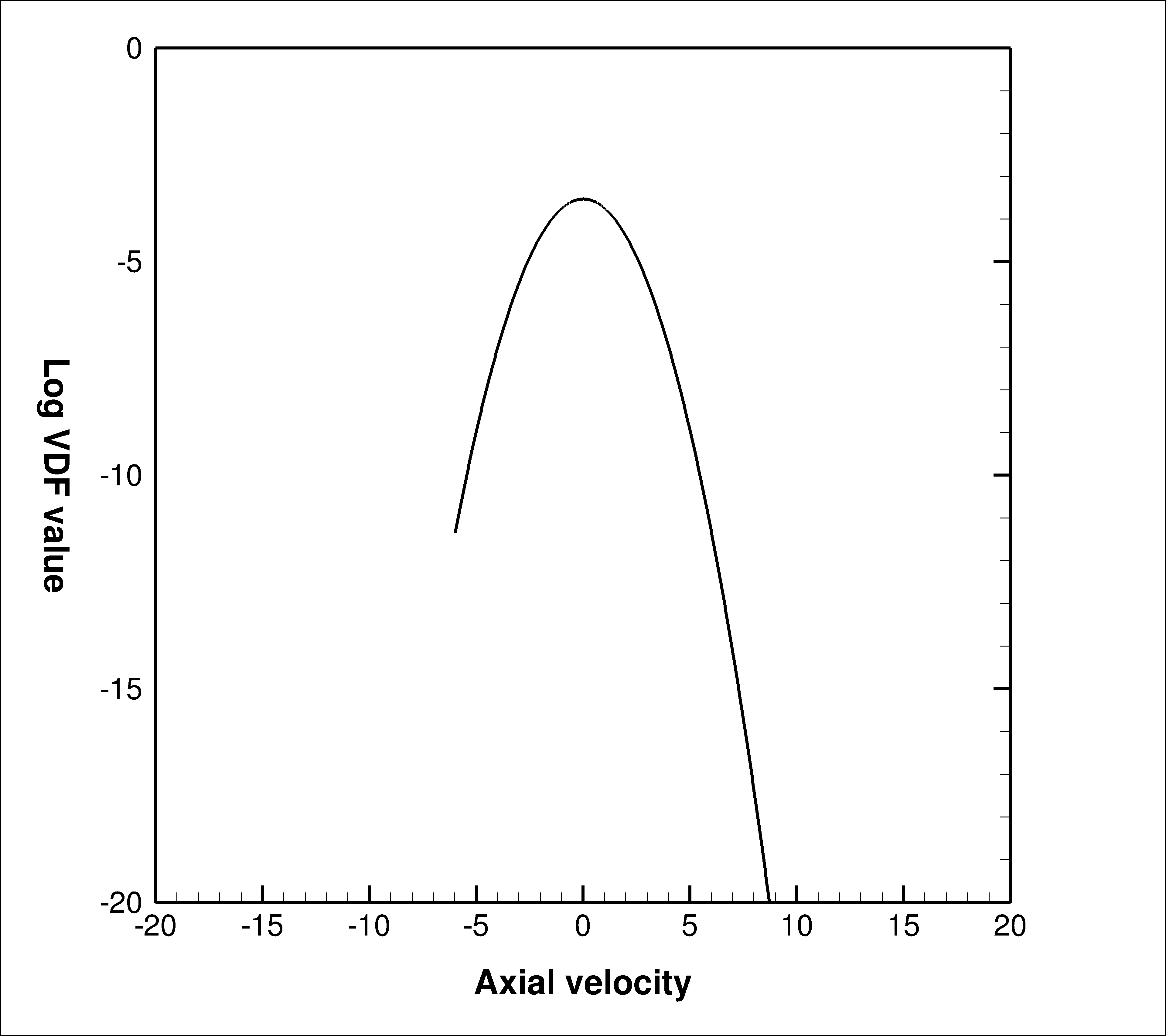}
    \quad
    (b)
    \includegraphics[trim=50 50 50 50, clip, width=0.42\linewidth,valign=t]{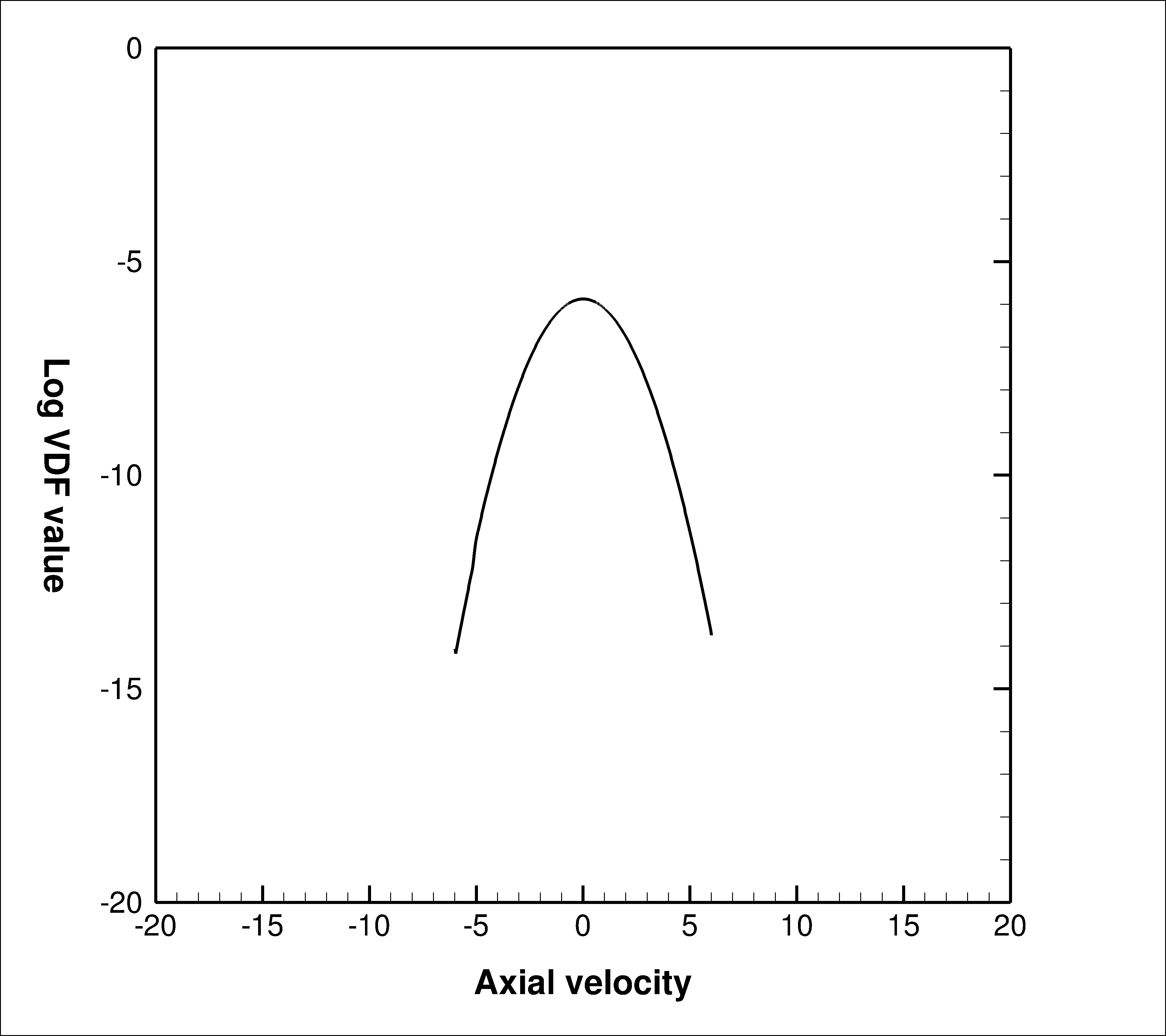}
    }
      \centerline{
    (c)
    \includegraphics[trim=50 50 50 50, clip, width=0.42\linewidth,valign=t]{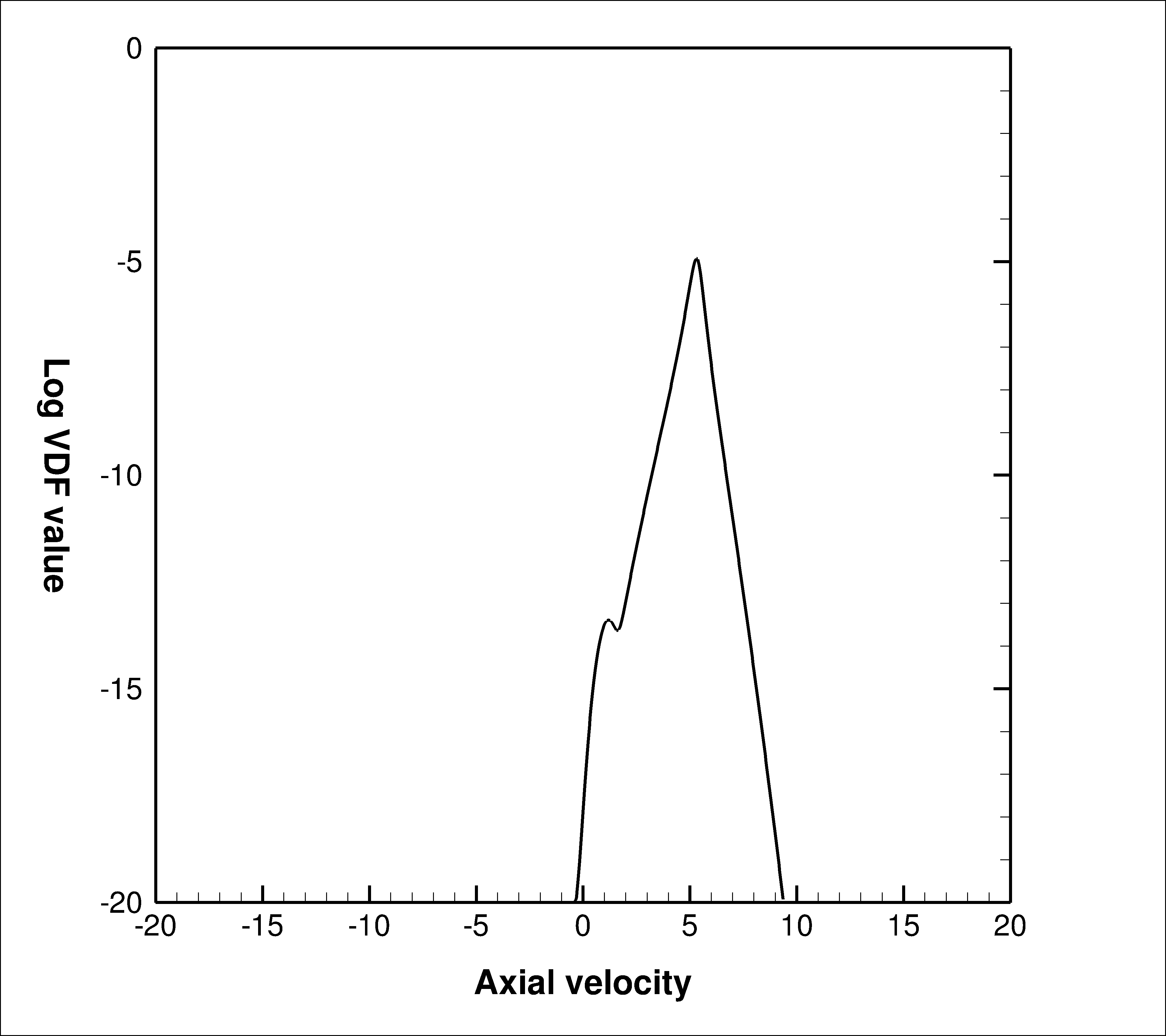}
    \quad
    (d)
    \includegraphics[trim=50 50 50 50, clip, width=0.42\linewidth,valign=t]{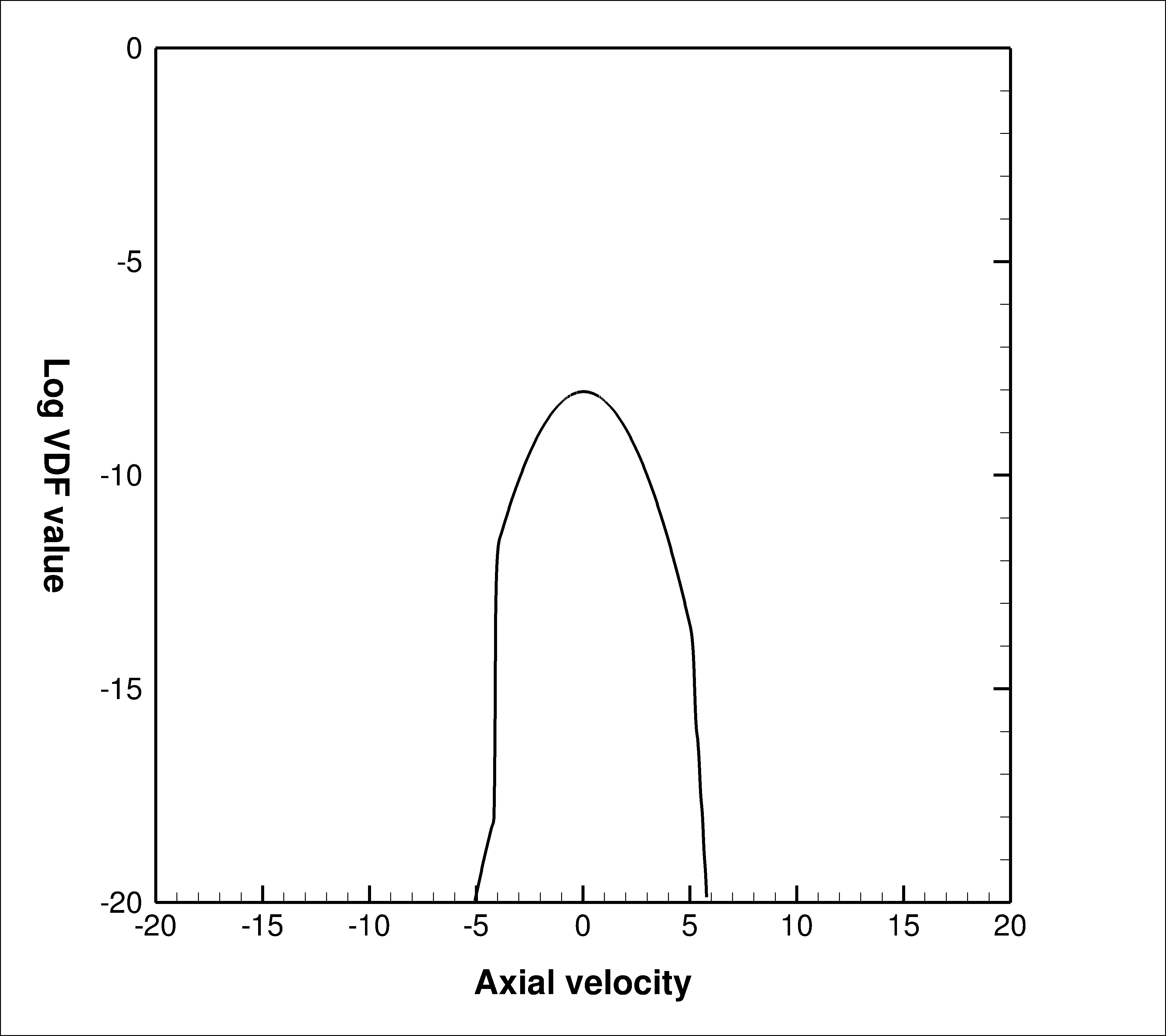}
    }
      \centerline{
    (e)
    \includegraphics[trim=50 50 50 50, clip, width=0.42\linewidth,valign=t]{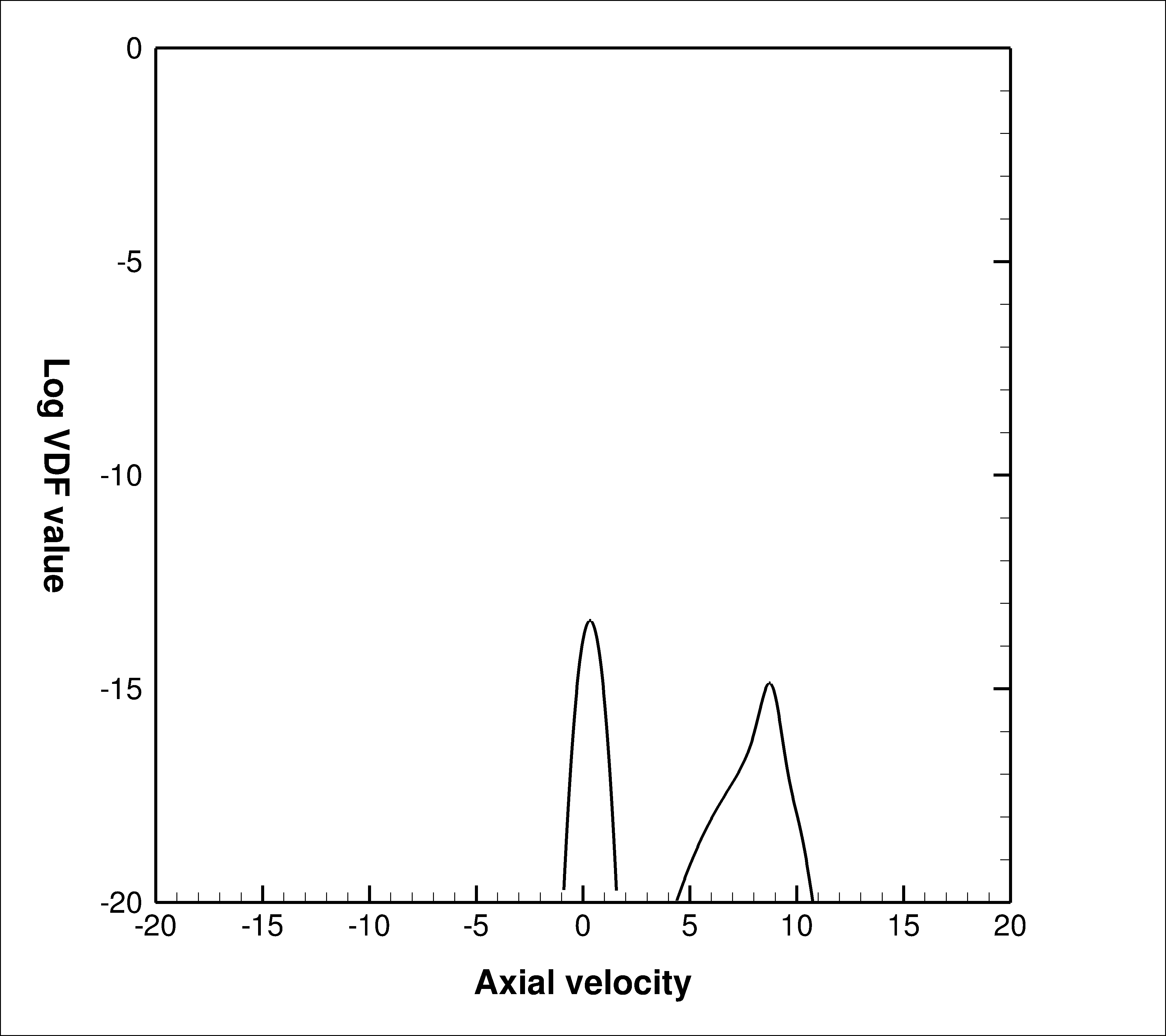}
    \quad
    (f)
    \includegraphics[trim=50 50 50 50, clip, width=0.42\linewidth,valign=t]{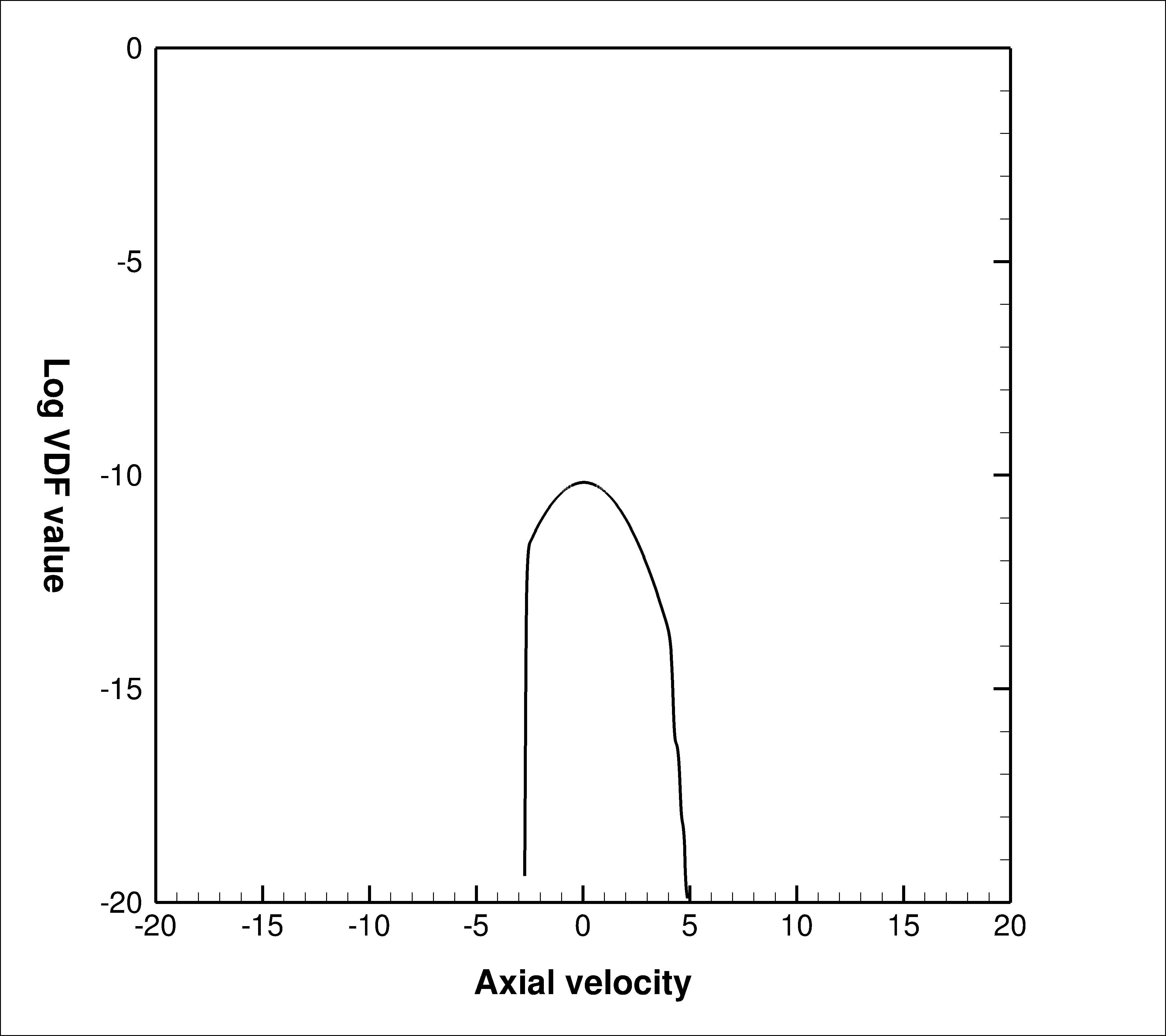}
    }
      \caption{Ion (a,c,e) and electron (b,d,f) distribution functions at $x=400$ (a,b), $x=1{,}100$ (c,d), and $x=1{,}400$ (e,f), all at $t=72$.}
    \label{fig:fv_early}
    \end{figure}
    
    \begin{figure}
      \centerline{
    (a)
    \includegraphics[trim=50 50 50 50, clip, width=0.42\linewidth,valign=t]{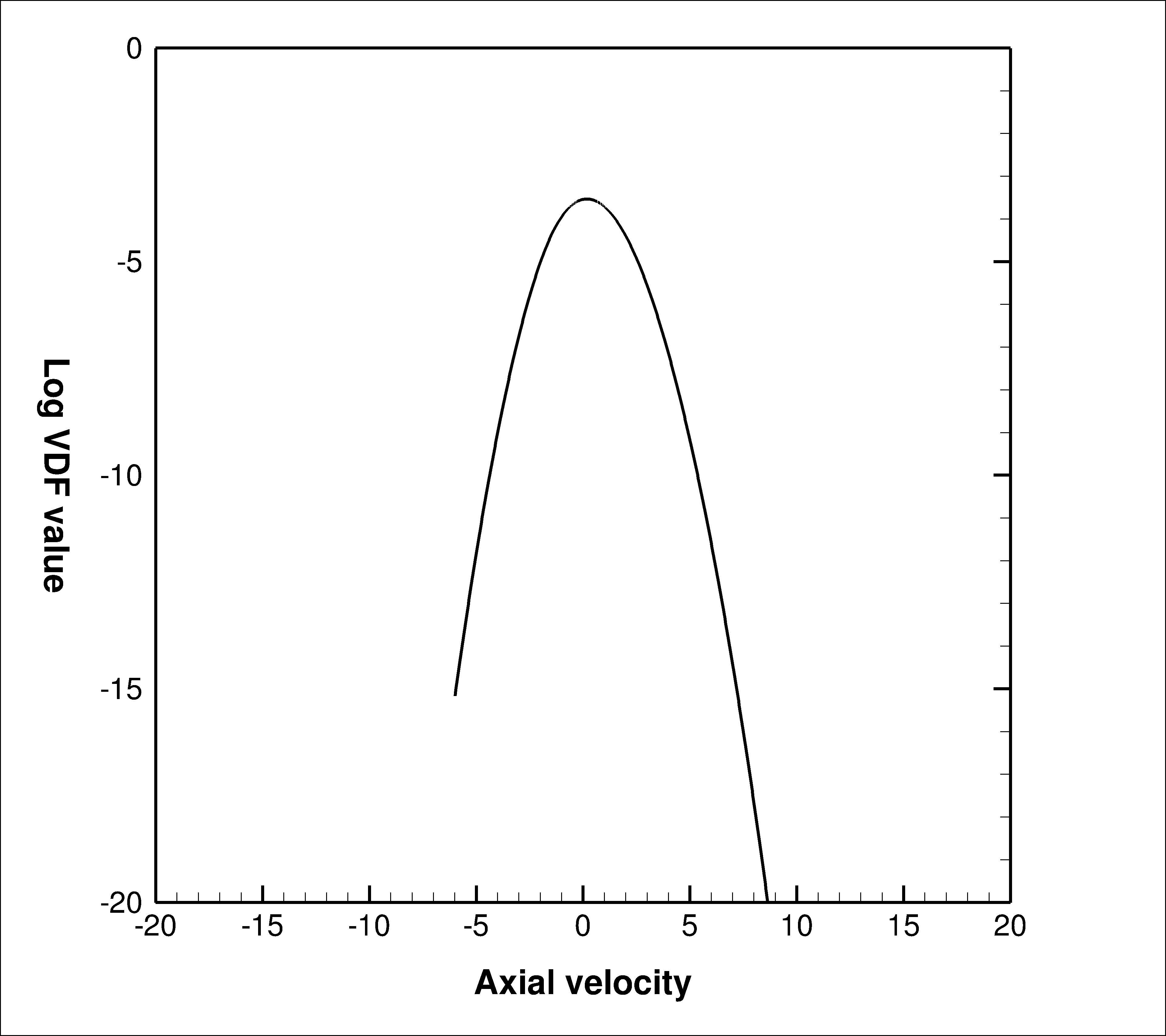}
    \quad
    (b)
    \includegraphics[trim=50 50 50 50, clip, width=0.42\linewidth,valign=t]{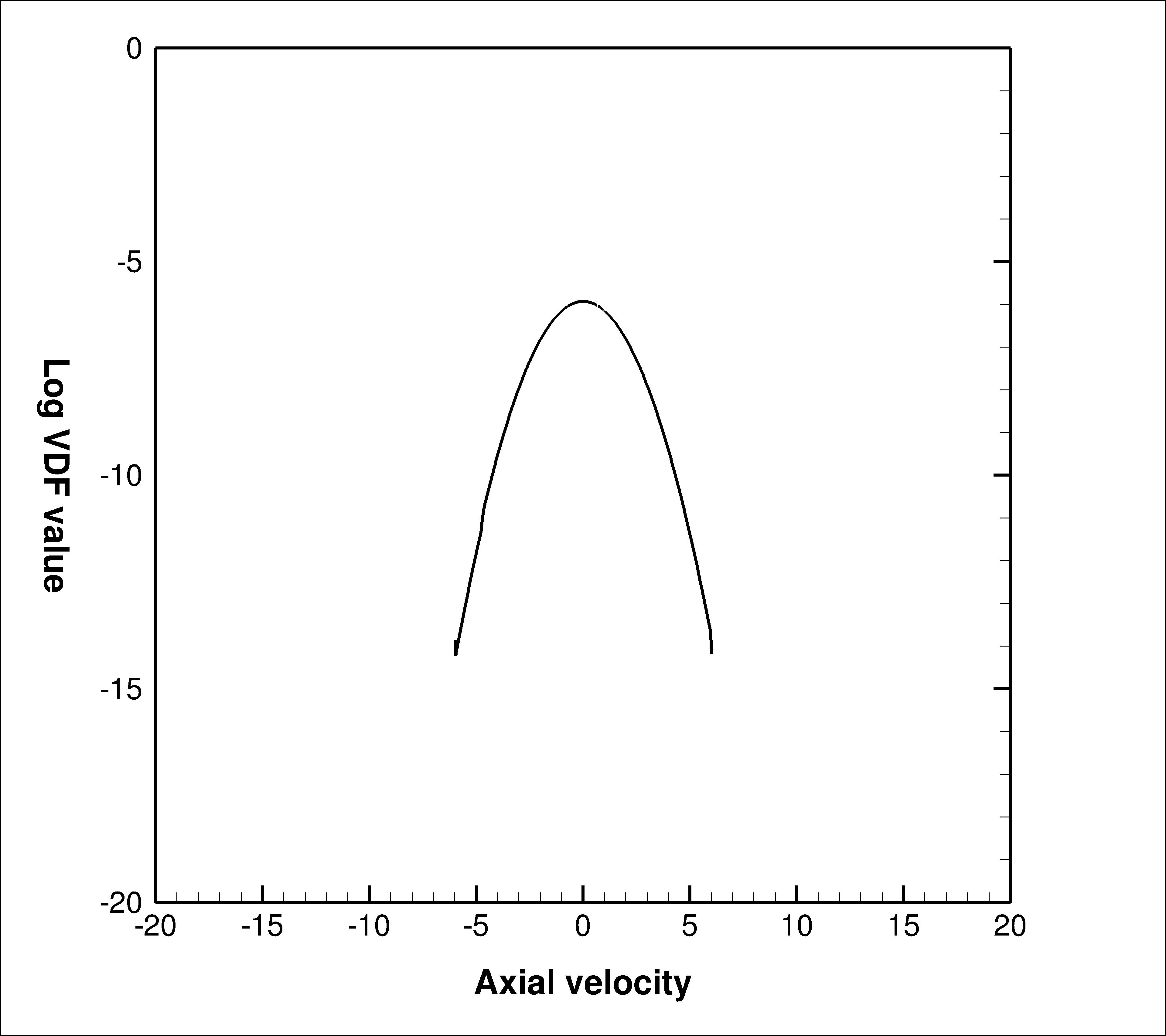}
    }
      \centerline{
    (c)
    \includegraphics[trim=50 50 50 50, clip, width=0.42\linewidth,valign=t]{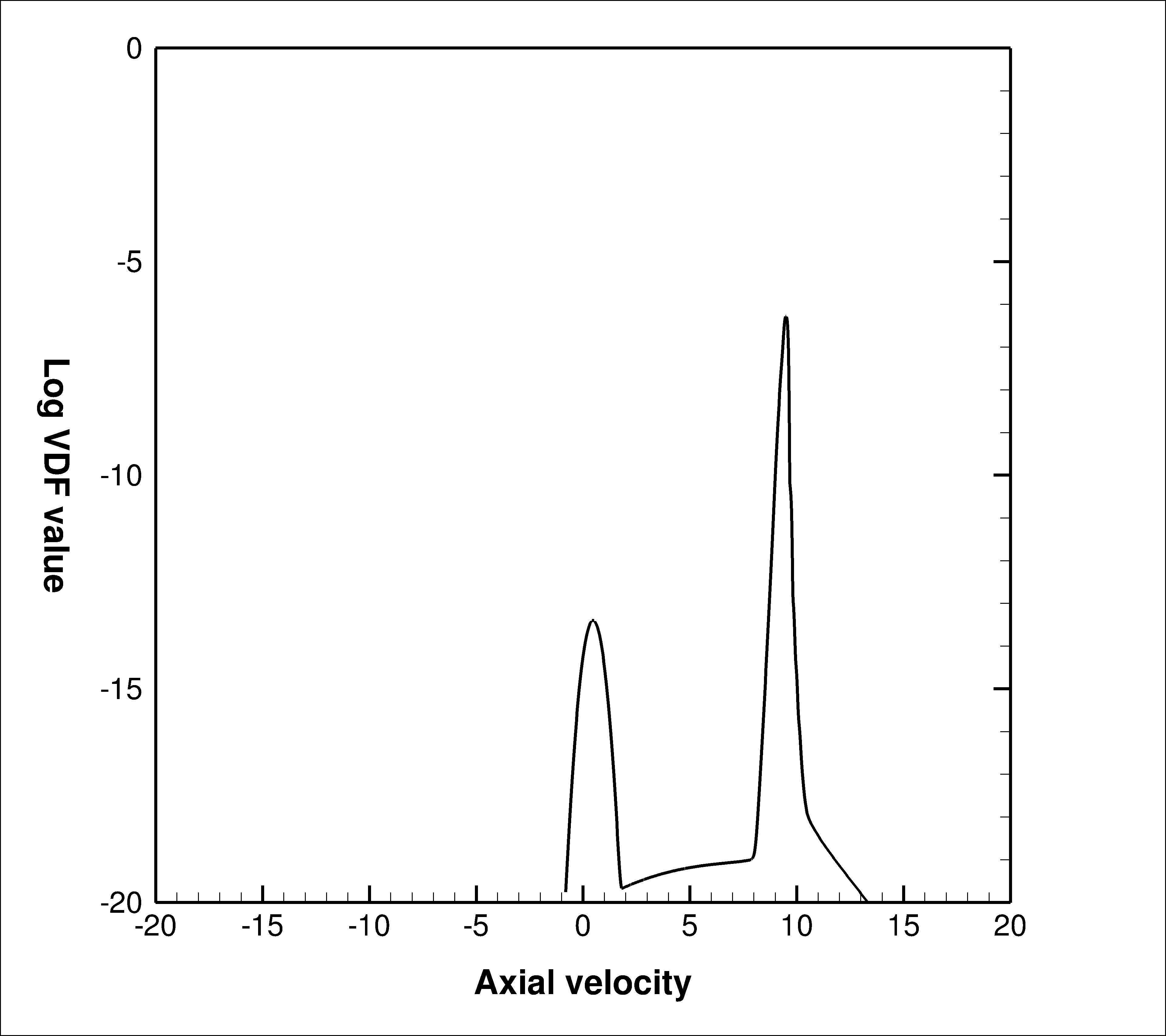}
    \quad
    (d)
    \includegraphics[trim=50 50 50 50, clip, width=0.42\linewidth,valign=t]{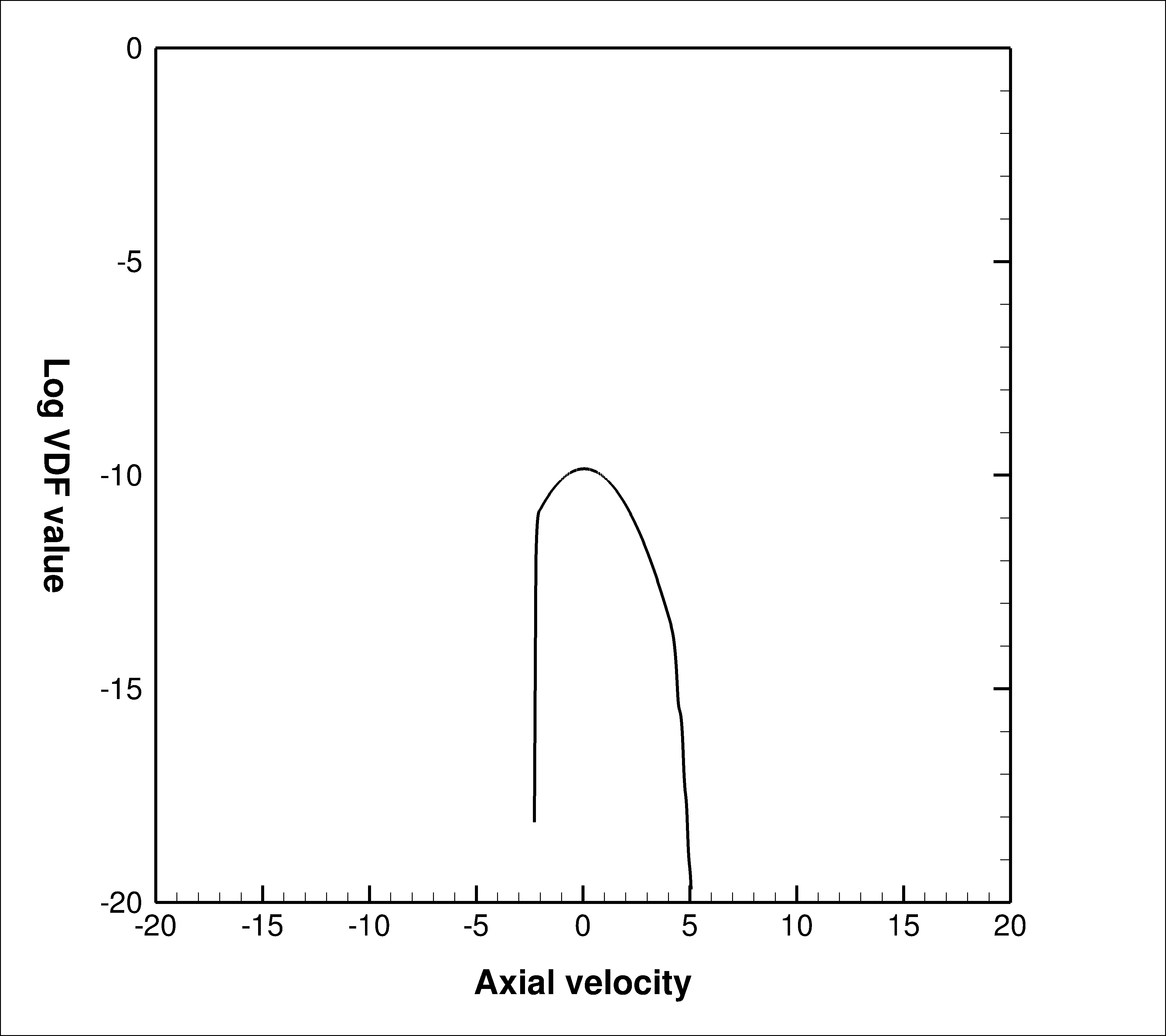}
    }
      \centerline{
    (e)
    \includegraphics[trim=50 50 50 50, clip, width=0.42\linewidth,valign=t]{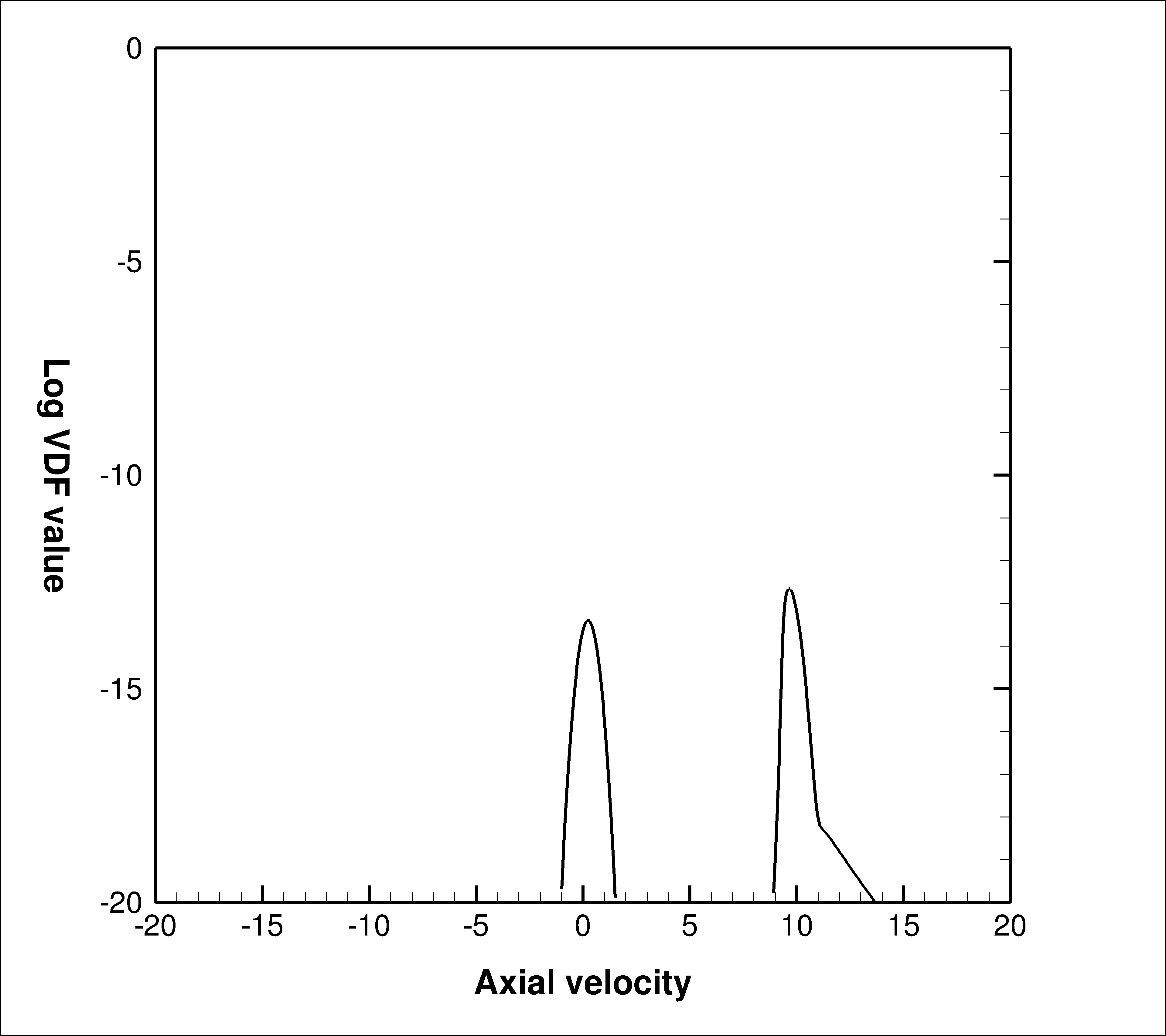}
    \quad
    (f)
    \includegraphics[trim=50 50 50 50, clip, width=0.42\linewidth,valign=t]{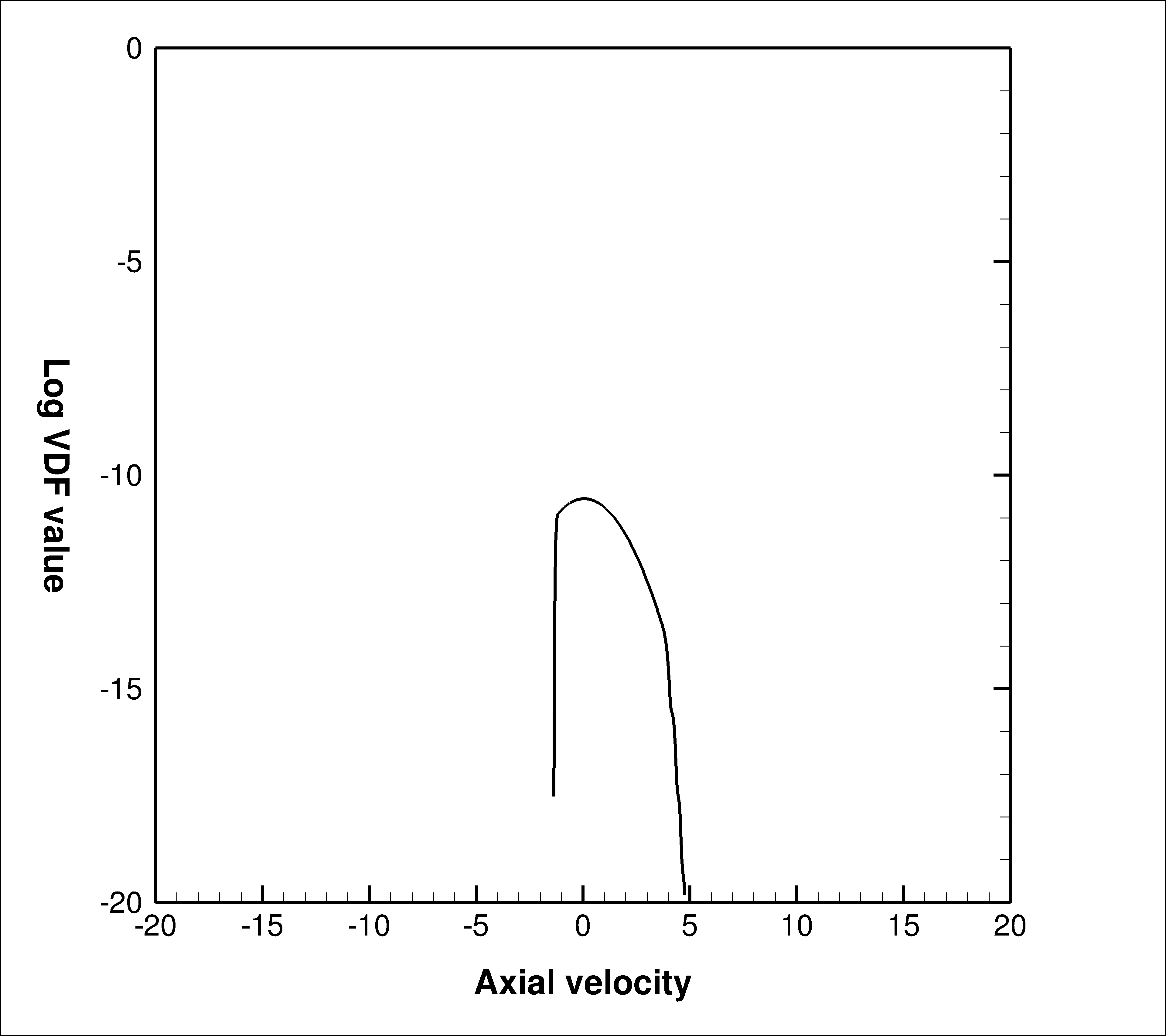}
    }
      \caption{Ion (a,c,e) and electron (b,d,f) distribution functions at $x=400$ (a,b), $x=3{,}000$ (c,d), and $x=3{,}300$ (e,f), all at $t=288$.}
    \label{fig:fv_late}
    \end{figure}
    
    The determination of thermodynamic nonequilibrium begins with an assessment of the deviation of the particle distribution function from a Maxwellian. Figures~\ref{fig:fv_early} and \ref{fig:fv_late} plot $x$-slices of the ion and electron distribution functions at an upstream axial location, near the double-layer location, and slightly downstream of the double layer at the same two time instances considered in \S~\ref{sec:dynamics}. On these linear--logarithmic axes, a Maxwellian distribution presents as a parabola. Panels (a) and (b) of both figures indicate that the distribution remains largely Maxwellian in the plume interior as expected. Panels (c) and (e) of both figures confirm that the downstream ion distribution is highly non-Maxwellian, with the left and right peaks arising from the background and fast ion populations, respectively, alongside the presence of broadened tails due to interparticle collisions. Panels (d) and (f) of both figures indicates that while the electron distribution remains close to a Maxwellian as suggested by Figs.~\ref{fig:vdf_early} and \ref{fig:vdf_late}, one may observe non-Maxwellian features arising from the preference for forward-moving electrons at the expansion front, as well as the retardation of fast electrons by the fast ions close behind. The presence of these nonequilibrium ion and electron distribution features suggests that a simulation approach that is not completely kinetic in nature may not fully predict such nuances in the ion and electron dynamics, as well as their downstream physical implications.

    \begin{figure}
      \centerline{
    (a)
    \includegraphics[trim=50 50 50 50, clip, width=0.42\linewidth,valign=t]{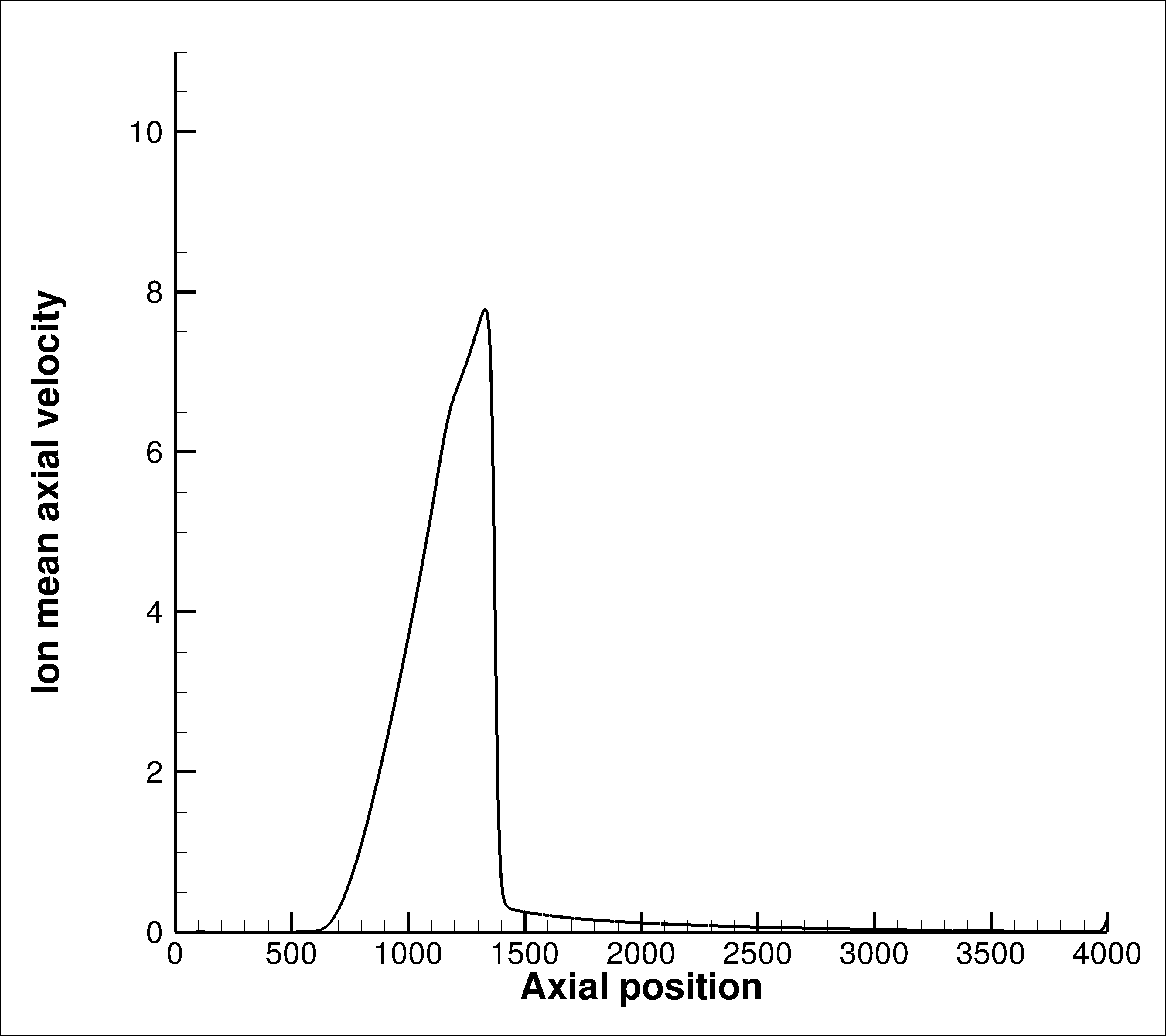}
    \quad
    (b)
    \includegraphics[trim=50 50 50 50, clip, width=0.42\linewidth,valign=t]{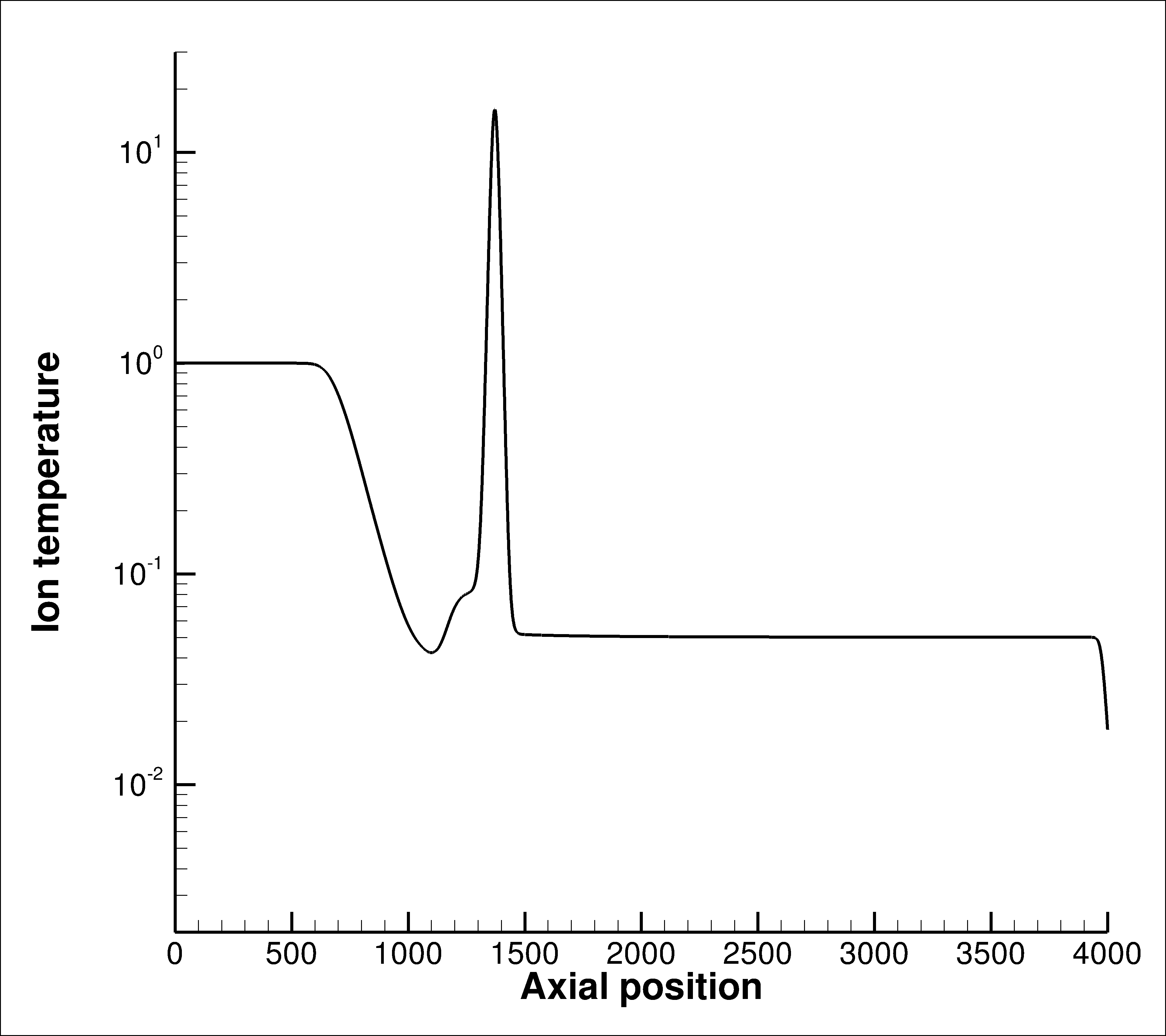}
    }
      \caption{Ion mean axial velocity (a) and temperature (b) profiles at $t=72$. The latter profile is always proportional to the random kinetic energy but corresponds to the thermodynamic temperature only in regions of local thermodynamic equilibrium, as discussed in the text.}
    \label{fig:vT_early}
    \end{figure}
    
    \begin{figure}
      \centerline{
    (a)
    \includegraphics[trim=50 50 50 50, clip, width=0.42\linewidth,valign=t]{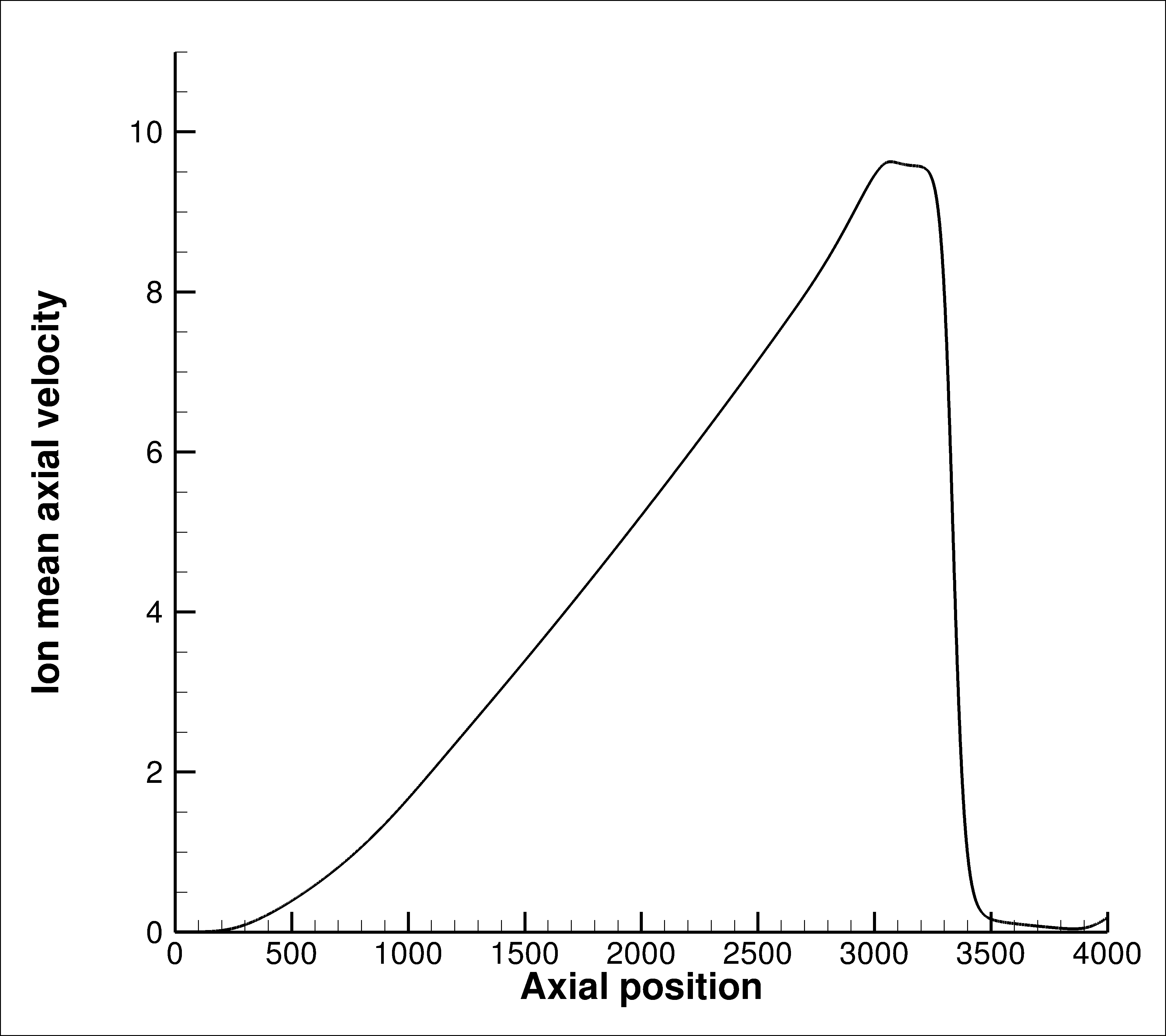}
    \quad
    (b)
    \includegraphics[trim=50 50 50 50, clip, width=0.42\linewidth,valign=t]{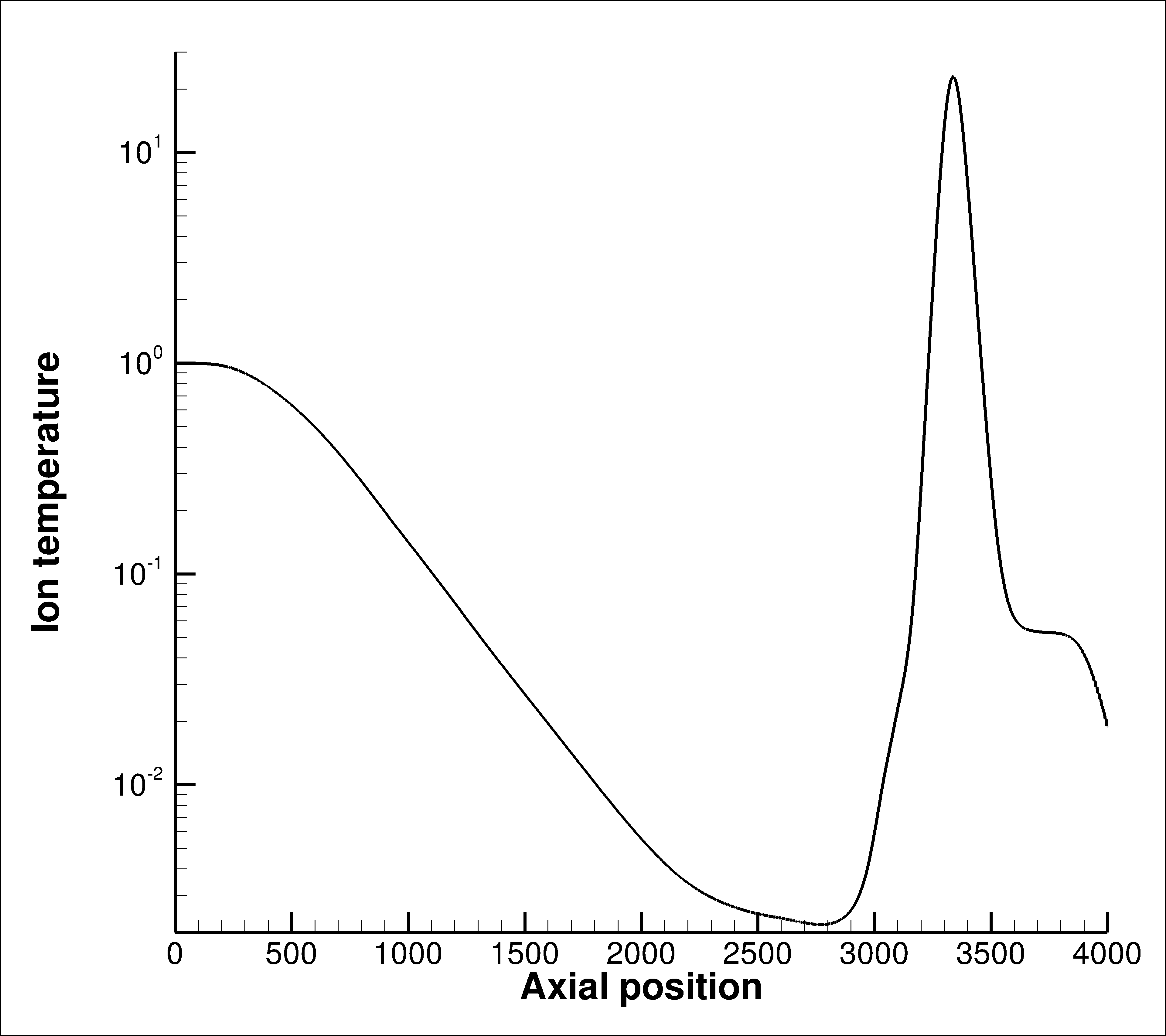}
    }
      \caption{Ion mean axial velocity (a) and temperature (b) profiles at $t=288$. See the caption of Fig.~\ref{fig:vT_early} for a disclaimer on the interpretation of the temperature.}
    \label{fig:vT_late}
    \end{figure}
    
    To gain more insight into the nonequilibrium nature of plume expansion, we examine the higher-order moments of the ion distribution function. Figures~\ref{fig:vT_early} and \ref{fig:vT_late} plot the ion mean axial velocity and temperature profiles at the same two time instances. Bearing in mind the non-Maxwellian nature of the ion distribution beyond the plume interior, the ``temperature" is more rigorously interpreted as a quantity proportional to the difference between the total and bulk (coherent) kinetic energies of the ions and thus represents their random (thermal) kinetic energy. Their mean axial velocity increases downstream and reaches a value of $O(10)$ thermal velocities near the expansion front. Conversely, their random kinetic energy decreases downstream by one to three orders of magnitude, although the ion ``temperature" then reaches a peak of $O(10)$ times the left reference temperature beyond the expansion front. These observations suggest that preceding the expansion front, the increased coherent kinetic energy due to ambipolar acceleration has not had sufficient time to be thermalized into random kinetic energy through interparticle collisions, bearing in mind that $\lambda_\text{mfp}\sim L$ and $t_\text{mfp}\sim t_\text{sim}$, and a fluid approximation assuming local thermodynamic equilibrium is evidently inappropriate in this region. Ahead of the expansion front, ambipolar acceleration is no longer sufficient to accelerate a significant number of ions to a speed of the magnitude of several thermal velocities, while the remaining accelerated ions still constitute a nonnegligible amount of incoherent kinetic energy that is considerable relative to their coherent kinetic energy. Note that panels (c) and (d) in Figs.~\ref{fig:fv_early} and \ref{fig:fv_late} are close to the location of the velocity peak and temperature trough, while panels (e) and (f) in the same figures are beyond the velocity peak and close to the temperature peak. Panel (e), in particular, suggests that the large ratio of incoherent to coherent kinetic energy may be a consequence of the locally bi-Maxwellian nature of the ion velocity distribution function. Once again, the artifacts at the right domain edge are due to the lack of backflow and do not significantly impact the results in the domain interior at the presented times.
    
    \begin{figure}
      \centerline{
    (a)
    \includegraphics[trim=50 50 50 50, clip, width=0.42\linewidth,valign=t]{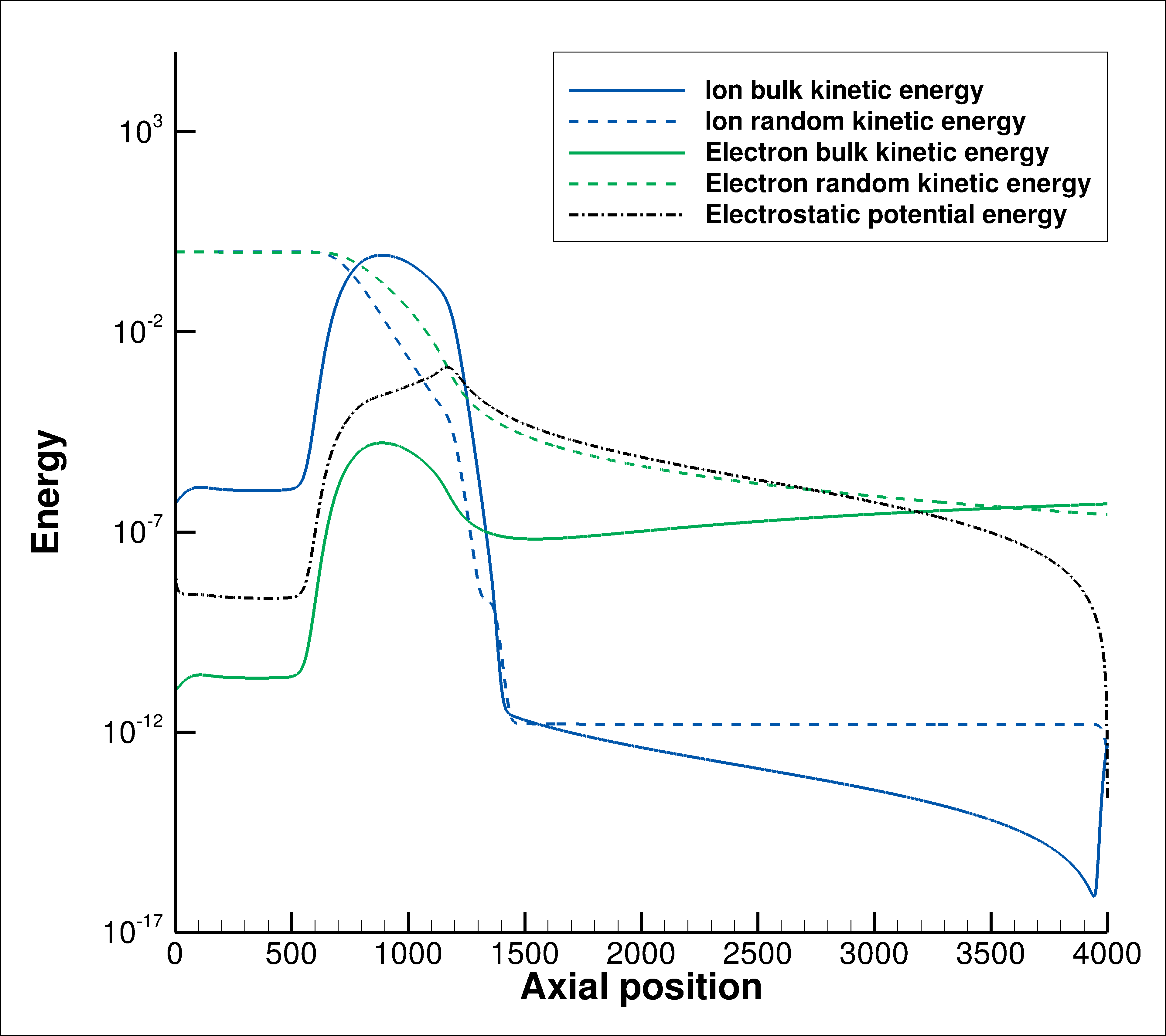}
    \quad
    (b)
    \includegraphics[trim=50 50 50 50, clip, width=0.42\linewidth,valign=t]{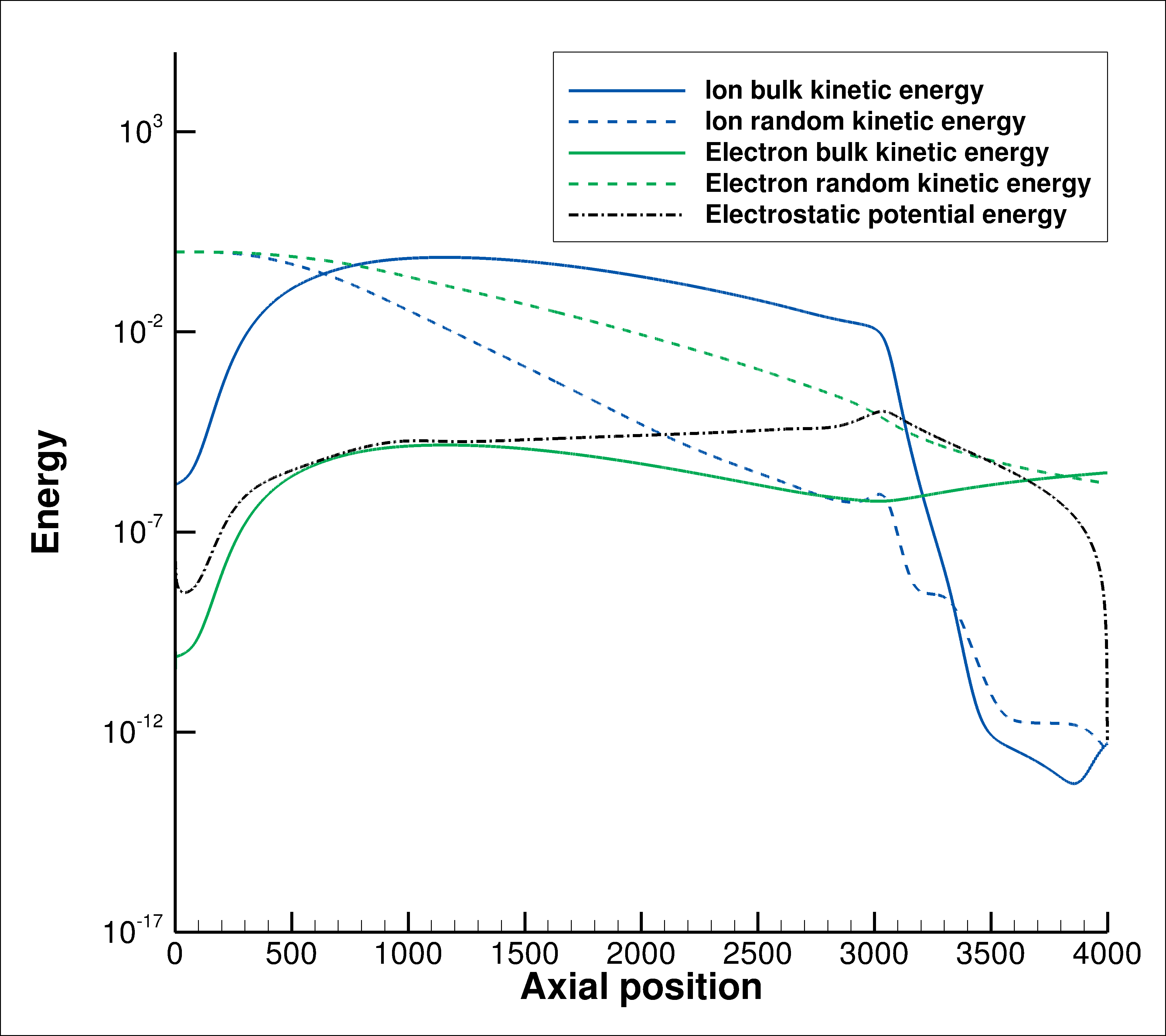}
    }
      \caption{Energetics at $t=72$ (a) and $t=288$ (b). All energies are normalized by the inlet random ion kinetic energy, $n_{i,1}k_B T_1/2$. The bulk and random kinetic energies are computed as $n_* m_* v_*^2/2$ and $n_* k_B T_*/2$, respectively, while the electrostatic potential energy is computed as $\varepsilon_0 E^2/2$.}
    \label{fig:energy}
    \end{figure}
    
    We conclude this discussion of thermodynamic nonequilibrium with an analysis of the complete energetics of the system. The ion and electron bulk and random kinetic energy profiles, as well as the electrostatic potential energy profile, are plotted in Fig.~\ref{fig:energy}. Three observations may be made. First, the bulk kinetic energy of the ions is much larger than their random kinetic energy in a region between the inlet and expansion front, corresponding approximately to the region of thermal nonequilibrium identified earlier. Second, the bulk kinetic energy of the electrons never overwhelms their random kinetic energy unlike in the case of ions, which corroborates the greater tendency of electrons to thermodynamic equilibrium throughout the domain represented by their near-Maxwellian distribution. Third, the electrostatic potential energy is relatively insignificant prior to the expansion front but is a dominant contributor after the front, suggesting that electrostatic forces are a primary driver of the energetics as one would expect. This energy is matched by the electron random kinetic energy immediately ahead of the front as these forces are themselves a direct consequence of the faster thermal velocity of electrons relative to ions. This congruence diverges downstream as one departs further from the location of maximum charge separation and the most dominant electrostatic forces become more long-range in nature. The rich interplay of different energies at various plume locations lends itself naturally to a discussion of the spatial inhomogeneity of the system, which is also reflected in the momentum flux.

    \subsection{Momentum flux and spatial inhomogeneity}\label{sec:momflux}
    
    \begin{figure}
      \centerline{
    (a)
    \includegraphics[trim=50 50 50 50, clip, width=0.42\linewidth,valign=t]{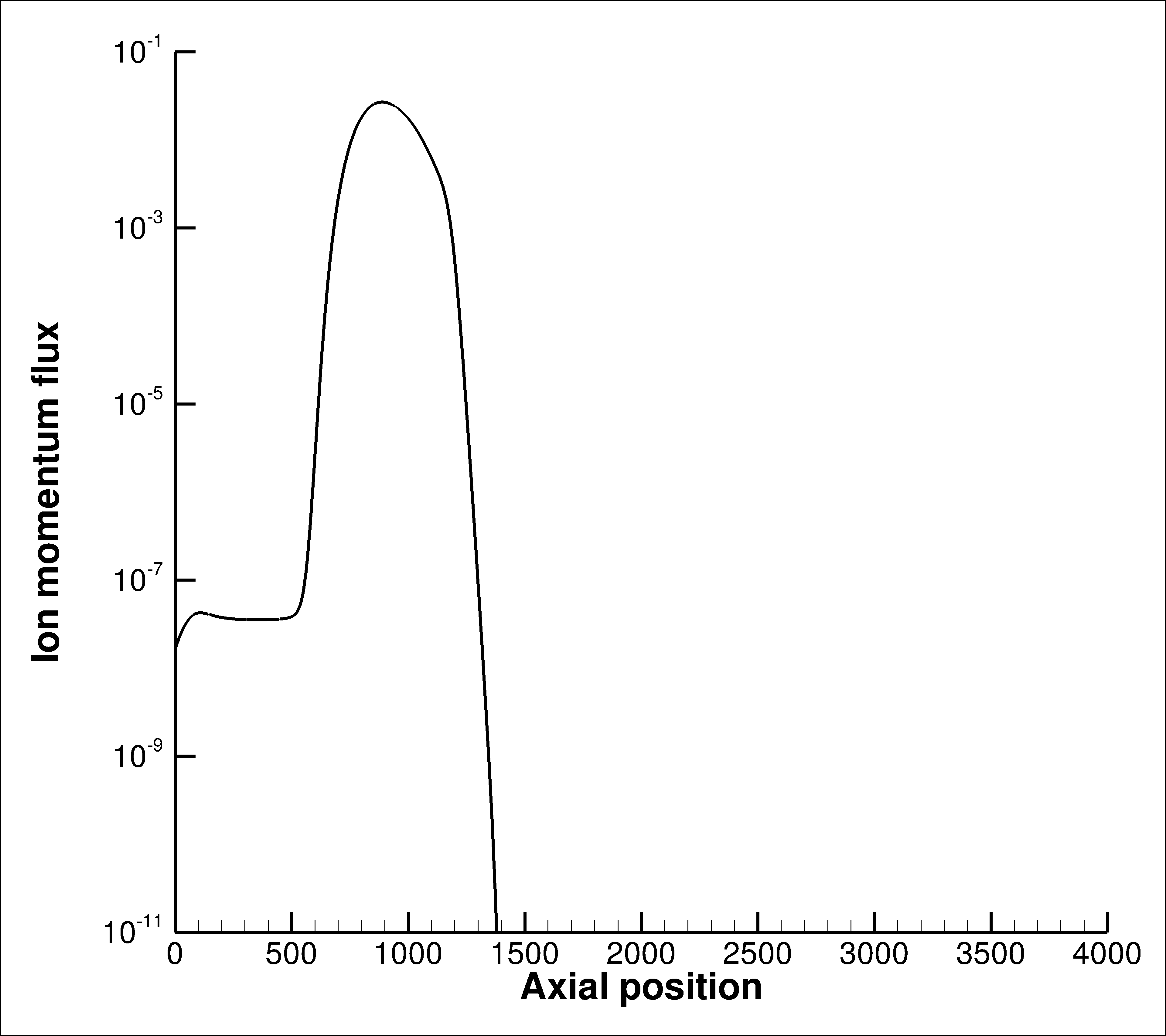}
    \quad
    (b)
    \includegraphics[trim=50 50 50 50, clip, width=0.42\linewidth,valign=t]{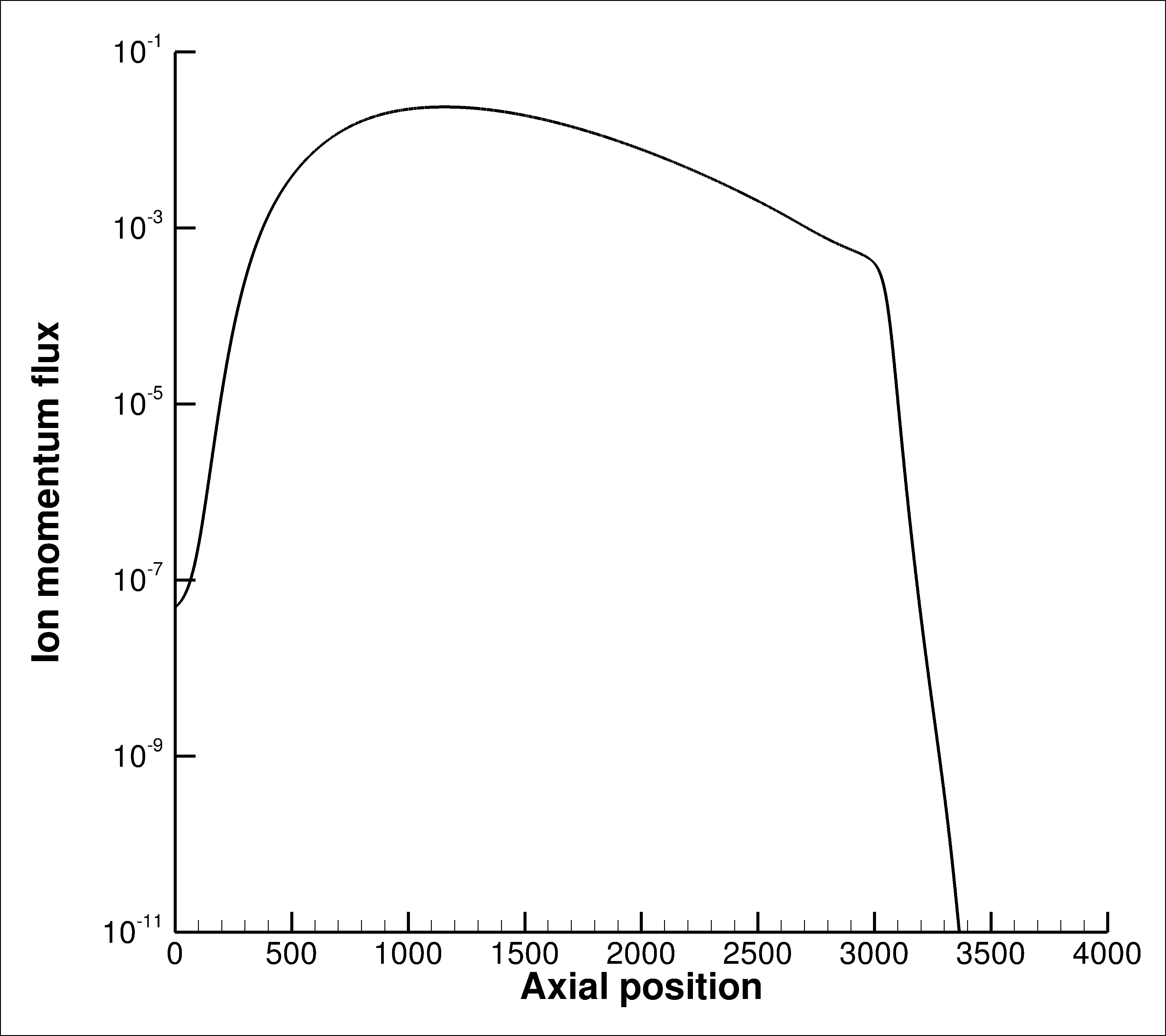}
    }
      \caption{Ion momentum flux, $n_iv_i^2$, at $t=72$ (a) and $t=288$ (b). These are proportional to the solid dark blue curves (ion bulk kinetic energy) in Fig.~\ref{fig:energy}.}
    \label{fig:nvv}
    \end{figure}
    
    In addition to the particle distribution function, mean axial velocity, and various energies of the system discussed above, the momentum flux also exhibits significant inhomogeneity. This flux is of especial interest since it is directly related to the produced thrust and thus thruster performance. The ion momentum flux, $n_i v_i^2$, is plotted in Fig.~\ref{fig:nvv} at the same two time instances considered earlier, dropping the ion mass in the flux expression since all ions are of the same identity. It reaches a peak magnitude of about $10^{-2}$ in the region of thermodynamic nonequilibrium but otherwise spans several orders of magnitude in spatial variation. (Note that this corresponds to a dimensional thrust on the order of micronewtons assuming a laser spot radius of $O(100\text{ \textmu m})$, in corroboration with the $O(1\text{ \textmu N s})$ impulse bit and $O(1\text{ s})$ pulse duration discussed by~\citet{keidar2004plasma}.) This suggests that thruster performance is intimately related to the characteristic time taken for bulk kinetic energy, which provides the basis for thruster acceleration, to relax to random kinetic energy and thus increase the internal energy of the expelled ions. The computed momentum flux provides an upper bound to the achievable momentum flux in actual thrusters since plume divergence necessitates some of this flux to occur radially in a physical multidimensional plume.
    
    The spatial inhomogeneity identified in \S~\ref{sec:noneq} and \ref{sec:momflux} suggests that the grid resolution may be relaxed downstream in locations with lower number density, although the relaxation cannot be performed indiscriminately since the major contributions to thrust occur throughout the region of thermodynamic nonequilibrium. This leads to corresponding considerations in the design of a nonuniform spatial grid to enable thruster-scale simulations.

    \subsection{Nonuniform grid design based on grid-point requirements}
    
    \begin{figure}
      \centerline{
    (a)
    \includegraphics[trim=50 50 50 50, clip, width=0.42\linewidth,valign=t]{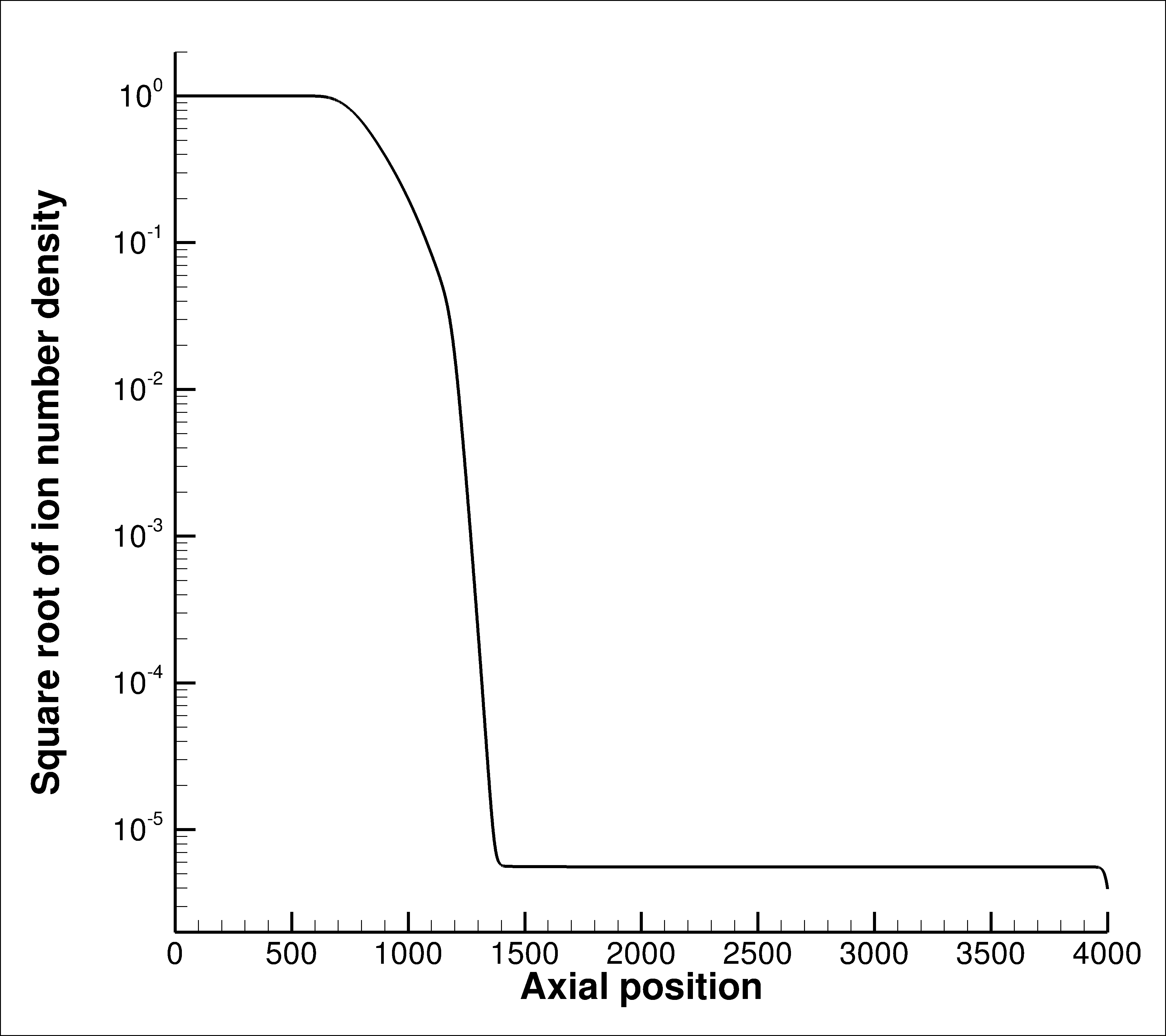}
    \quad
    (b)
    \includegraphics[trim=50 50 50 50, clip, width=0.42\linewidth,valign=t]{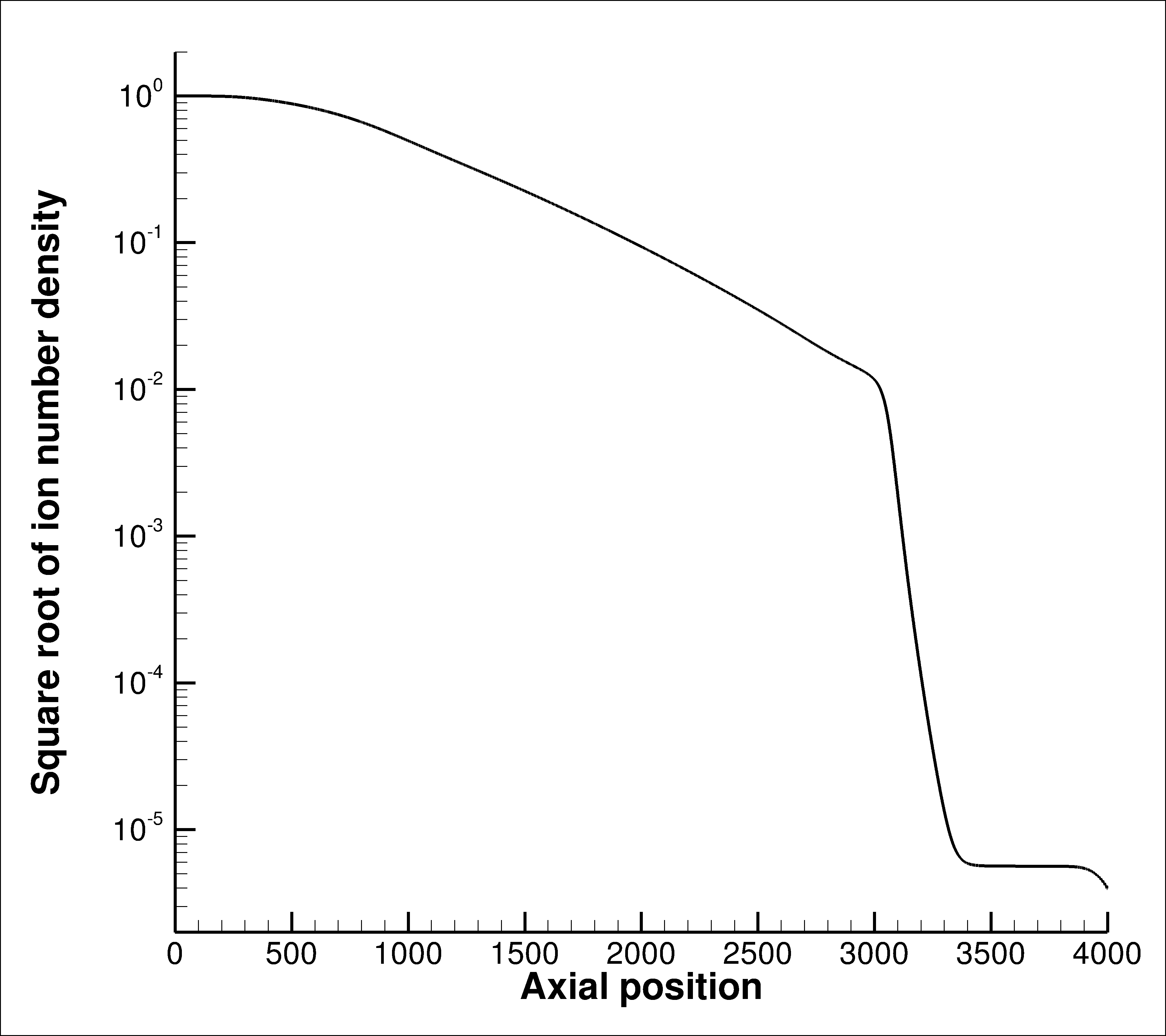}
    }
      \caption{$\sqrt{n_i}$ at $t=72$ (a) and $t=288$ (b).}
    \label{fig:sqrtn}
    \end{figure}
    
    \begin{figure}
      \centerline{
    (a)
    \includegraphics[trim=50 50 50 50, clip, width=0.42\linewidth,valign=t]{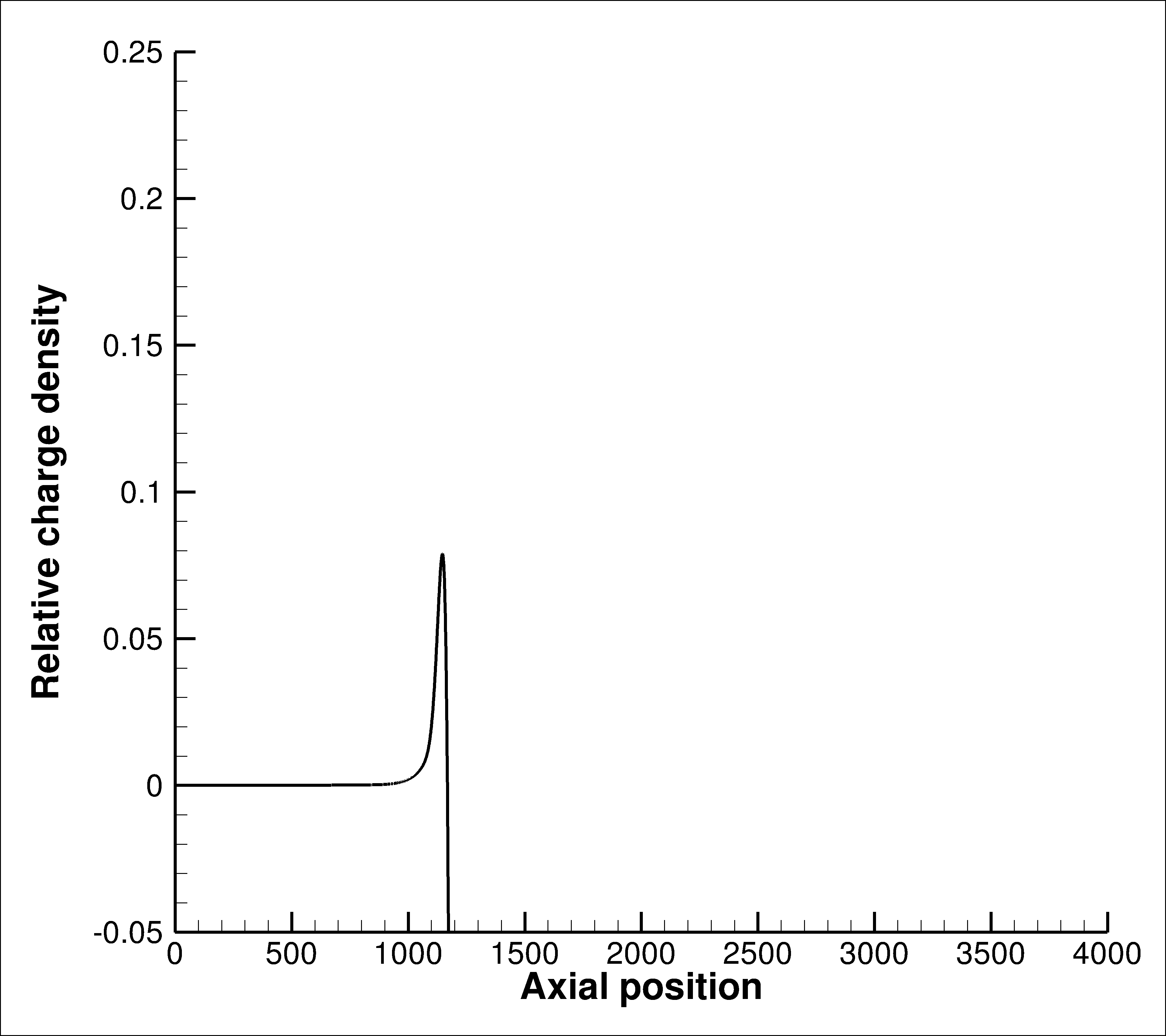}
    \quad
    (b)
    \includegraphics[trim=50 50 50 50, clip, width=0.42\linewidth,valign=t]{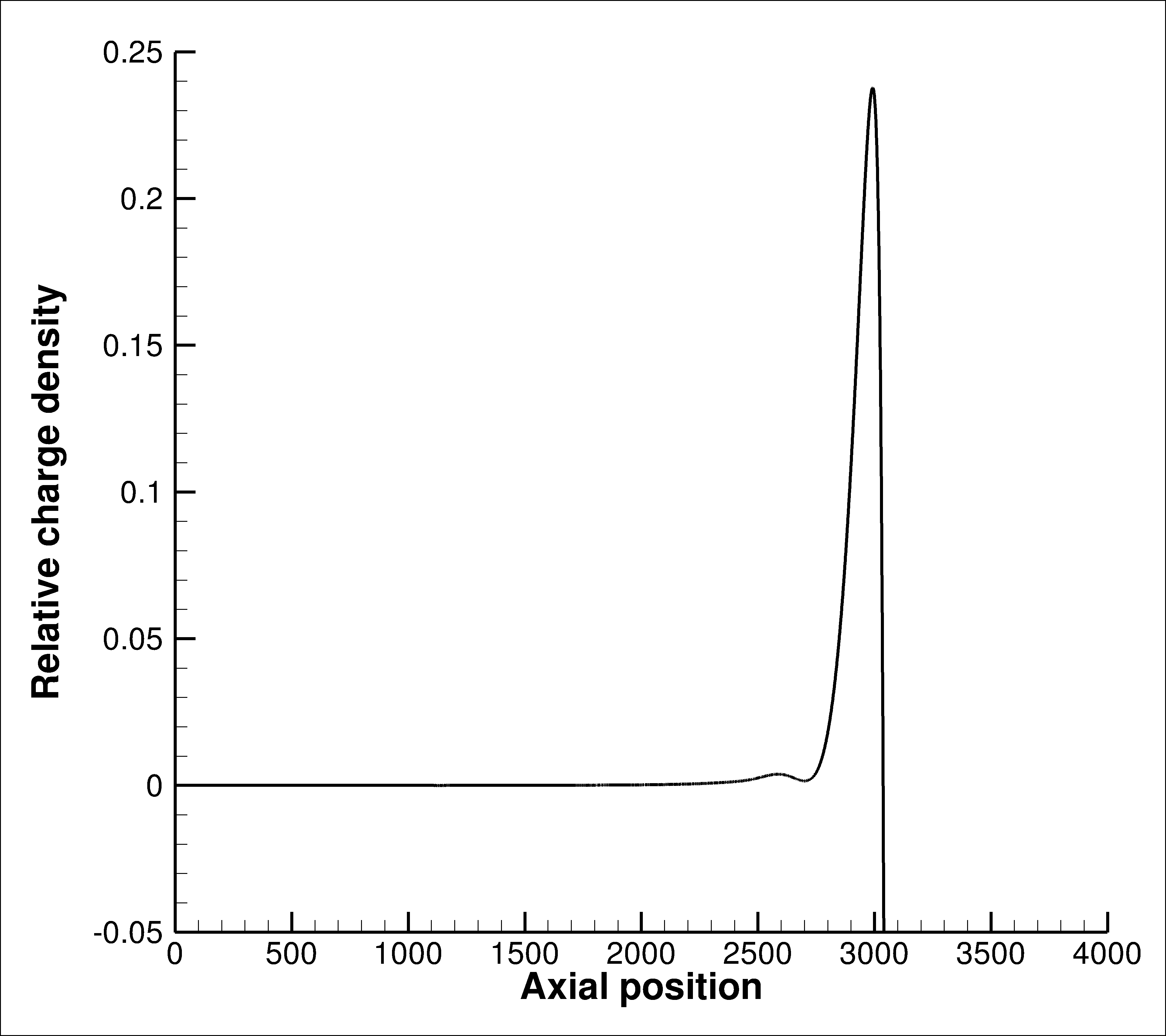}
    }
      \caption{Relative charge density, $(n_i-n_e)/n_i$, at $t=72$ (a) and $t=288$ (b).}
    \label{fig:charge}
    \end{figure}
    
    In Ref.~\citep{chan2022grid}, we observed that the grid-point requirements of a uniform-grid DK simulation scale with the thermal velocity associated with the coldest characteristic temperature, as well as the most pertinent Debye length. The former is dictated in this case by Eq.~\eqref{eqn:Tratio}, in particular by the background temperature of $300\text{ K}$, which is a constant of the system. Hence, in our considerations for nonuniform grid design, we focus on the requirements due to spatial variations in the latter, which is itself a function of the number density profiles outputted by the solver. Noting from the previous Section that the bulk ion kinetic energy derived from electrostatic origins has mostly not had the time to be thermalized into random kinetic energy, we assume that the characteristic temperature governing the ion dynamics is effectively $T_1$ preceding the expansion front and $T_2$ ahead of it. A gross estimate of the spatial variation of the local Debye length then chiefly depends on the corresponding variation of $\sqrt{n_i}$, which is plotted in Fig.~\ref{fig:sqrtn}. The relative charge density, $(n_i-n_e)/n_i$, is plotted in Fig.~\ref{fig:charge} for reference, since it dictates the axial location where consideration of the local Debye length is most pertinent. Figures~\ref{fig:sqrtn} and \ref{fig:charge} suggest that the local Debye length increases by up to two orders of magnitude relative to the left reference value at the location of peak charge density, with a slight increase as one proceeds downstream. Hence, in this work, we target a nonuniform grid whose spatial grid size increases by up to two orders of magnitude from the left end of the computational domain to the right for the baseline domain length.

    \subsection{Comparison of uniform and nonuniform grid simulations}\label{sec:nonuniform}
    
    While the solver grid could be made nonuniform along both the spatial and velocity coordinates, we focus on nonuniformity in space in this work for simplicity, with a brief discussion of nonuniformity in velocity in \S~\ref{sec:velexp}.
    
    \subsubsection{Baseline domain length with spatial coarsening}
    
    \begin{figure}
      \centerline{
    (a)
    \includegraphics[trim=50 50 50 50, clip, width=0.42\linewidth,valign=t]{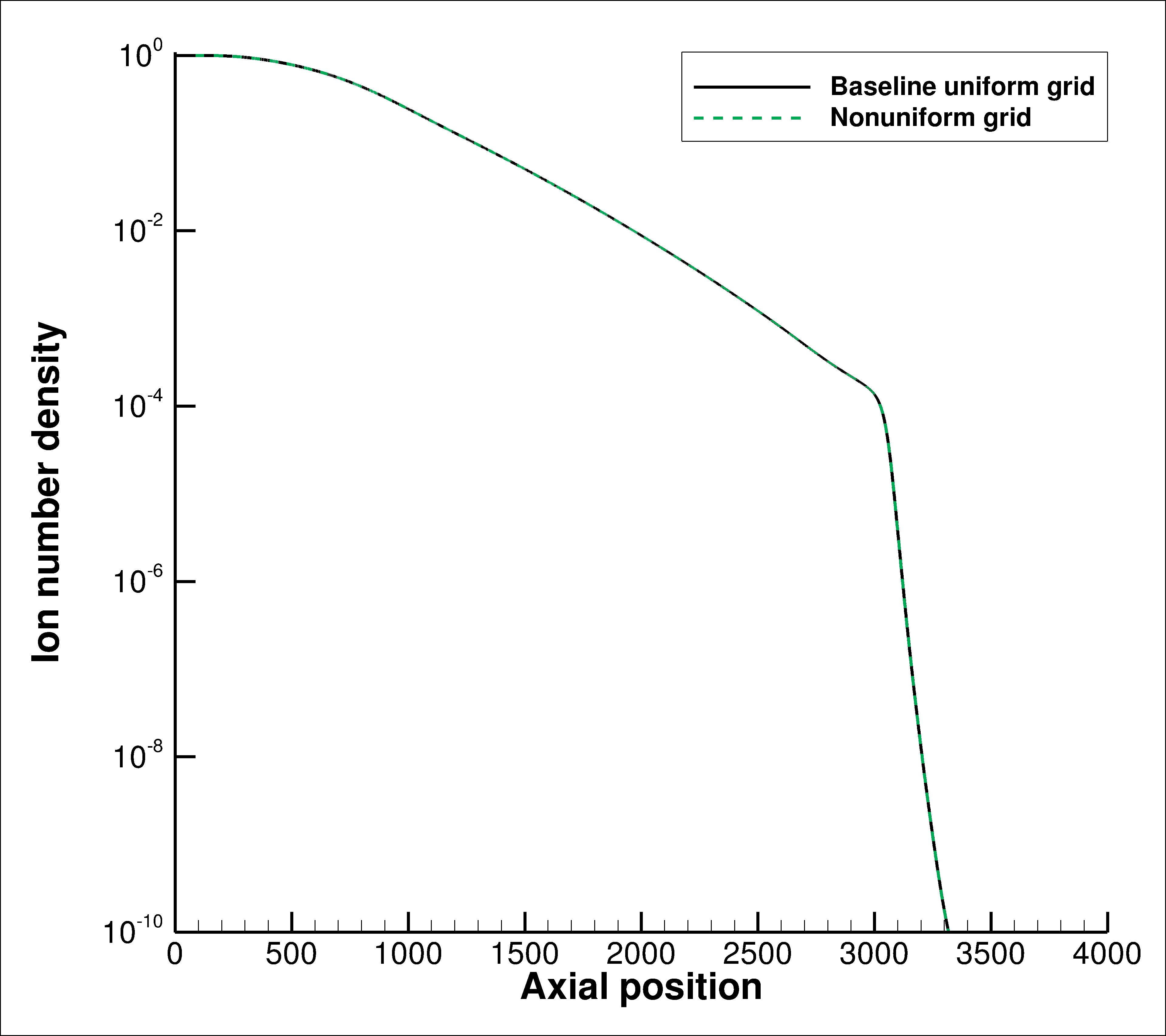}
    \quad
    (b)
    \includegraphics[trim=50 50 50 50, clip, width=0.42\linewidth,valign=t]{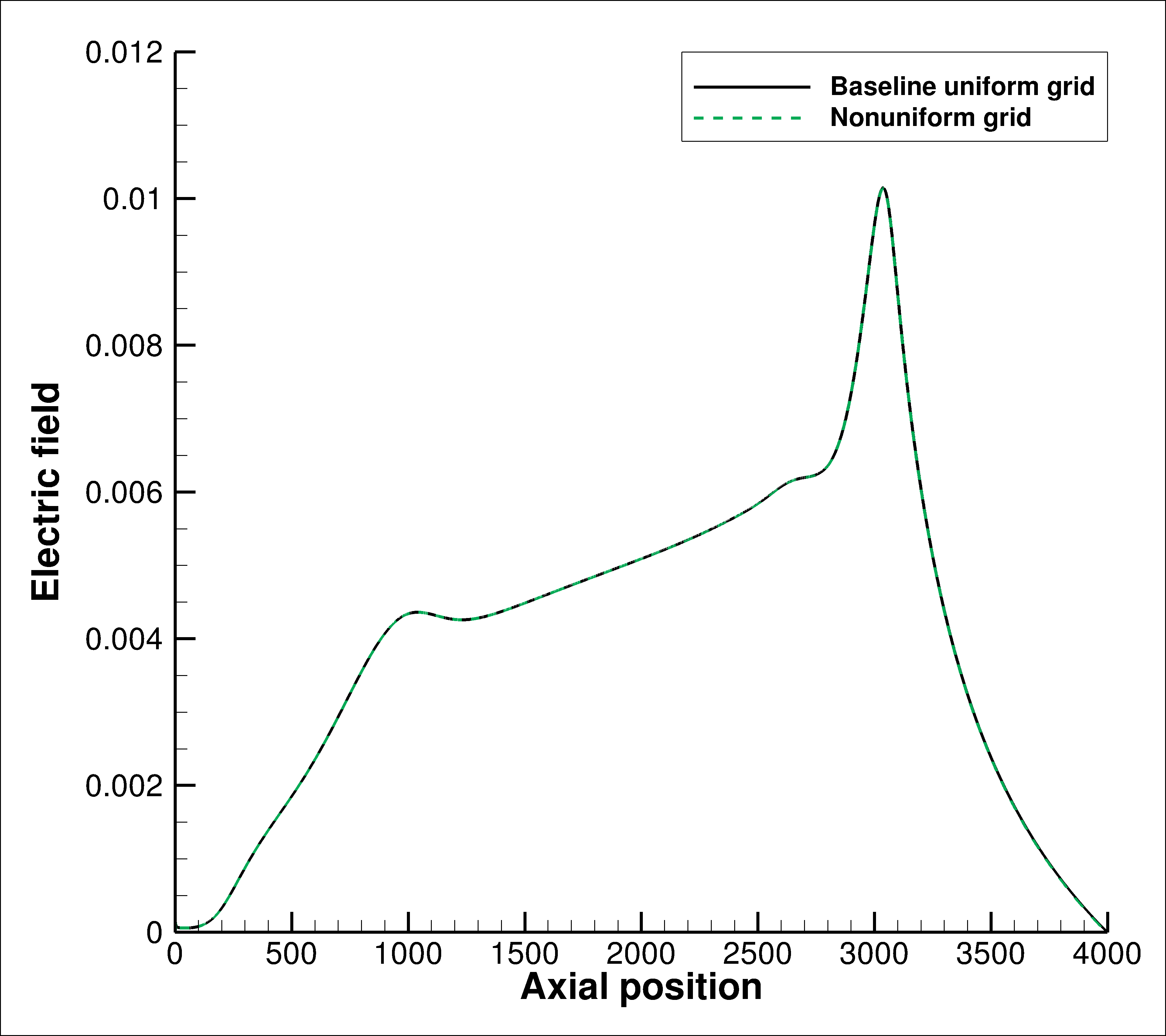}
    }
      \centerline{
    (c)
    \includegraphics[trim=50 50 50 50, clip, width=0.42\linewidth,valign=t]{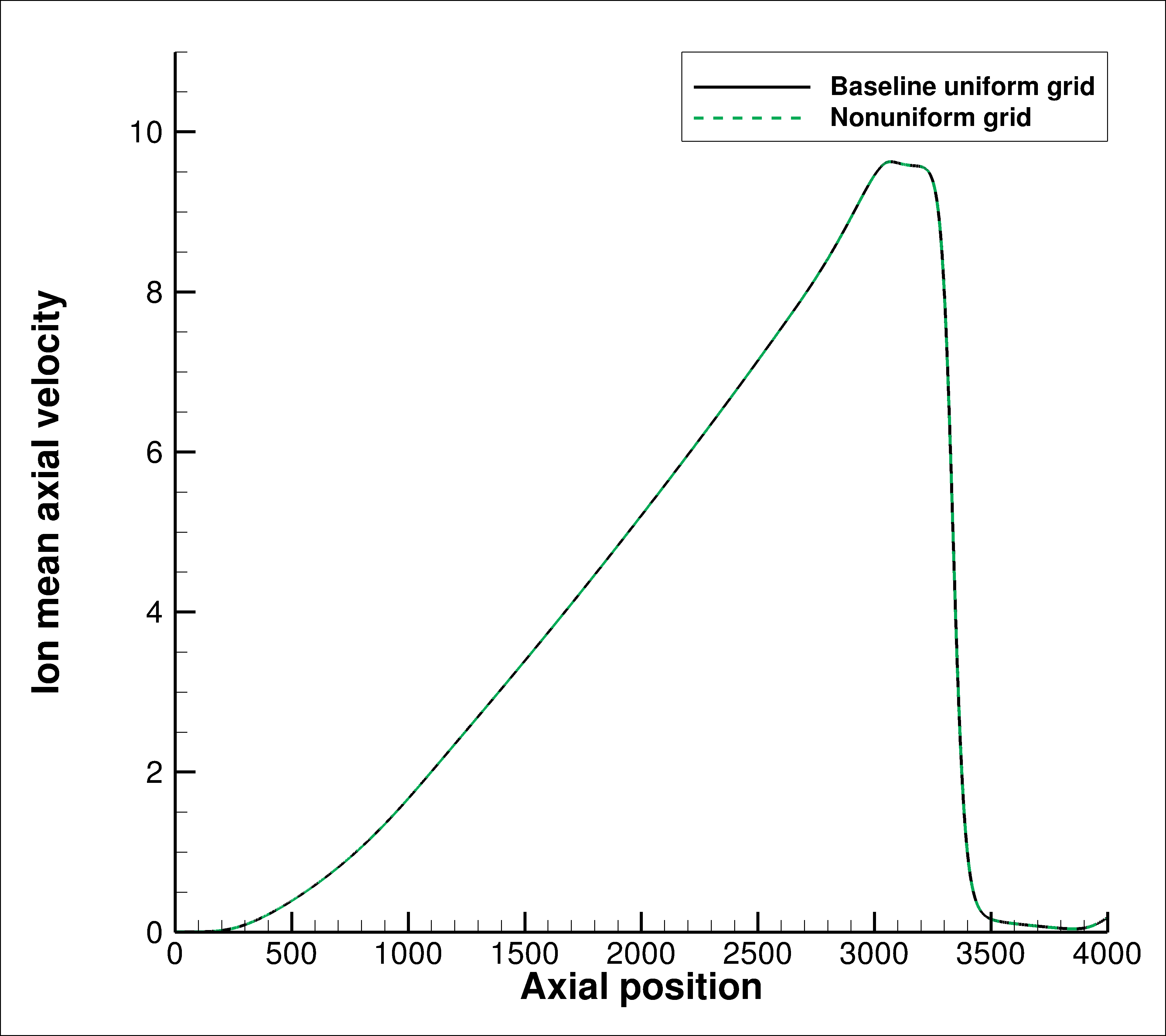}
    \quad
    (d)
    \includegraphics[trim=50 50 50 50, clip, width=0.42\linewidth,valign=t]{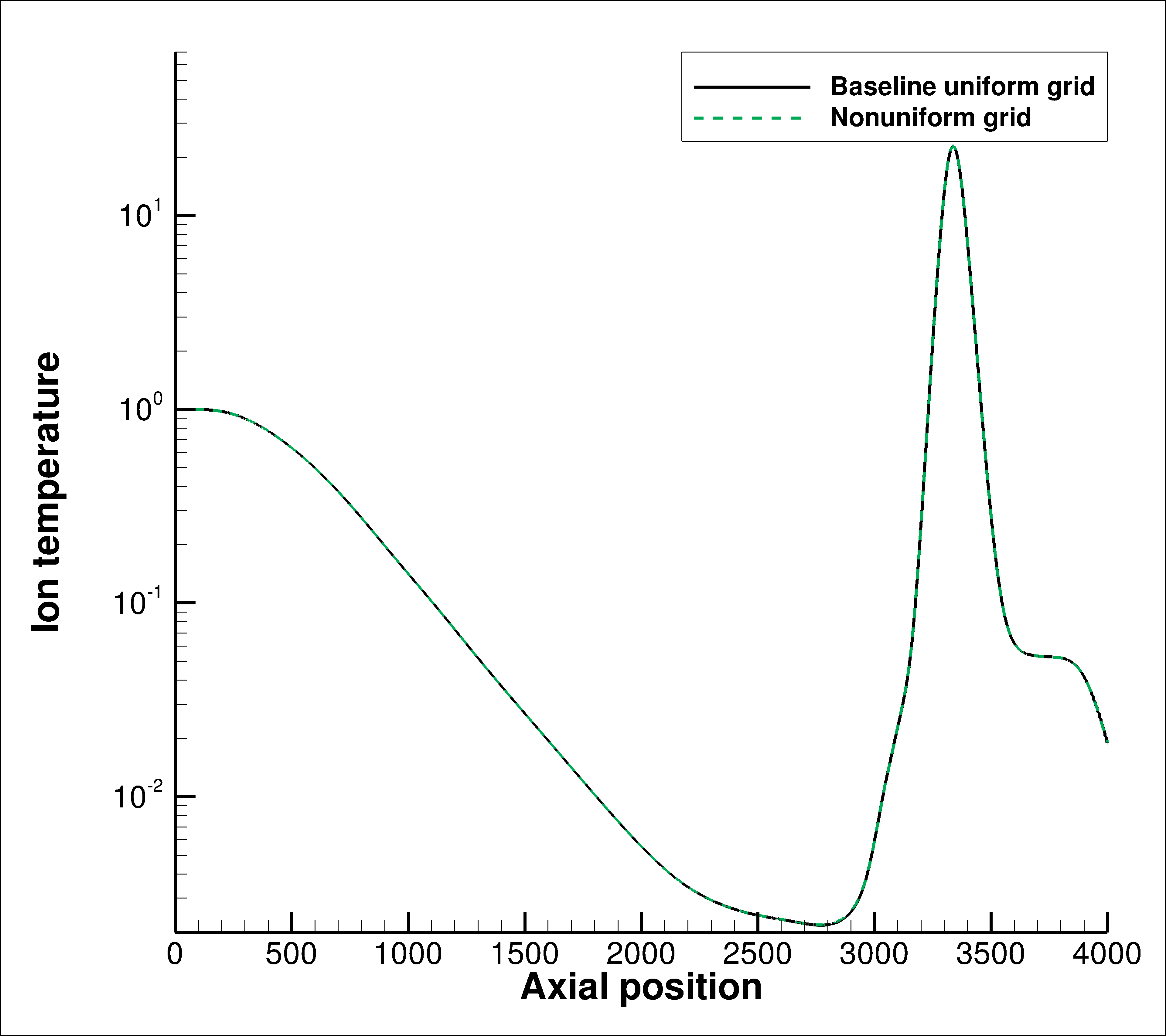}
    }
      \centerline{
    (e)
    \includegraphics[trim=50 50 50 50, clip, width=0.42\linewidth,valign=t]{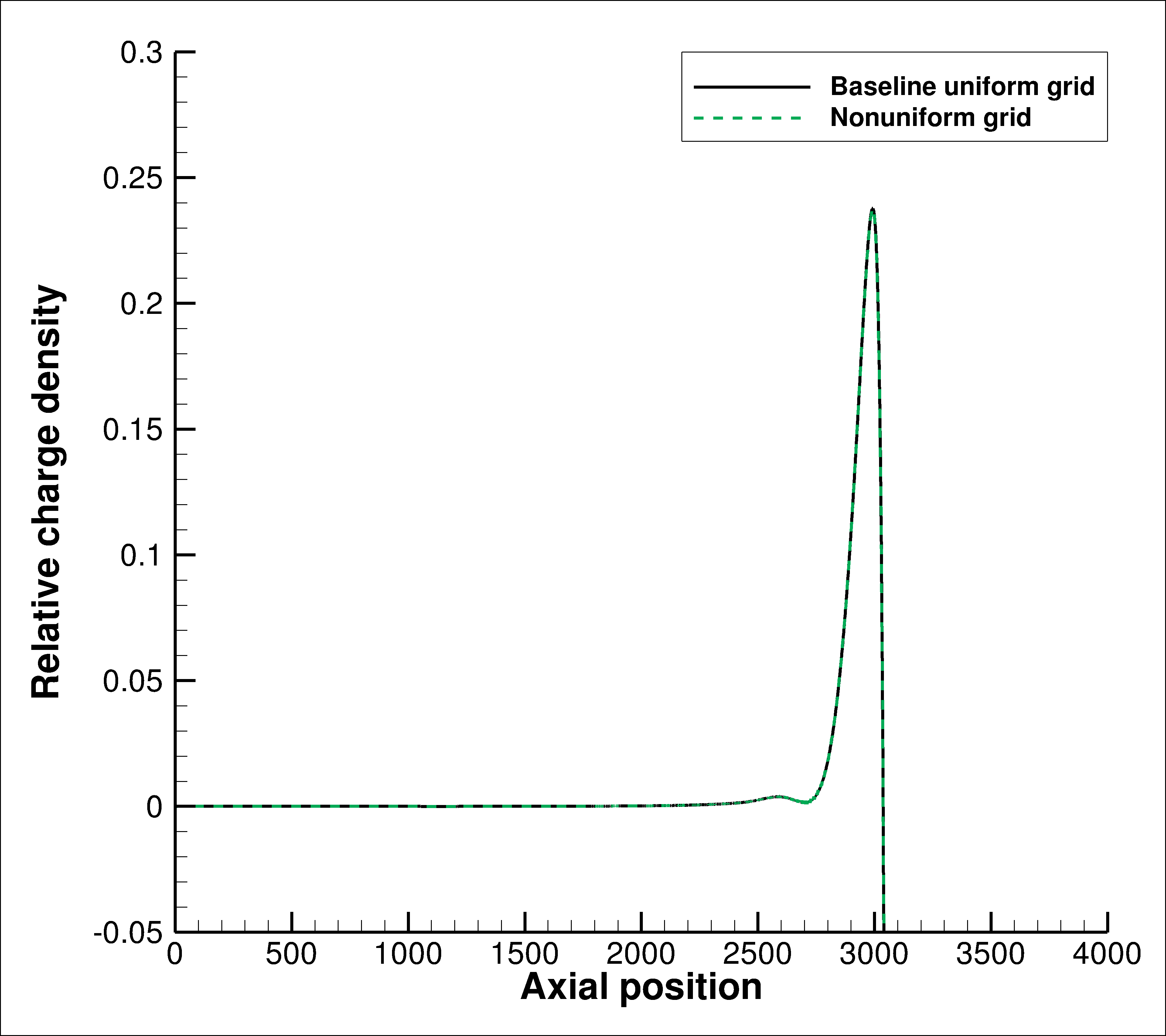}
    \quad
    (f)
    \includegraphics[trim=50 50 50 50, clip, width=0.42\linewidth,valign=t]{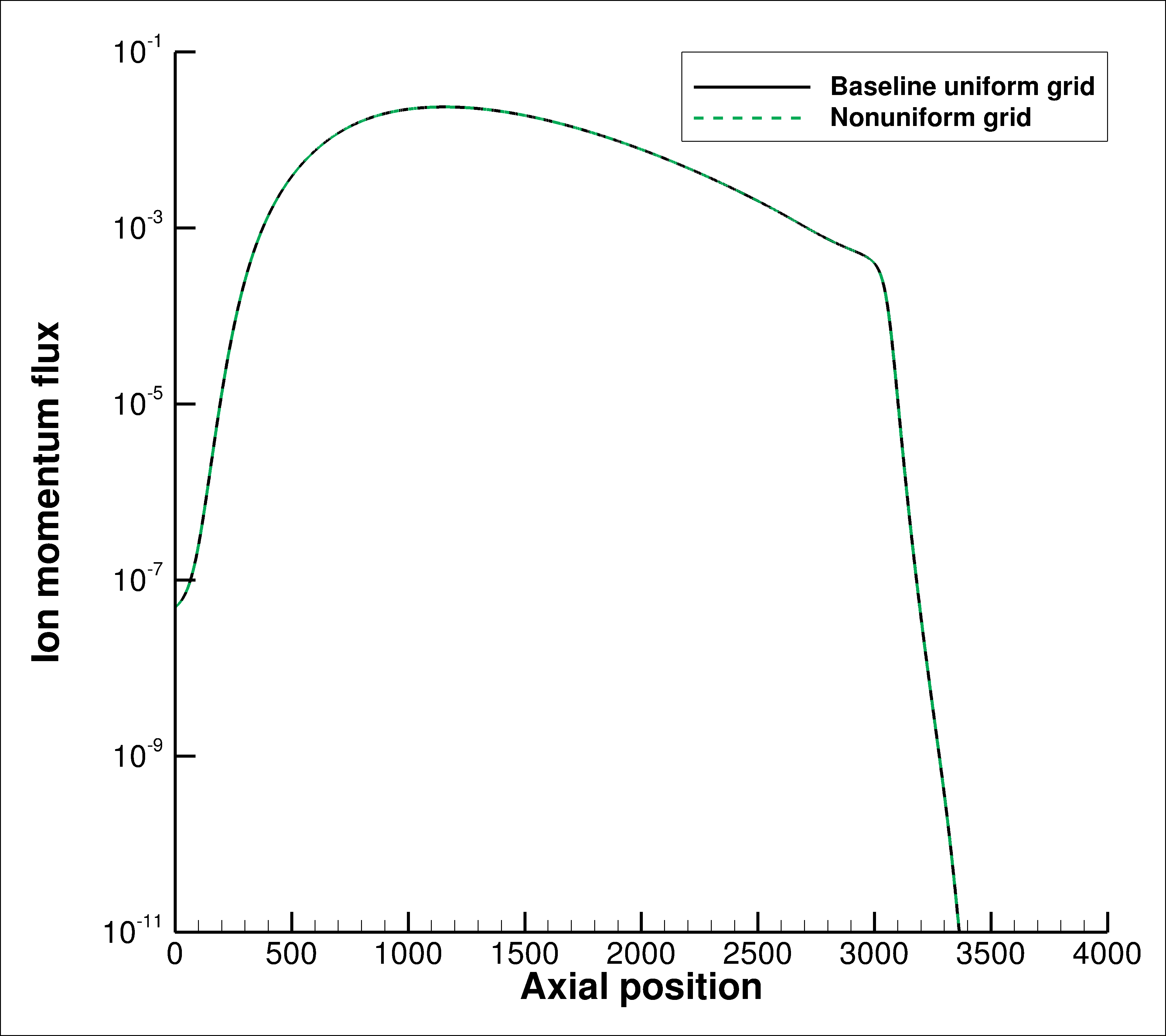}
    }
      \caption{Comparison of ion number density (a), electric field (b), ion mean axial velocity (c), ion temperature (d), relative charge density (e), and ion momentum flux (f) for the baseline domain length and a grid expansion ratio of $0.1\%$, all at $t=288$.}
    \label{fig:len1exp1}
    \end{figure}
    
\begin{figure}
      \centerline{
    (a)
    \includegraphics[trim=50 50 50 50, clip, width=0.42\linewidth,valign=t]{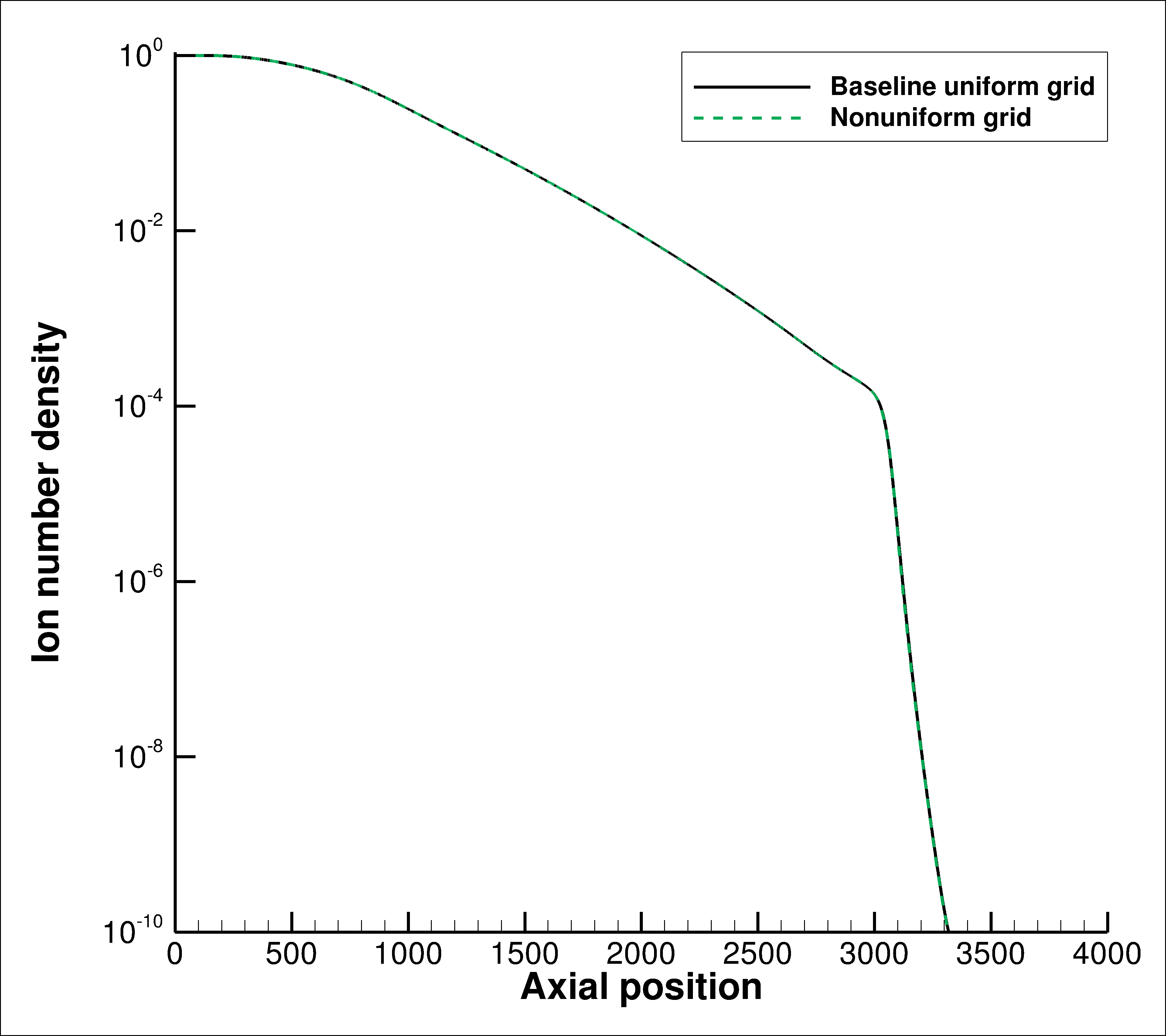}
    \quad
    (b)
    \includegraphics[trim=50 50 50 50, clip, width=0.42\linewidth,valign=t]{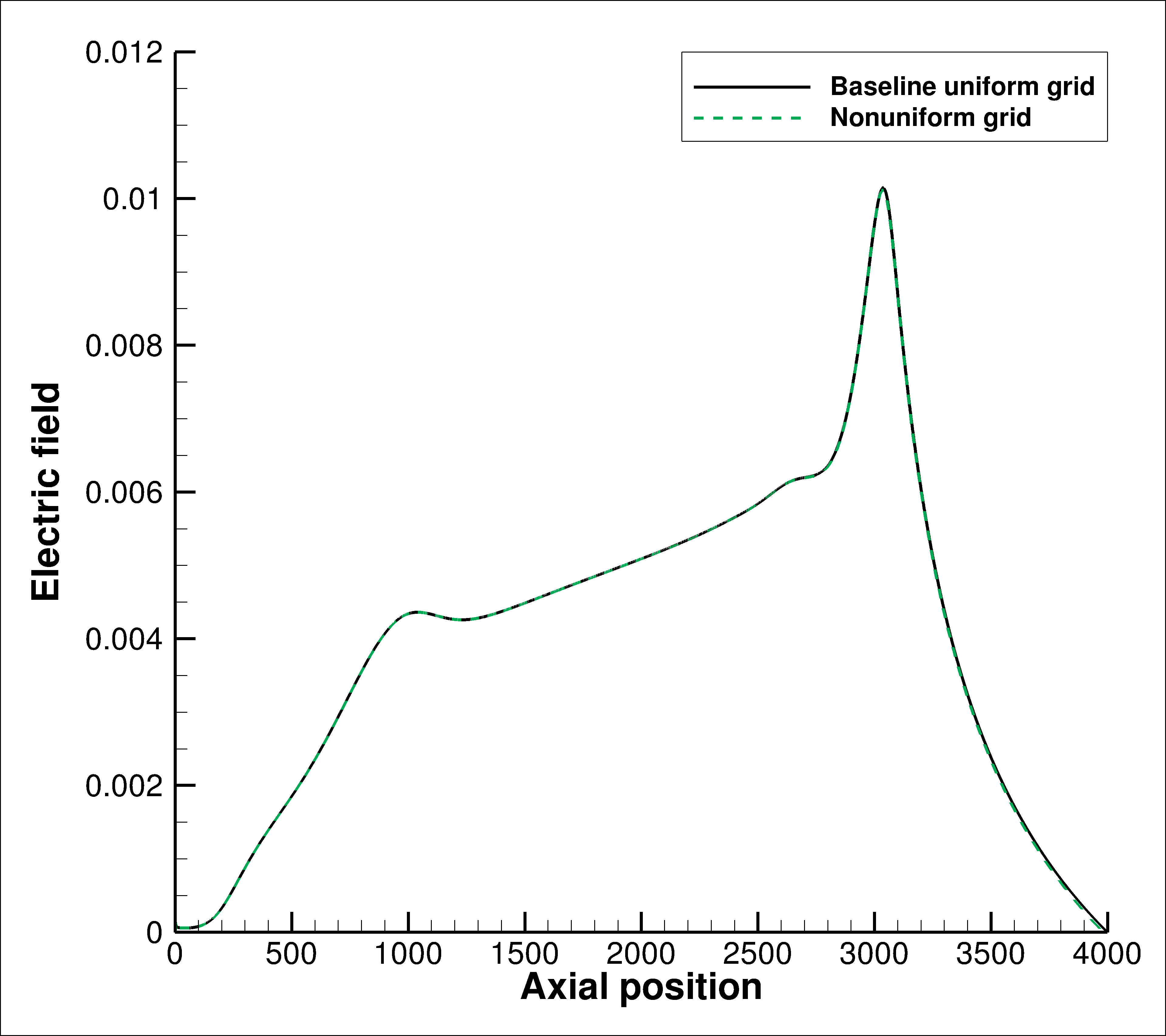}
    }
      \centerline{
    (c)
    \includegraphics[trim=50 50 50 50, clip, width=0.42\linewidth,valign=t]{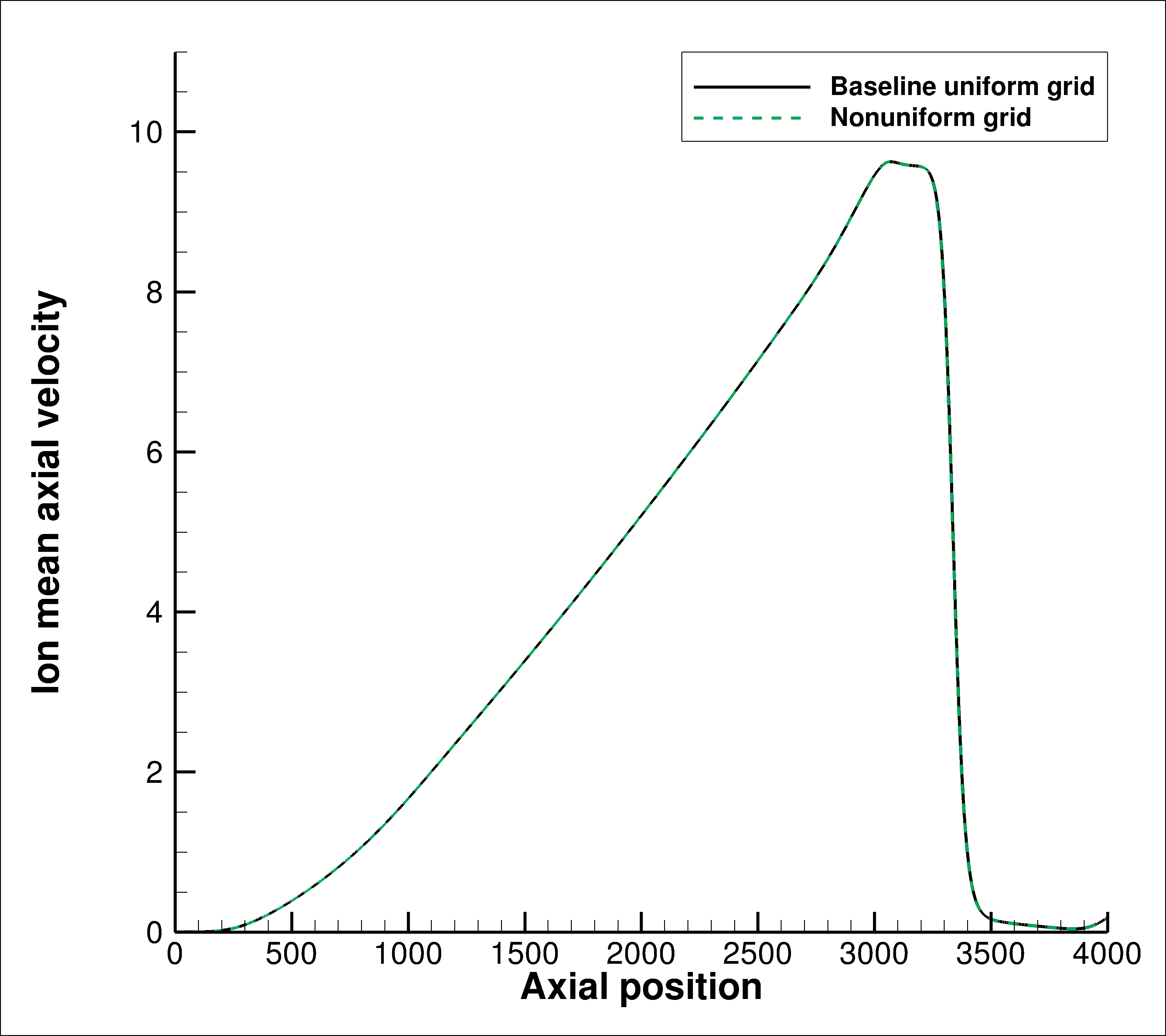}
    \quad
    (d)
    \includegraphics[trim=50 50 50 50, clip, width=0.42\linewidth,valign=t]{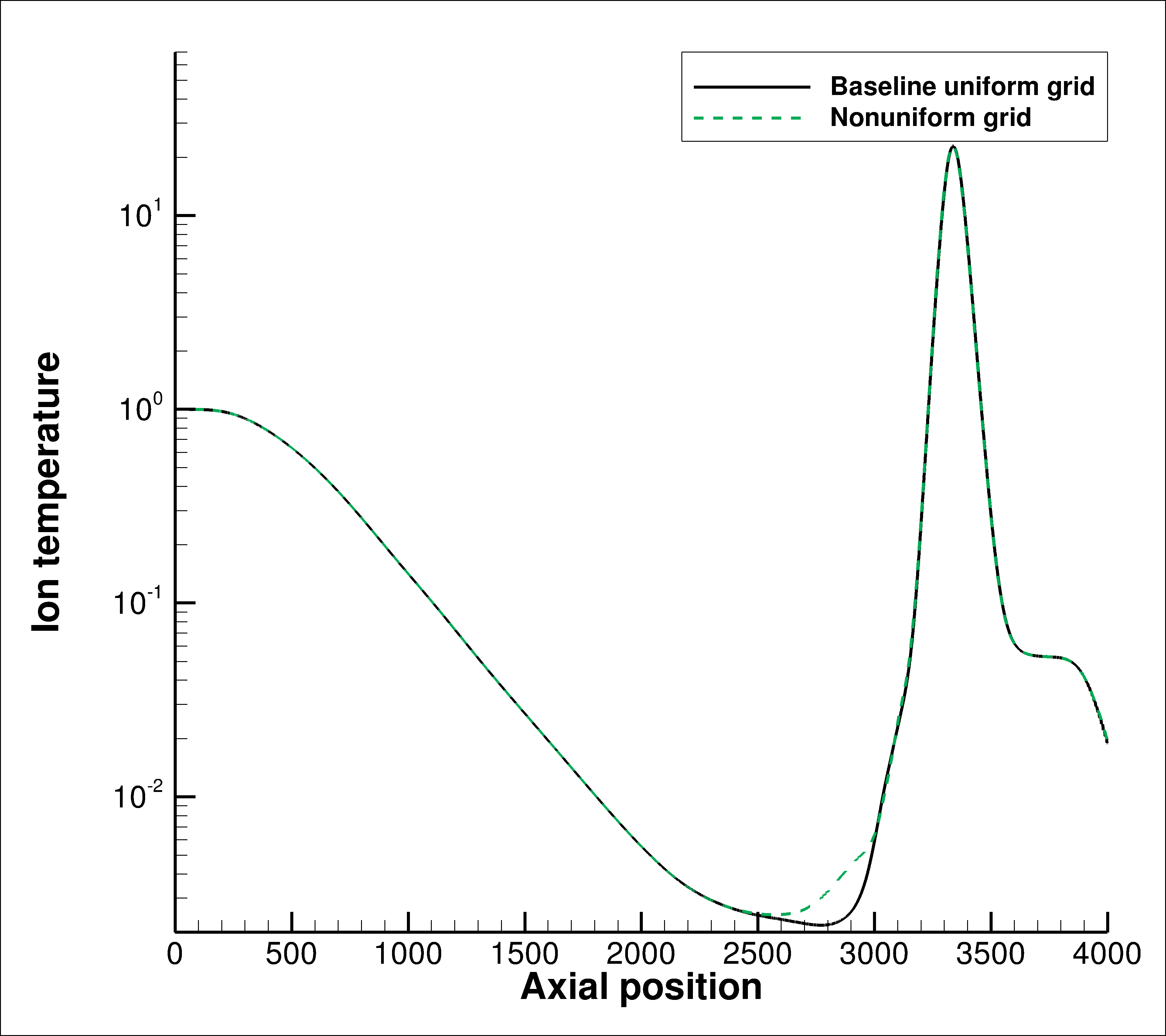}
    }
      \centerline{
    (e)
    \includegraphics[trim=50 50 50 50, clip, width=0.42\linewidth,valign=t]{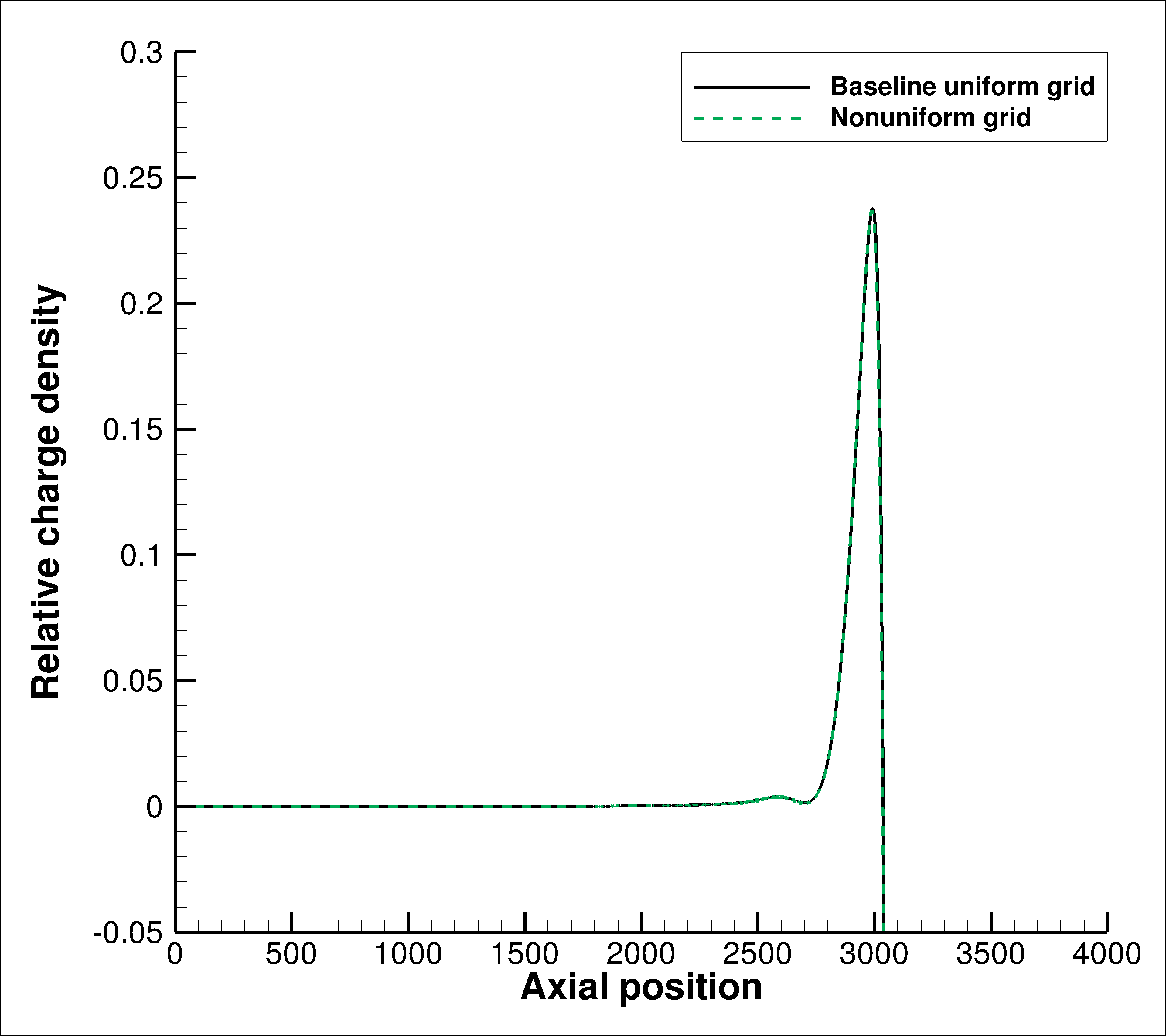}
    \quad
    (f)
    \includegraphics[trim=50 50 50 50, clip, width=0.42\linewidth,valign=t]{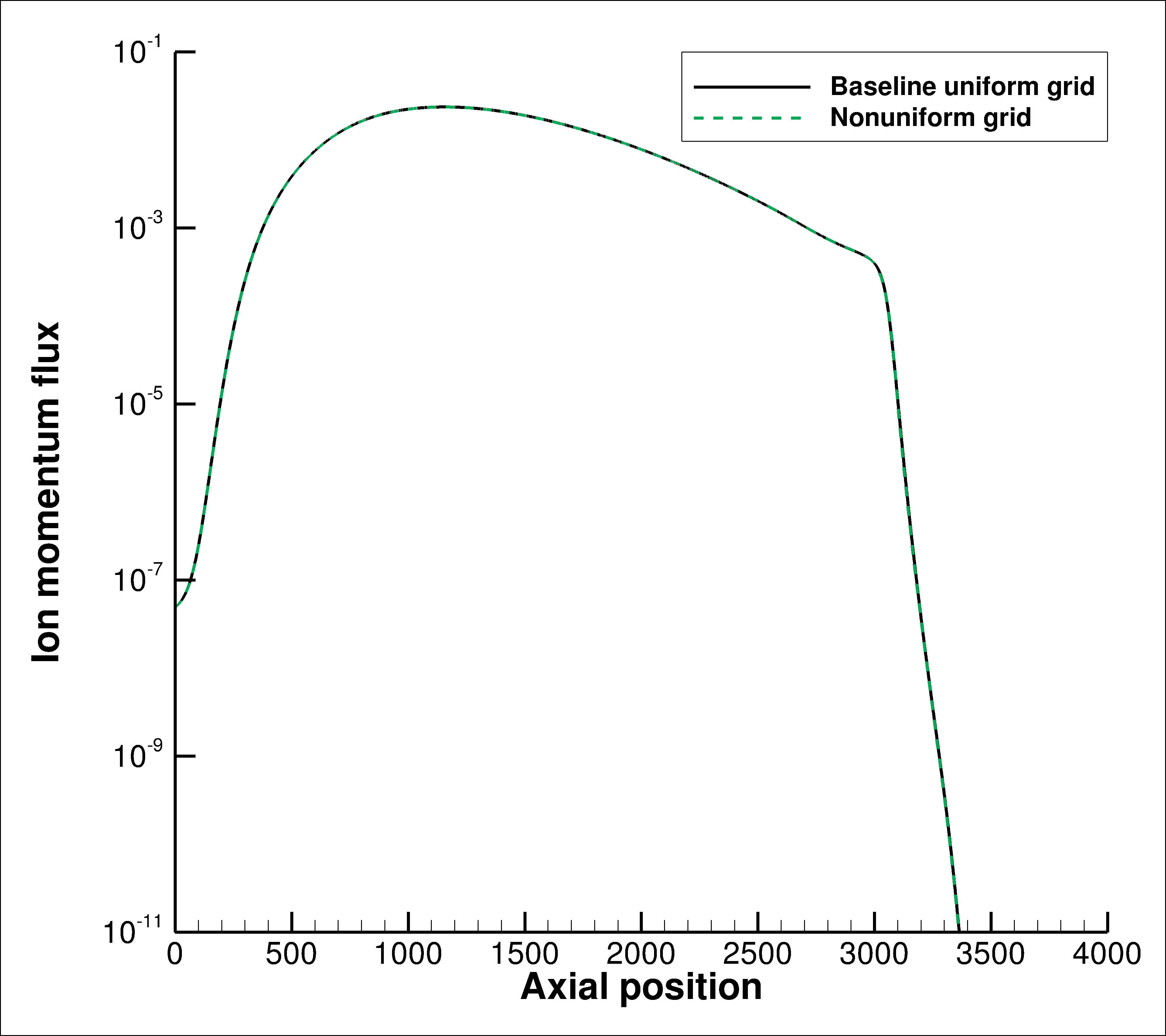}
    }
      \caption{Comparison of ion number density (a), electric field (b), ion mean axial velocity (c), ion temperature (d), relative charge density (e), and ion momentum flux (f) for the baseline domain length and a grid expansion ratio of $0.5\%$, all at $t=288$.}
    \label{fig:len1exp5}
    \end{figure}

    A nonuniform grid is constructed by increasing the grid size beyond the halfway point of the computational domain. The grid size successively increases by a fixed compounded percentage downstream beyond this point. Two expansion ratios, defined as $(\Delta x_{i+1}-\Delta x_i)/(\Delta x_i)$, are considered: $0.1\%$ and $0.5\%$. The number of spatial grid points for each expansion ratio is $5{,}534$ and $4{,}614$, respectively, compared to $8{,}000$ for the corresponding uniform grid. The ratio of the maximum to the minimum grid size for each expansion ratio is $5.4$ and $20.6$, respectively. Visual comparisons of key macroscopic quantities are provided for these two ratios in Figs.~\ref{fig:len1exp1} and \ref{fig:len1exp5}, respectively. Visual convergence of all macroscopic quantities is satisfactory at this resolution, apart from the ion temperature in the vicinity of the double layer for the $0.5\%$ grid, which fundamentally arises from resolution of the ion distribution peaks.
    
    \subsubsection{Extended domain length with spatial coarsening}\label{sec:longdomain}
    
    The dimensional domain length of the baseline simulations is $27\text{ \textmu m}$, which is somewhat removed from actual device dimensions. The nonuniform grid capability discussed above enables domain lengths that approach practical thruster dimensions more closely. Here, the domain length is extended to $L=40{,}000$ and $L=400{,}000$, keeping the transition location at $x_c=800$ and continuing to increase the grid size from $x=2{,}000$ onwards. These correspond to dimensional lengths of $0.27\text{ mm}$ and $2.7\text{ mm}$ and bring the domain length up to $O(10)$ and $O(100)$ mean free paths, respectively.
    
\begin{figure}
      \centerline{
    (a)
    \includegraphics[trim=50 50 50 50, clip, width=0.42\linewidth,valign=t]{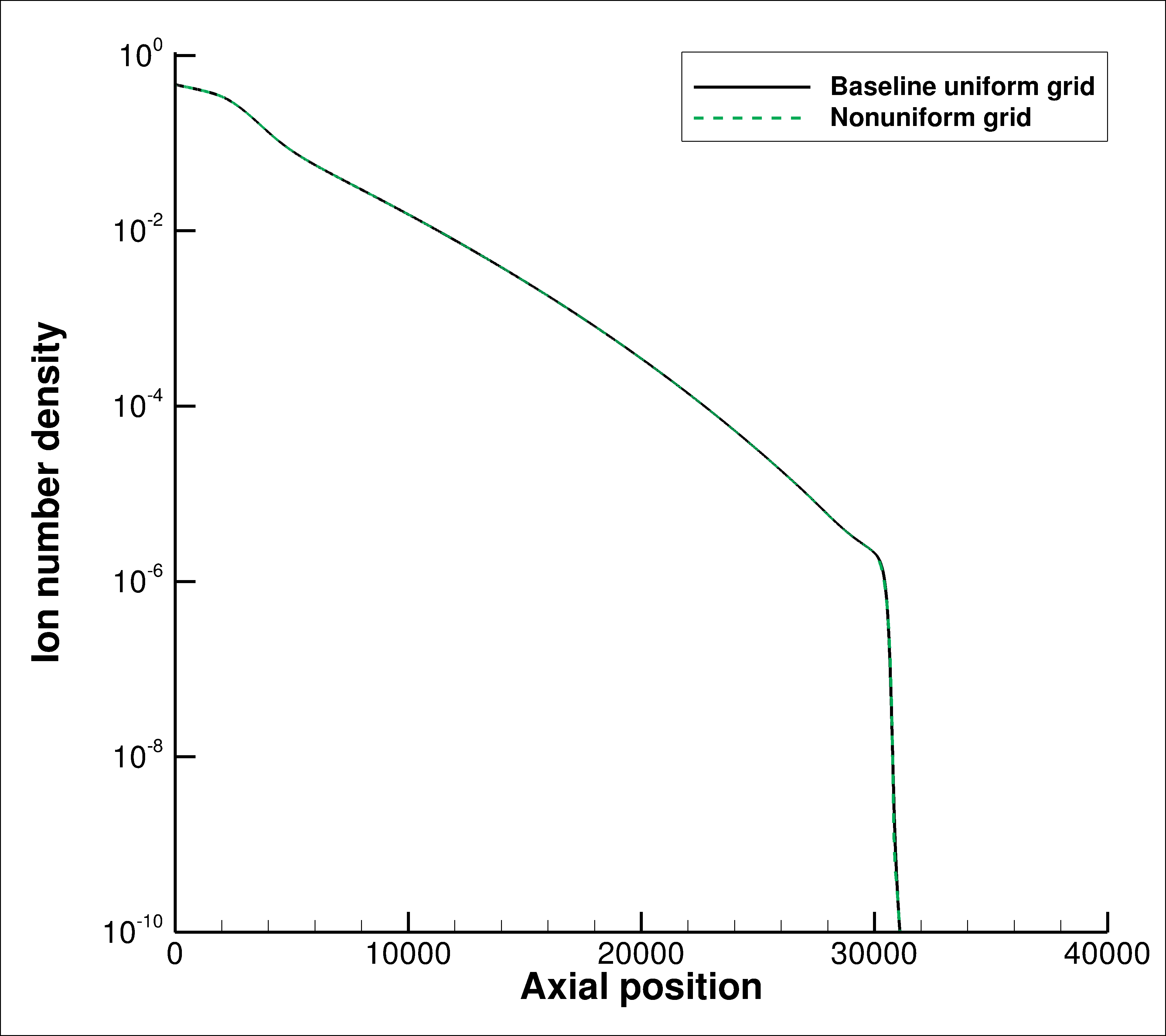}
    \quad
    (b)
    \includegraphics[trim=50 50 50 50, clip, width=0.42\linewidth,valign=t]{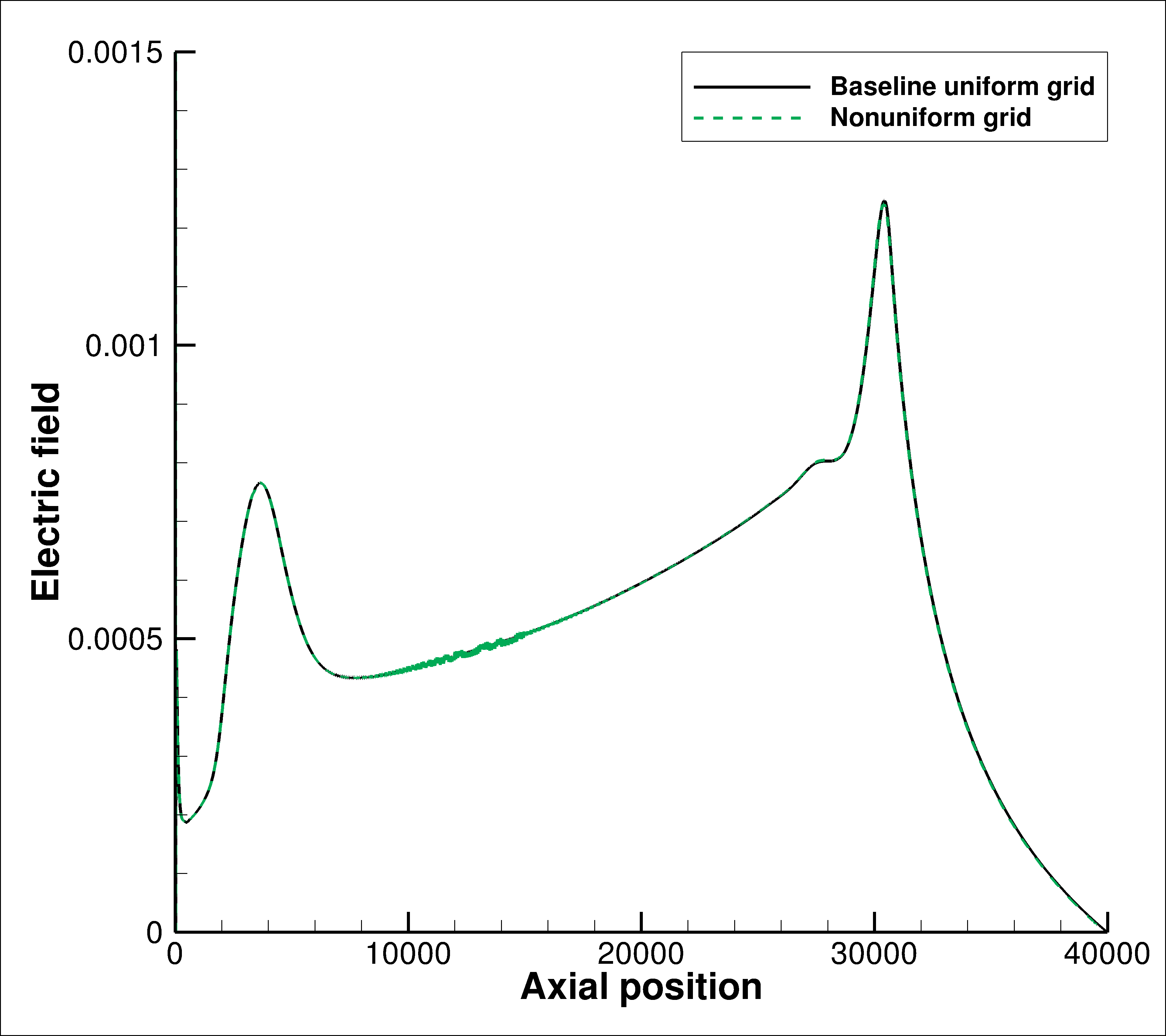}
    }
      \centerline{
    (c)
    \includegraphics[trim=50 50 50 50, clip, width=0.42\linewidth,valign=t]{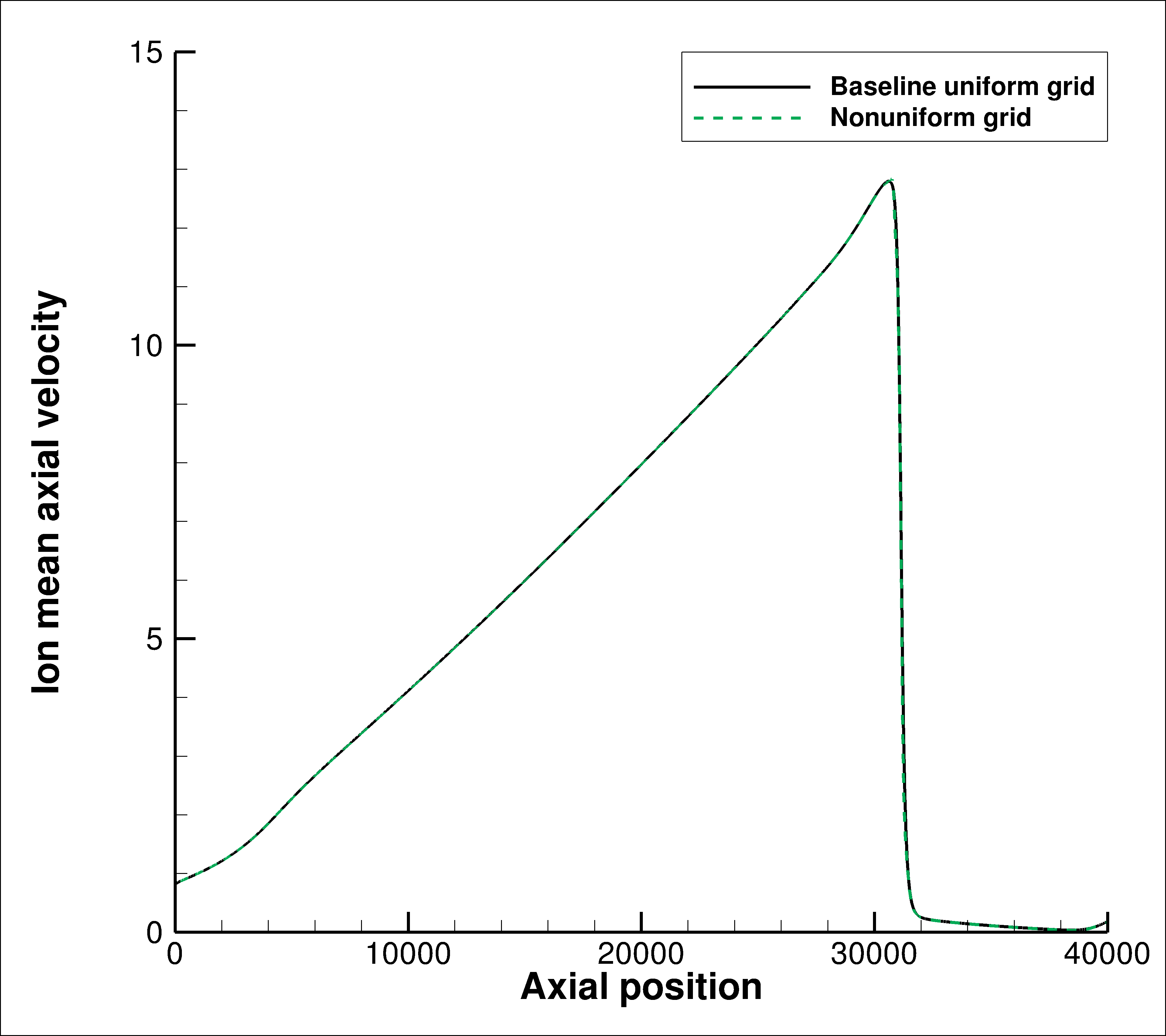}
    \quad
    (d)
    \includegraphics[trim=50 50 50 50, clip, width=0.42\linewidth,valign=t]{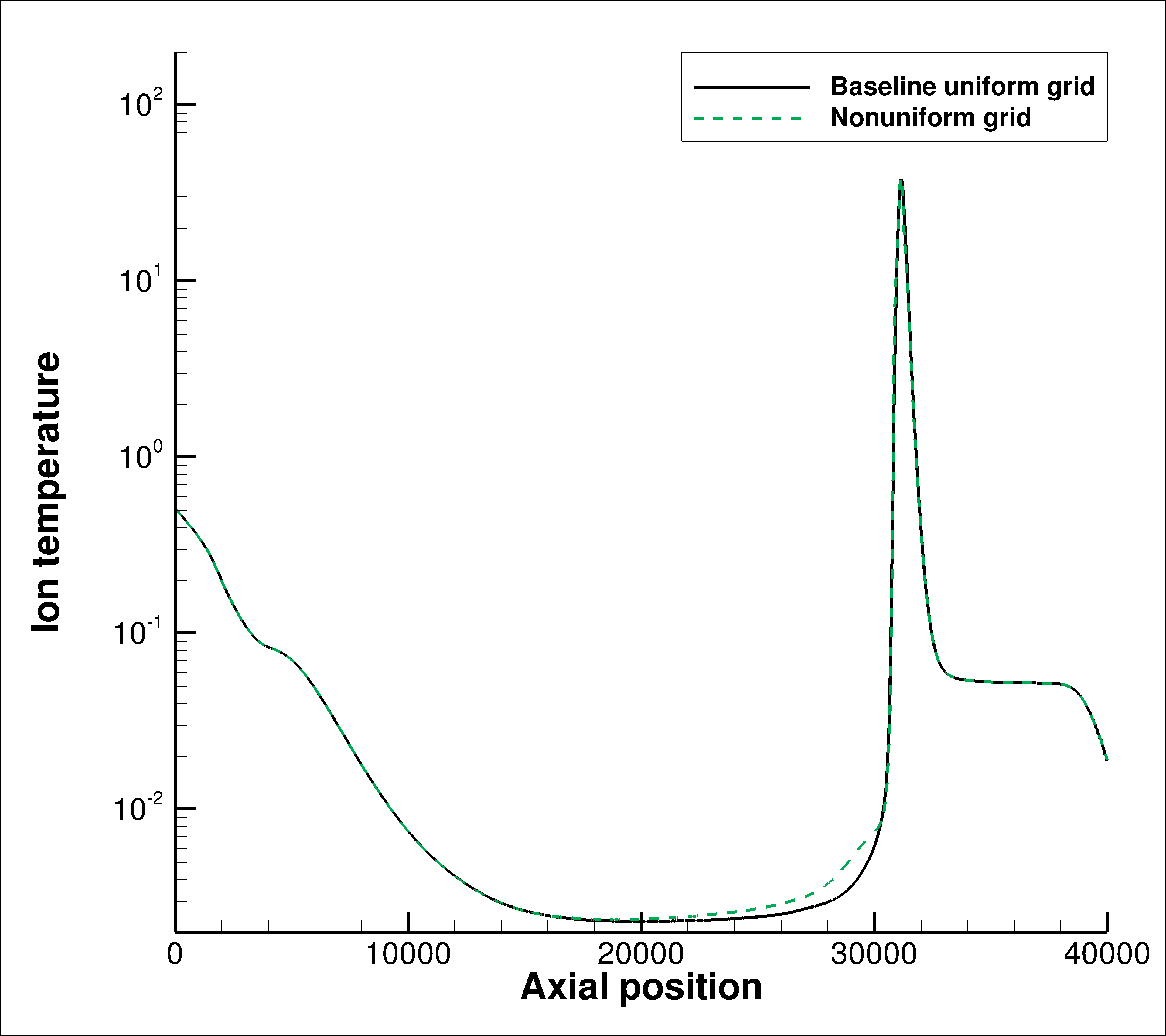}
    }
      \centerline{
    (e)
    \includegraphics[trim=50 50 50 50, clip, width=0.42\linewidth,valign=t]{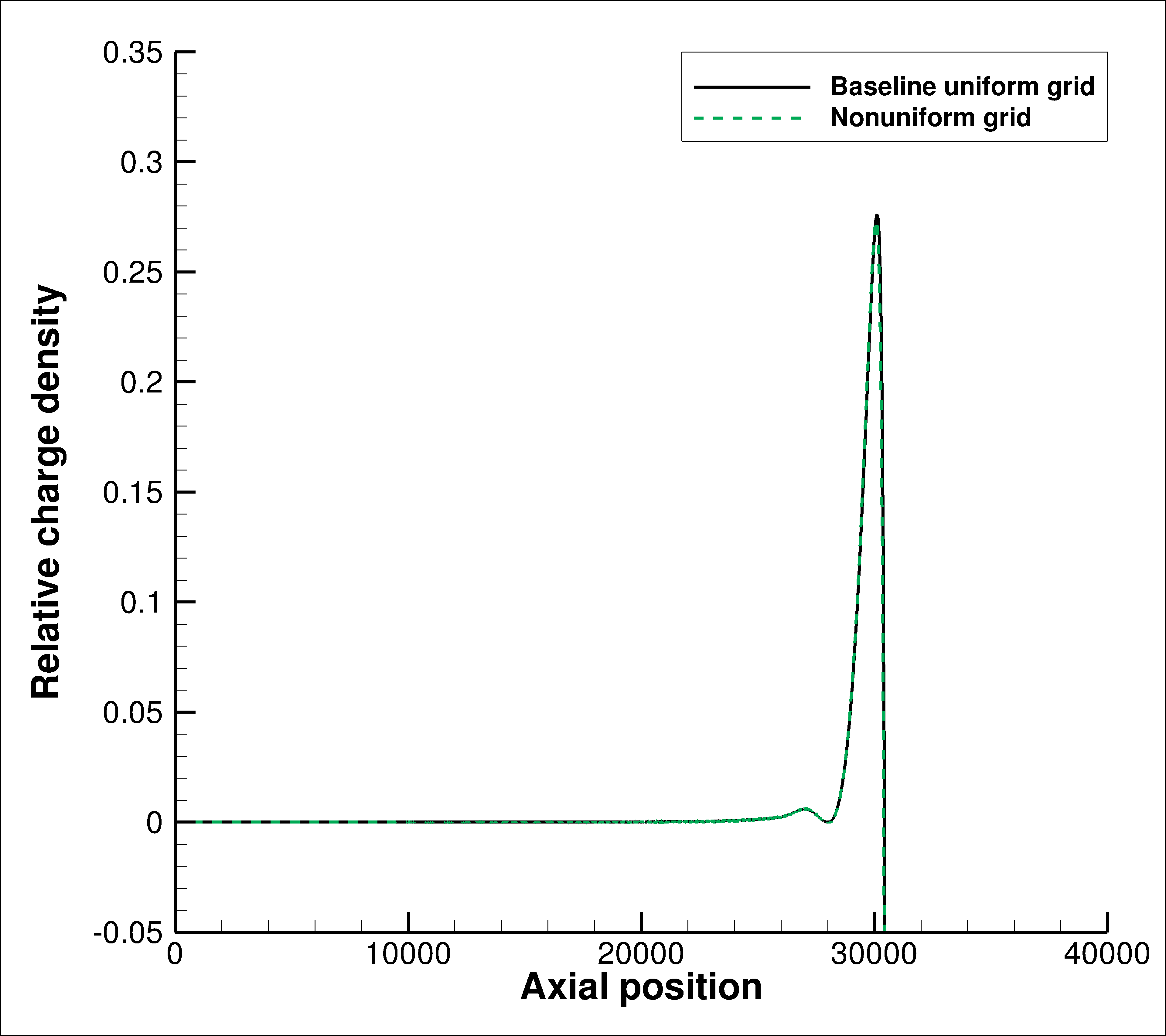}
    \quad
    (f)
    \includegraphics[trim=50 50 50 50, clip, width=0.42\linewidth,valign=t]{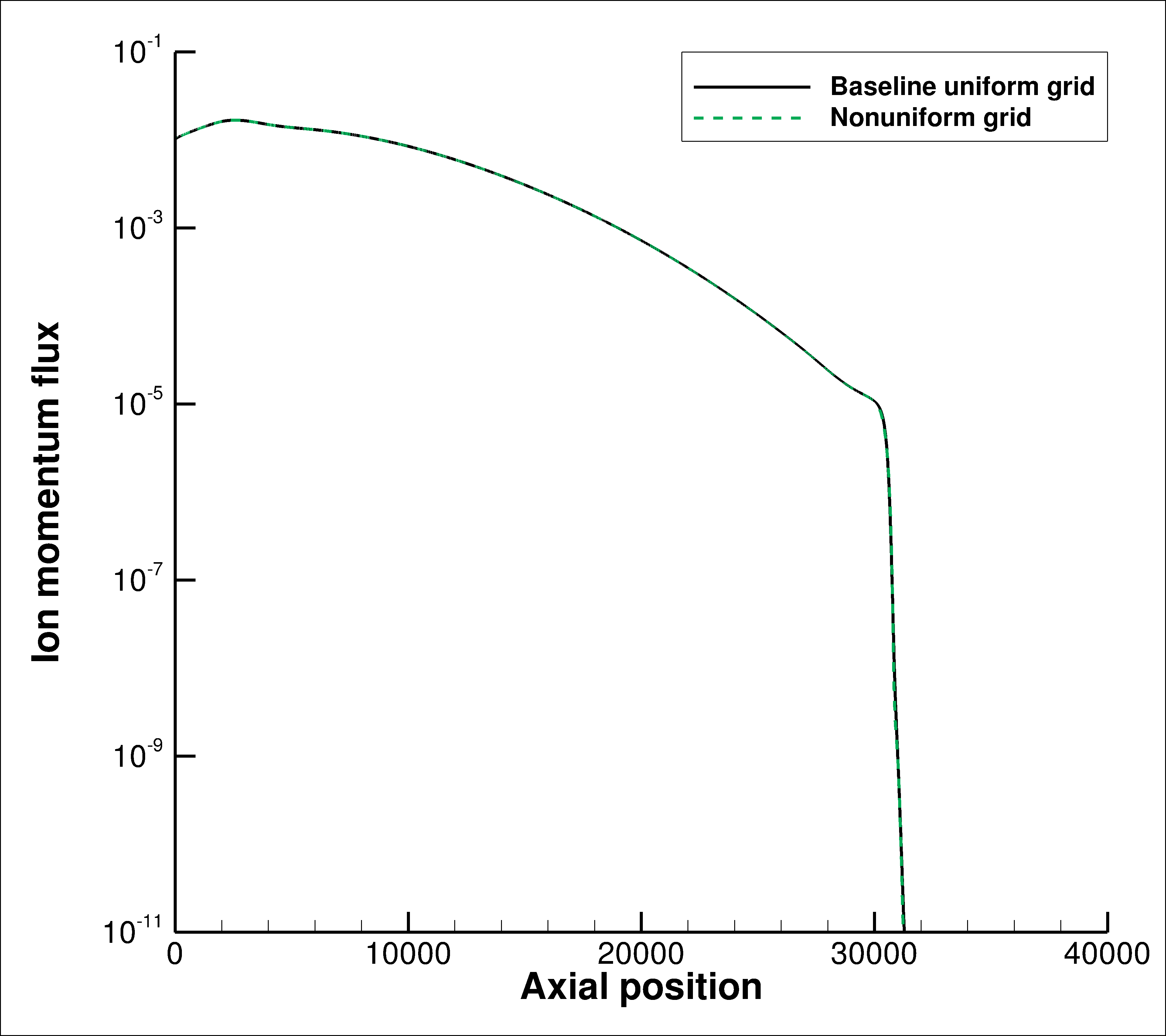}
    }
      \caption{Comparison of ion number density (a), electric field (b), ion mean axial velocity (c), ion temperature (d), relative charge density (e), and ion momentum flux (f) for a domain length 10 times the baseline and a grid expansion ratio of $0.1\%$, all at $t=2{,}880$.}
    \label{fig:len10exp1}
    \end{figure}

    To ensure that the corresponding uniform-grid simulation is tractable, a comparison between uniform and nonuniform grid results is performed only for the domain length of $L=40{,}000$. In addition, the uniform spatial grid resolution is relaxed to $\Delta x = 4$, at which low-order macroscopic quantities remain converged to $O(0.1\%)$ but not the particle distribution function itself. A visual comparison of various macroscopic quantities is provided in Fig.~\ref{fig:len10exp1} only for the expansion ratio of $0.1\%$, as the simulation corresponding to a $0.5\%$ ratio did not converge. The number of spatial grid points is $2{,}854$ in this case, compared with $10{,}000$ for the uniform grid, and the ratio of the maximum to the minimum grid size is $10.5$. Visual convergence of most of the quantities remains satisfactory, apart from visible oscillations in the relative charge density and electric field, as well as deviations in the temperature within the region of thermodynamic nonequilibrium. In particular, the key metric of momentum flux remains largely converged.
    
\begin{figure}
      \centerline{
    (a)
    \includegraphics[trim=50 50 17 50, clip, width=0.42\linewidth,valign=t]{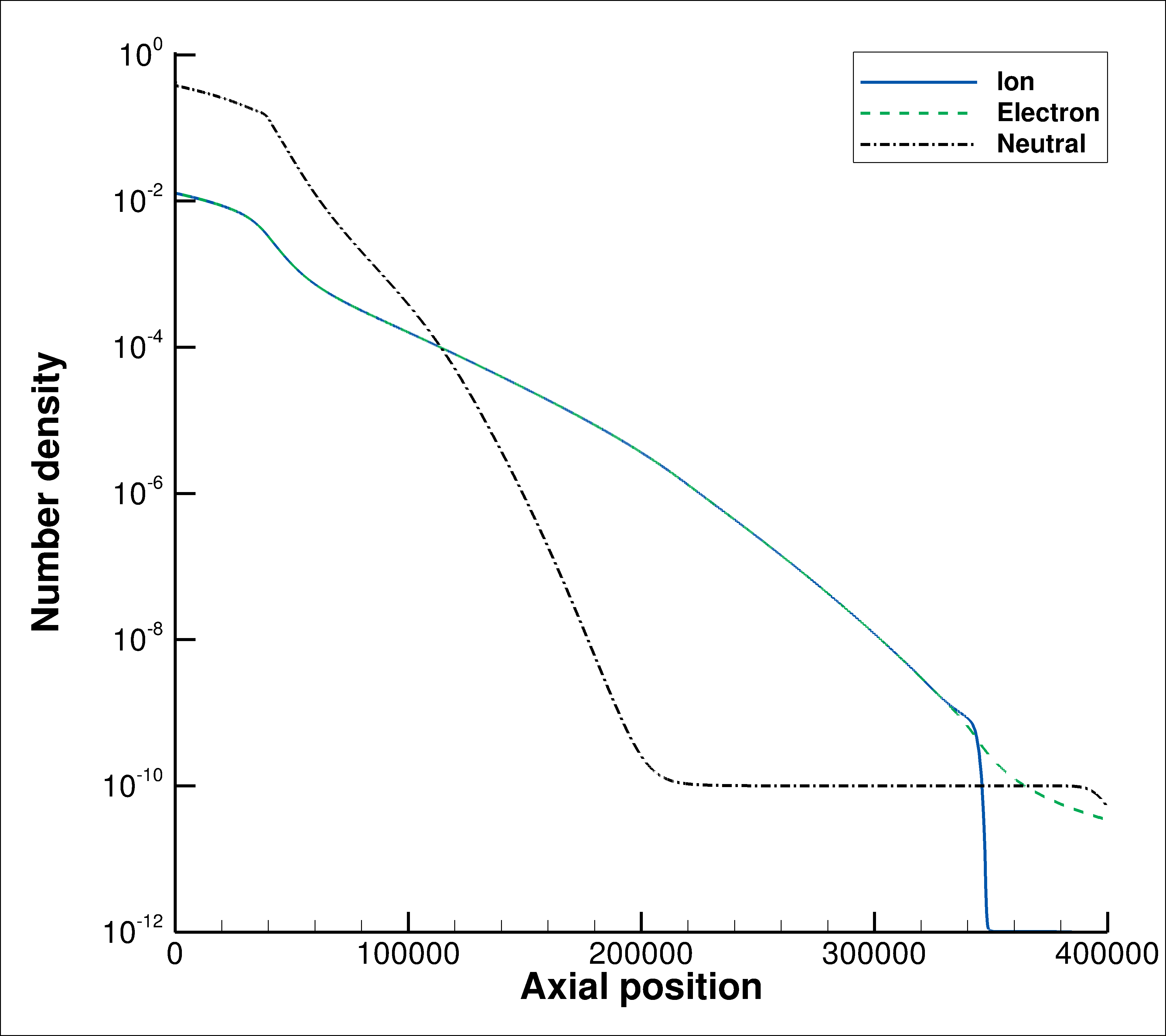}
    \quad
    (b)
    \includegraphics[trim=50 50 17 50, clip, width=0.42\linewidth,valign=t]{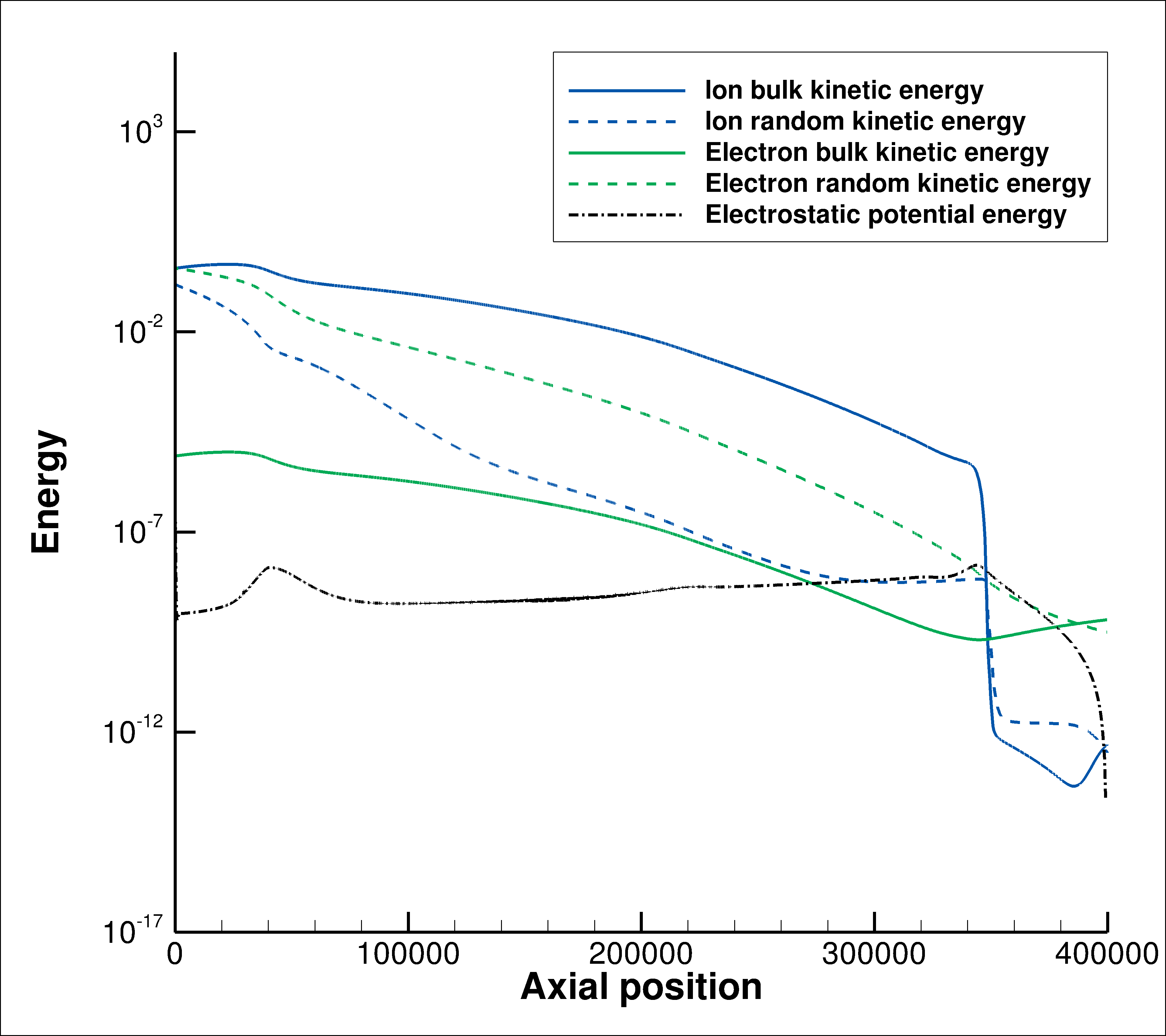}
    }
      \centerline{
    (c)
    \includegraphics[trim=50 50 17 50, clip, width=0.42\linewidth,valign=t]{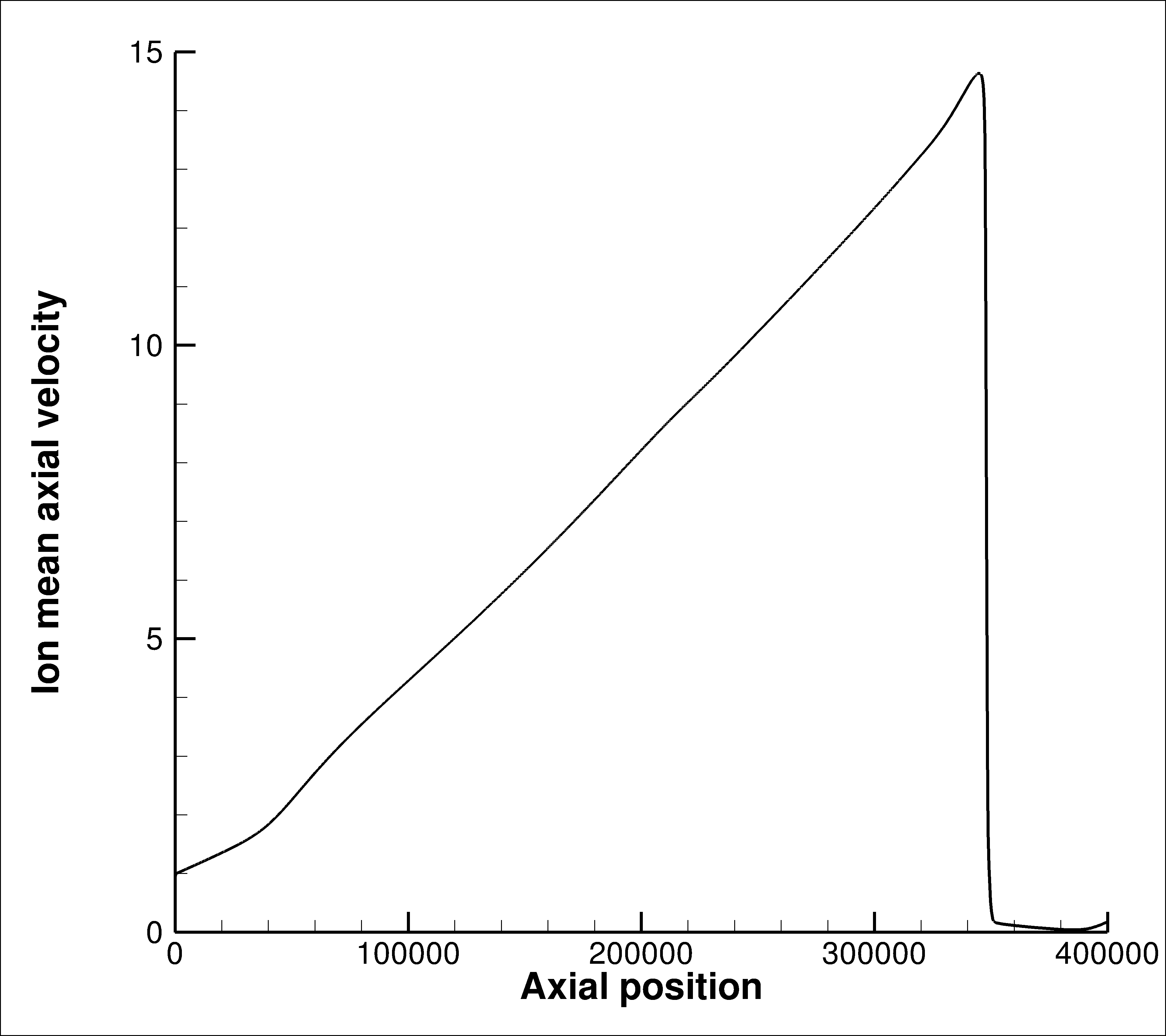}
    \quad
    (d)
    \includegraphics[trim=50 50 17 50, clip, width=0.42\linewidth,valign=t]{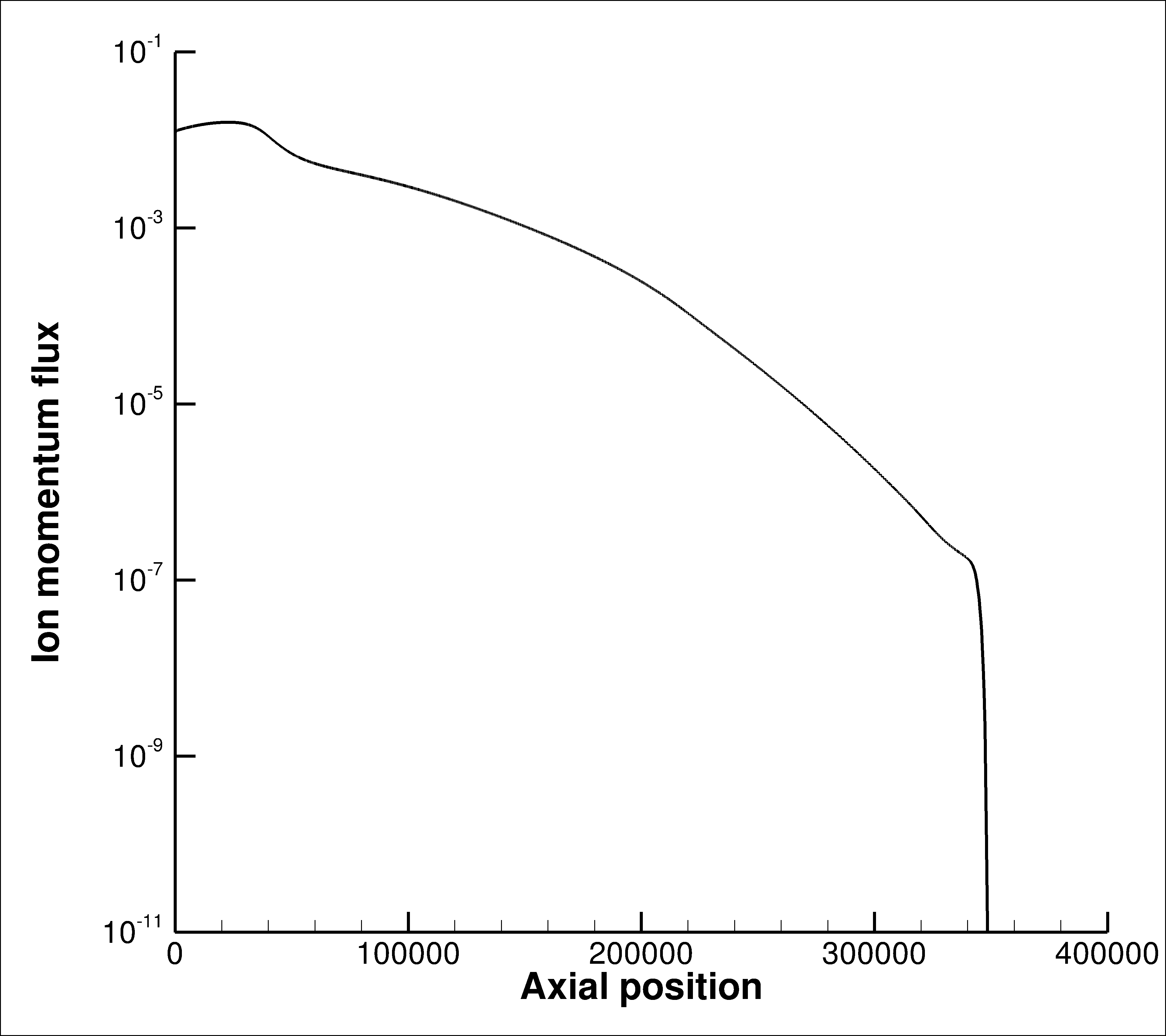}
    }
      \caption{Number densities (a), energies (b), ion mean axial velocity (c), and ion momentum flux (d) for a domain length 100 times the baseline and a grid expansion ratio of $0.1\%$, all at $t=28{,}800$.}
    \label{fig:len100exp1}
    \end{figure}

    With this satisfactory convergence in hand for an extended domain, we proceed to extend the domain further to $L=400{,}000$, also with an expansion ratio of $0.1\%$. The number of spatial grid points is $5{,}109$ in this case, compared with what would have been $100{,}000$ for the corresponding uniform grid, and the ratio of the maximum to the minimum grid size is $100.4$. The corresponding square root of the ion number density, which was plotted in Fig.~\ref{fig:sqrtn} for the baseline domain length, approaches a value of under $10^{-3}$ at the location of maximum charge separation for this domain length (not shown here), so the tolerable grid-size ratio appears to be empirically an order of magnitude larger than this quantity. Key macroscopic quantities are plotted in Fig.~\ref{fig:len100exp1}. Several distinctive features are present relative to the results from the baseline domain length simulations. While panel (a) indicates bunching of all particles in the left one-tenth of the domain where collisionality is expected to be most pronounced, panel (b) reflects the sustained depletion of ion random kinetic energy throughout a significant portion of the domain up to the inlet, indicating that incomplete thermalization is occurring even tens to hundreds of mean free paths behind the expansion front. The inlet mean axial velocity is nonzero as evidenced in panel (c), and the corresponding ion momentum flux also peaks near the inlet in panel (d), rather than near the expansion front in the baseline simulations. The flux still peaks at a magnitude of about $10^{-2}$, but it almost monotonically decays downstream by several orders of magnitude, since the growth of the mean axial velocity is linear while the decay of the number density is near-exponential. Although the broad strokes of the profiles in Figs.~\ref{fig:len10exp1} and \ref{fig:len100exp1} are similar, the varying importance of collisionality across the domain precludes complete self-similarity of the process.

    In principle, the DK solver can be extended to longer domain lengths, such as $L = 4\times10^6$, which would yield centimeter-scale profiles. However, a simulation employing an expansion ratio of $0.1\%$ did not converge, and work is ongoing to test less aggressive grid expansion ratios. Nevertheless, the rapid streamwise decay of the ion momentum flux in Figs.~\ref{fig:len10exp1}(f) and \ref{fig:len100exp1}(d) suggests that thruster characteristics may be broadly characterized in a reasonable fashion with the domain lengths considered in this manuscript.
    
    \subsubsection{Baseline domain length with spatial and velocity coarsening}\label{sec:velexp}
    
    \begin{figure}
      \centerline{
    (a)
    \includegraphics[trim=50 50 50 50, clip, width=0.42\linewidth,valign=t]{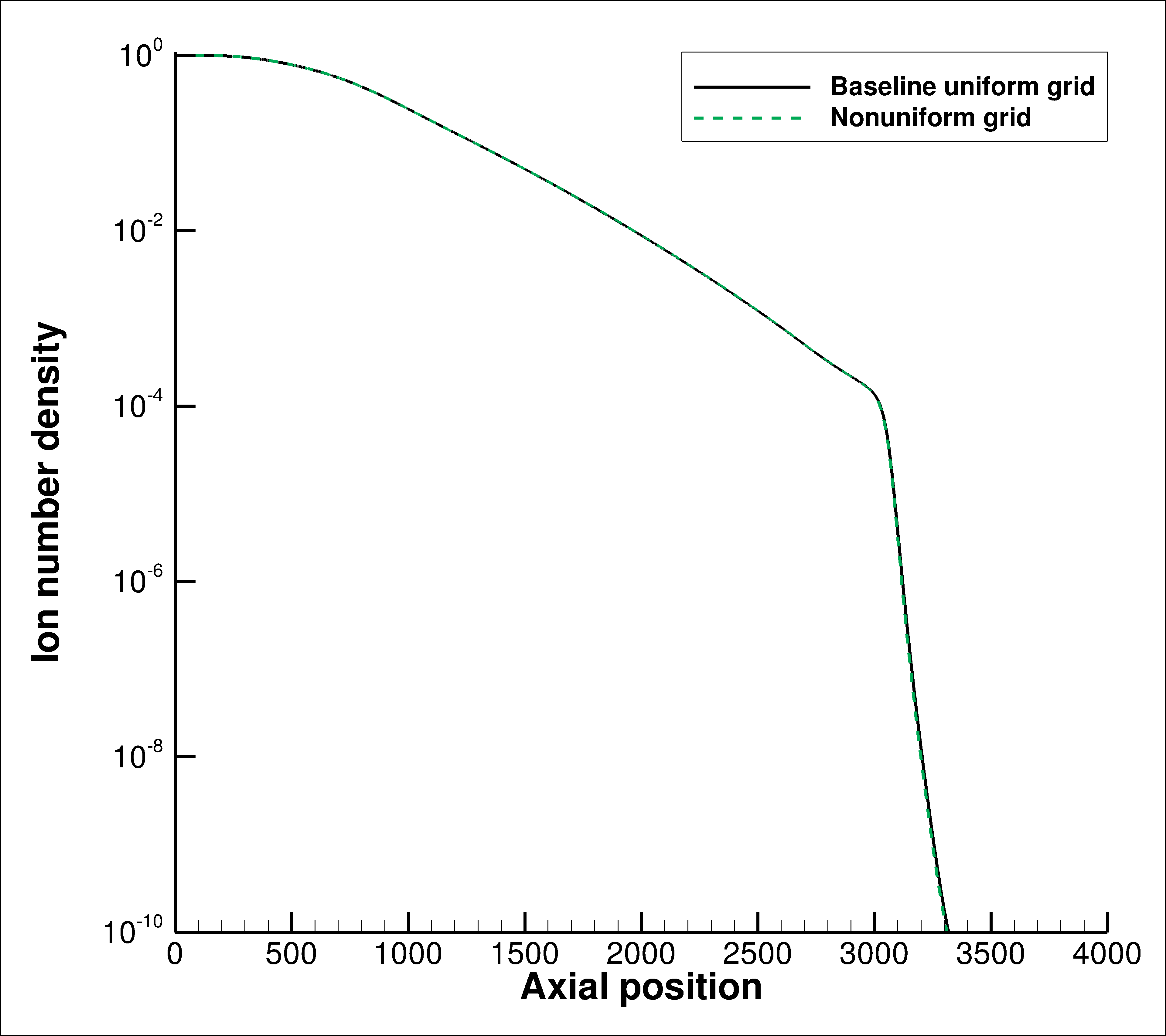}
    \quad
    (b)
    \includegraphics[trim=50 50 50 50, clip, width=0.42\linewidth,valign=t]{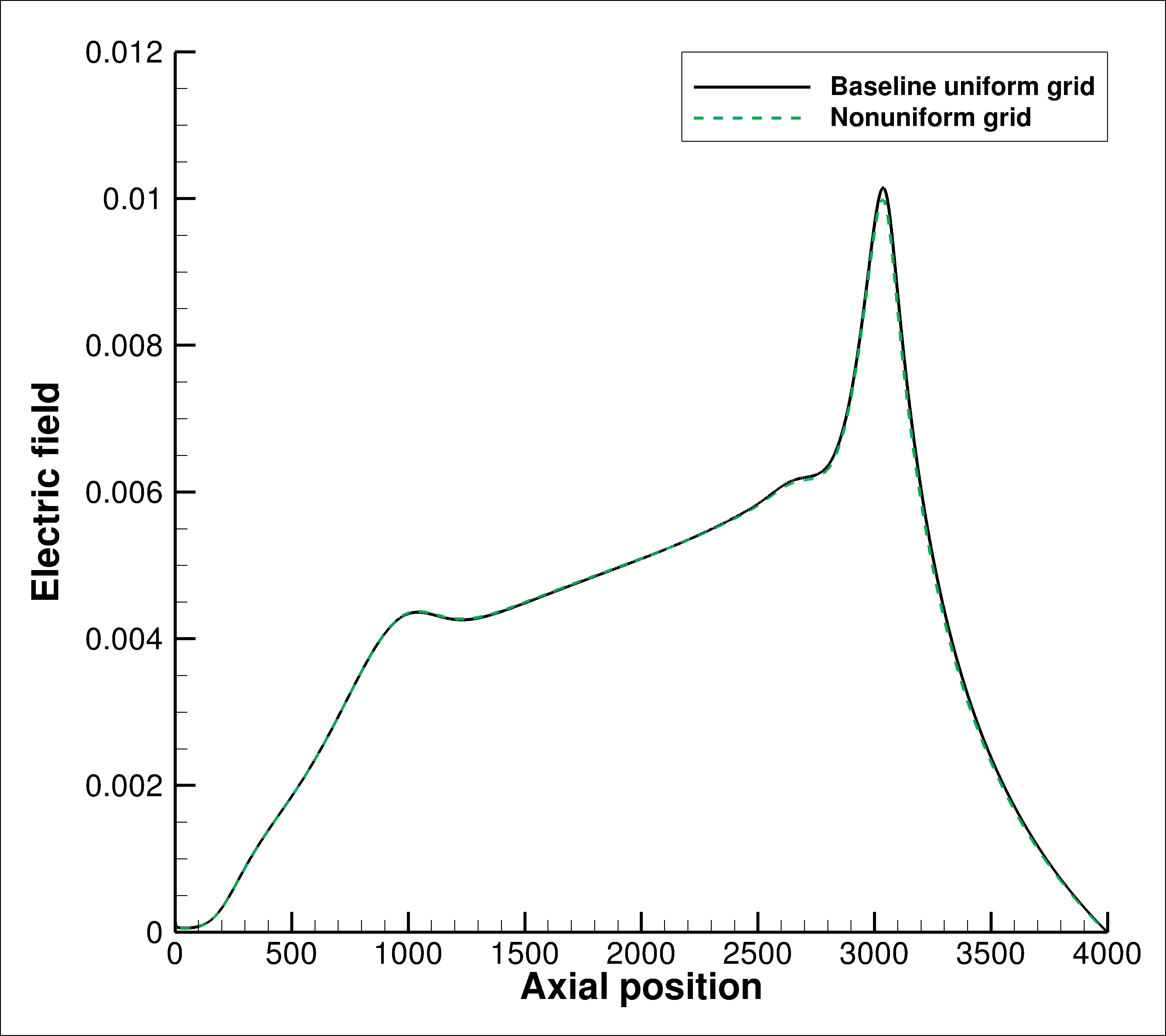}
    }
      \centerline{
    (c)
    \includegraphics[trim=50 50 50 50, clip, width=0.42\linewidth,valign=t]{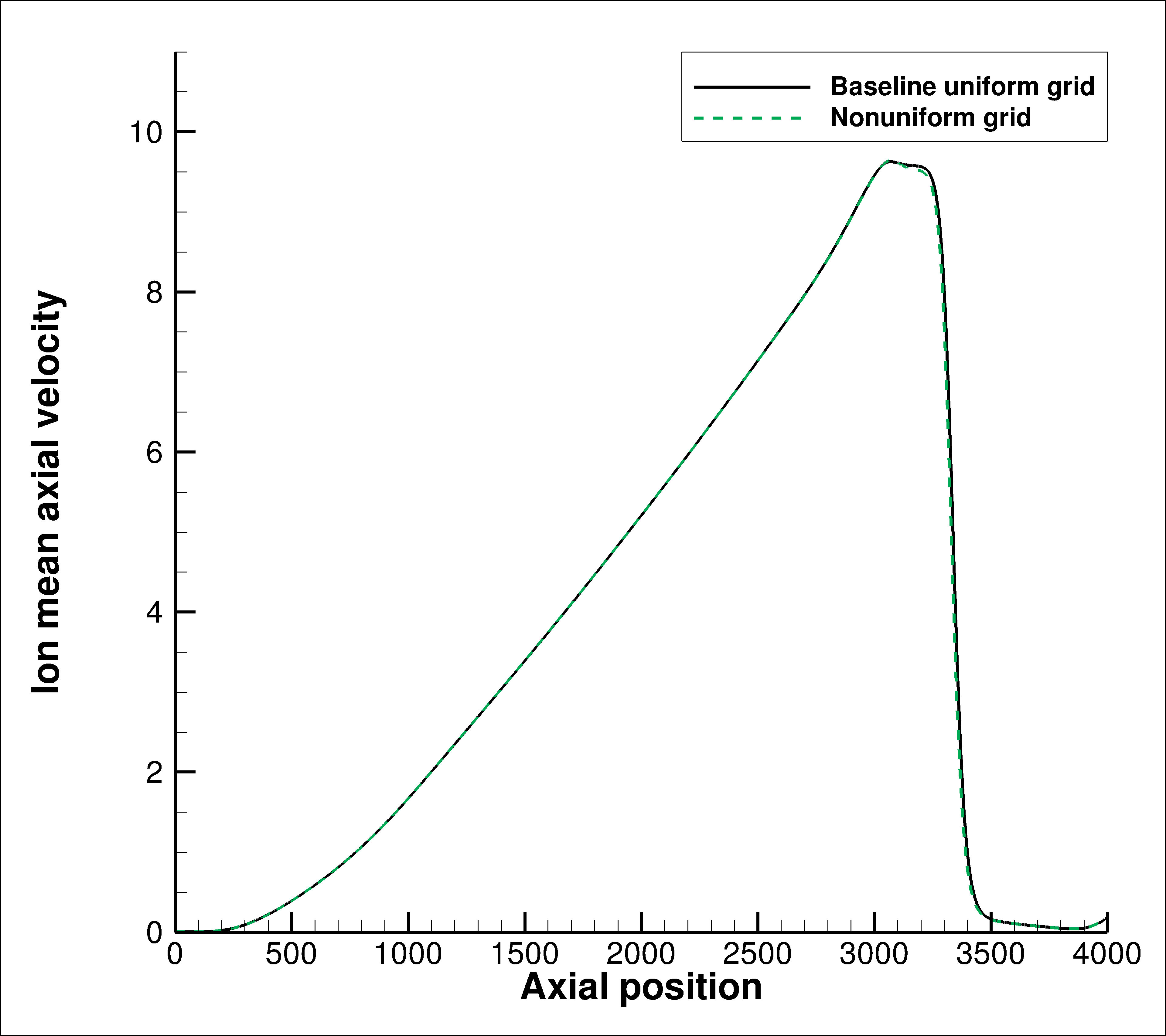}
    \quad
    (d)
    \includegraphics[trim=50 50 50 50, clip, width=0.42\linewidth,valign=t]{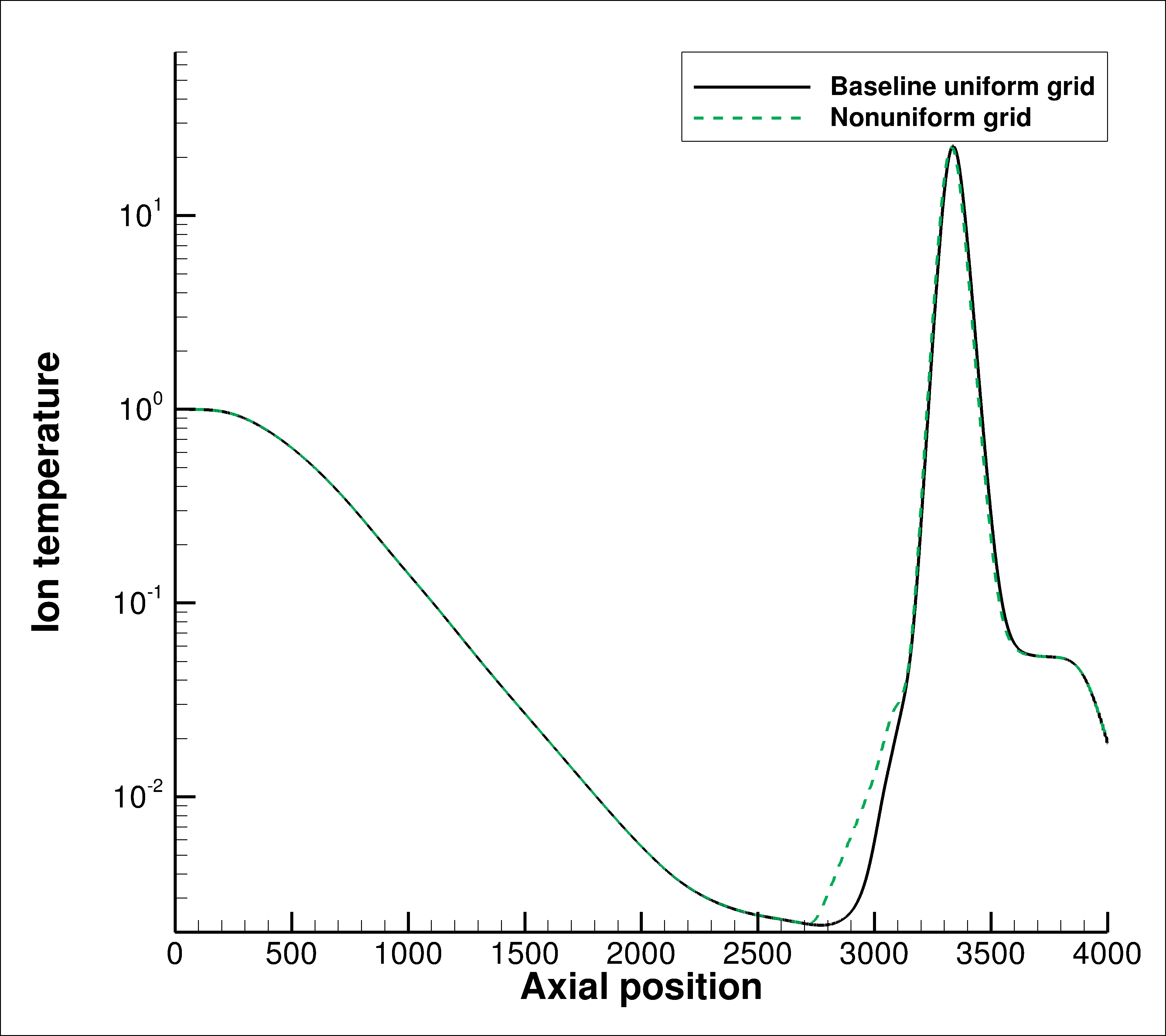}
    }
      \centerline{
    (e)
    \includegraphics[trim=50 50 50 50, clip, width=0.42\linewidth,valign=t]{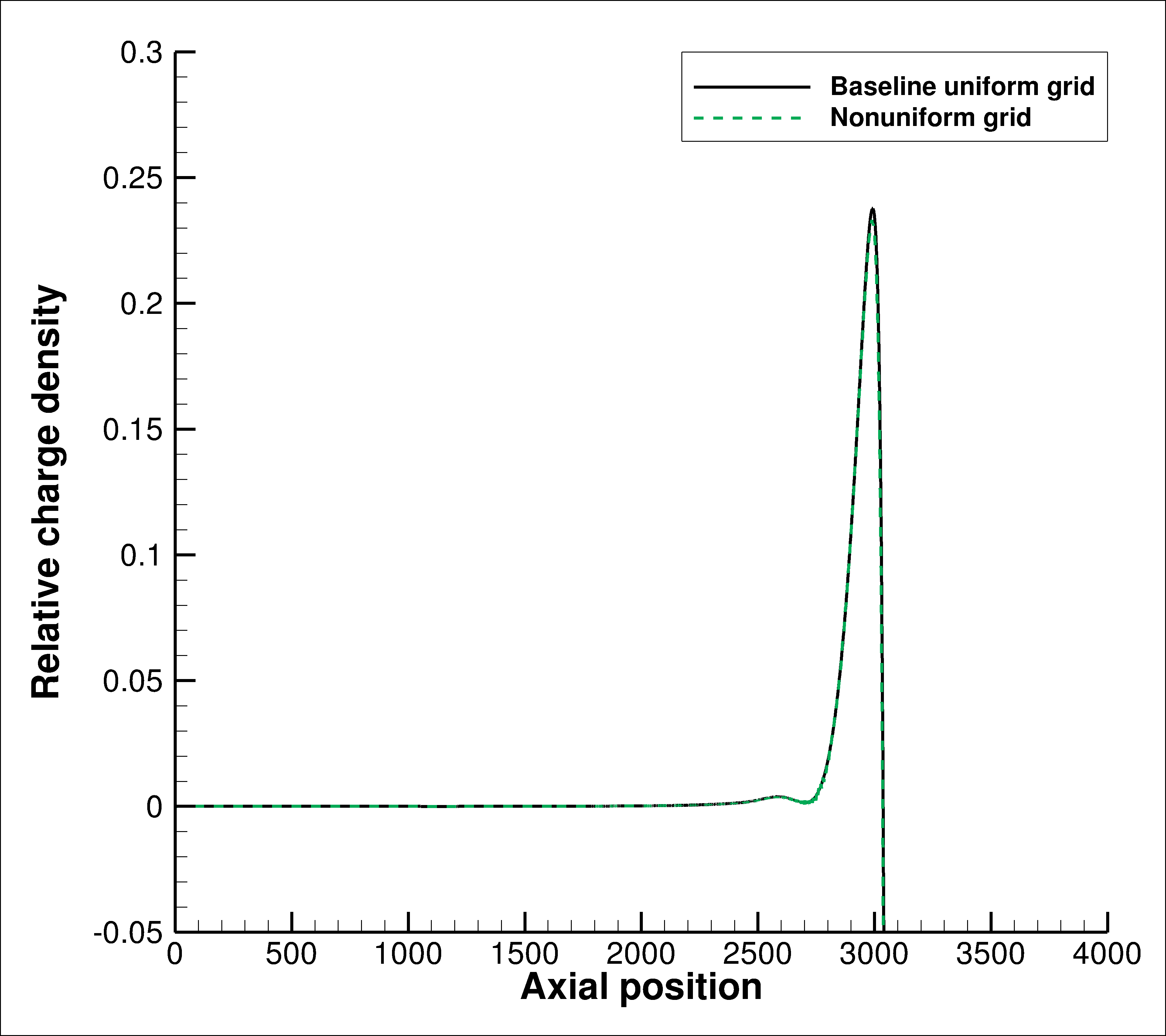}
    \quad
    (f)
    \includegraphics[trim=50 50 50 50, clip, width=0.42\linewidth,valign=t]{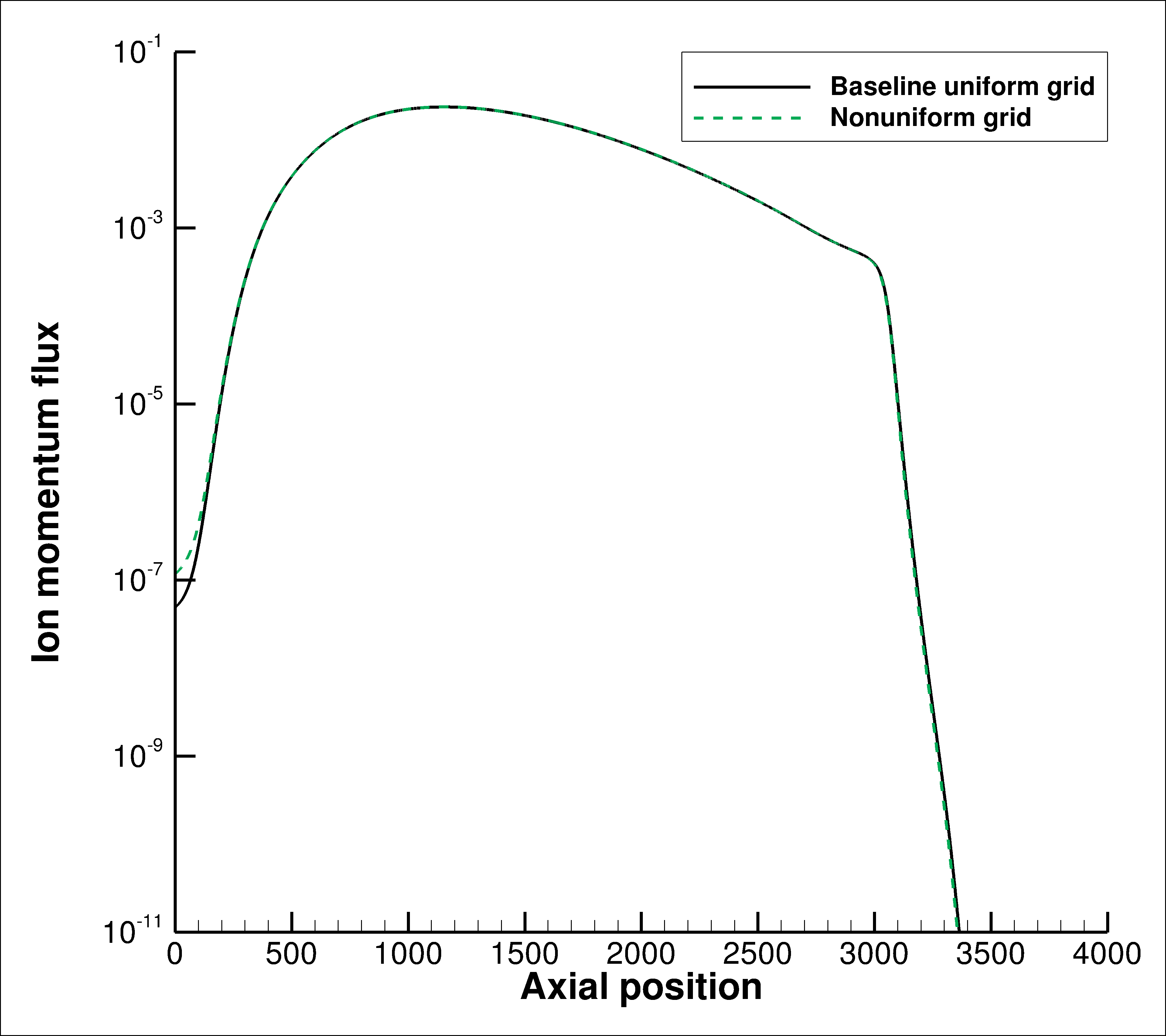}
    }
      \caption{Comparison of ion number density (a), electric field (b), ion mean axial velocity (c), ion temperature (d), relative charge density (e), and ion momentum flux (f) for the baseline domain length, a spatial grid expansion ratio of $0.1\%$, and a velocity grid expansion ratio of $5\%$, all at $t=288$. The baseline uniform grid here corresponds to the mesh with uniform velocity grid spacing and nonuniform spatial grid spacing ($0.1\%$ grid expansion ratio) considered in the dashed curves of Fig.~\ref{fig:len1exp1}.}
    \label{fig:velexp}
    \end{figure}
    
    In addition to spatial coarsening, velocity coarsening may also be used to reduce the number of computational cells. While the characteristic velocity scale to be resolved is predetermined by the prescribed species temperatures, the velocity grid spacing can be coarsened at the domain edges since the distribution typically falls off quickly in magnitude at these extremes and comparable resolution often becomes less crucial. Here, the electron grid size is increased in both directions beyond the zero-velocity centerline, while the ion grid size is increased for all negative velocities, as well as positive velocities above six thermal velocities. An expansion ratio, defined as $(\Delta v_{i+1}-\Delta v_i)/(\Delta v_i)$, of $5\%$ is considered, reducing the number of electron grid points from 432 to 102 and the number of neutral and ion grid points from 936 to 403. The ratio of the maximum to the minimum grid size for electrons and neutrals/ions is $10.7$ and $20.1$, respectively. Visual comparisons of key macroscopic quantities are provided in Fig.~\ref{fig:velexp}. Visual convergence of all macroscopic quantities is mostly satisfactory at this resolution with observable deviations in the momentum flux near the inlet, as well as the electric field and particularly the ion temperature near the expansion front.

    \subsubsection{Characterization of facility effects}
    
\begin{figure}
      \centerline{
    (a)
    \includegraphics[trim=50 50 50 50, clip, width=0.42\linewidth,valign=t]{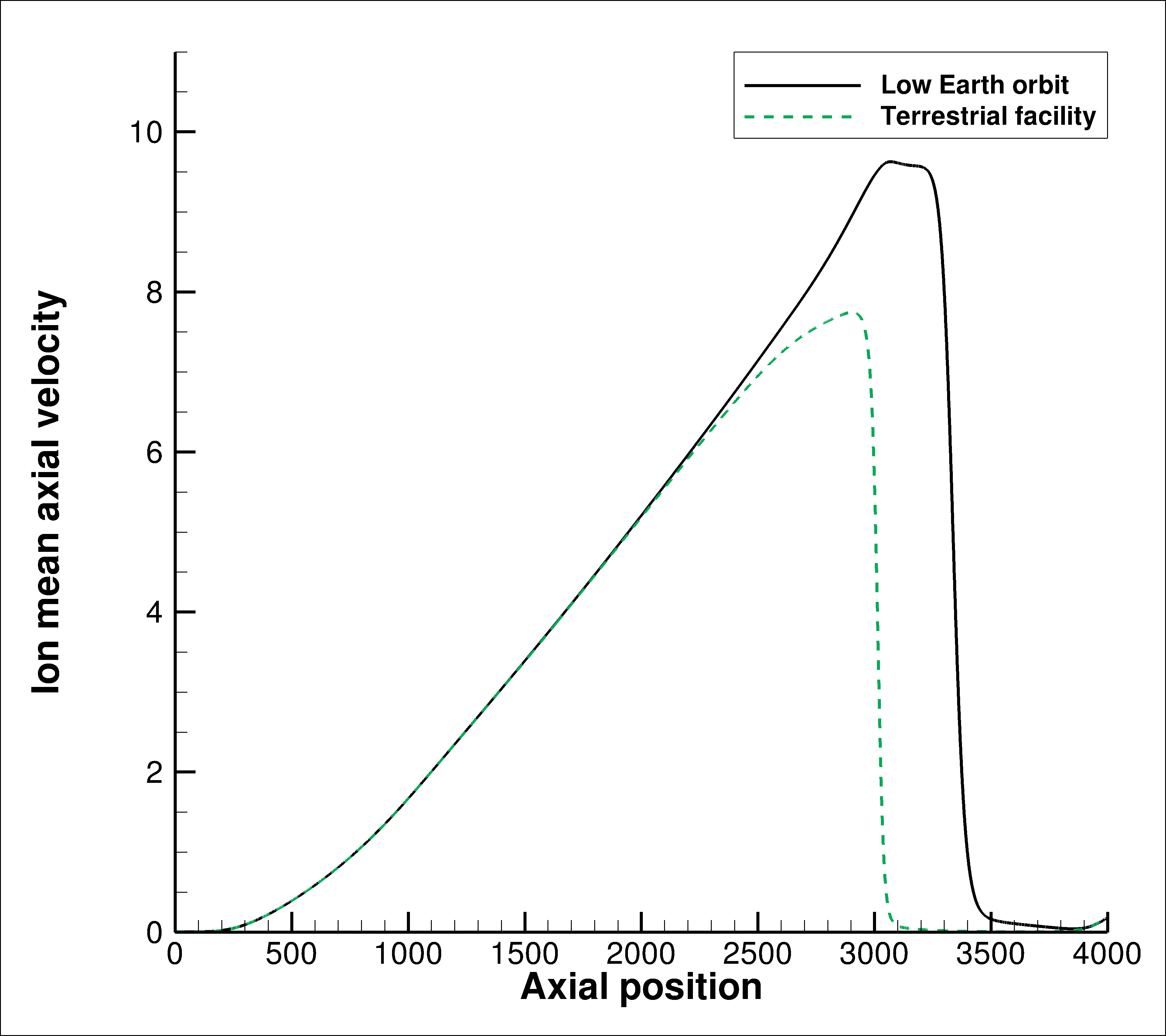}
    \quad
    (b)
    \includegraphics[trim=50 50 50 50, clip, width=0.42\linewidth,valign=t]{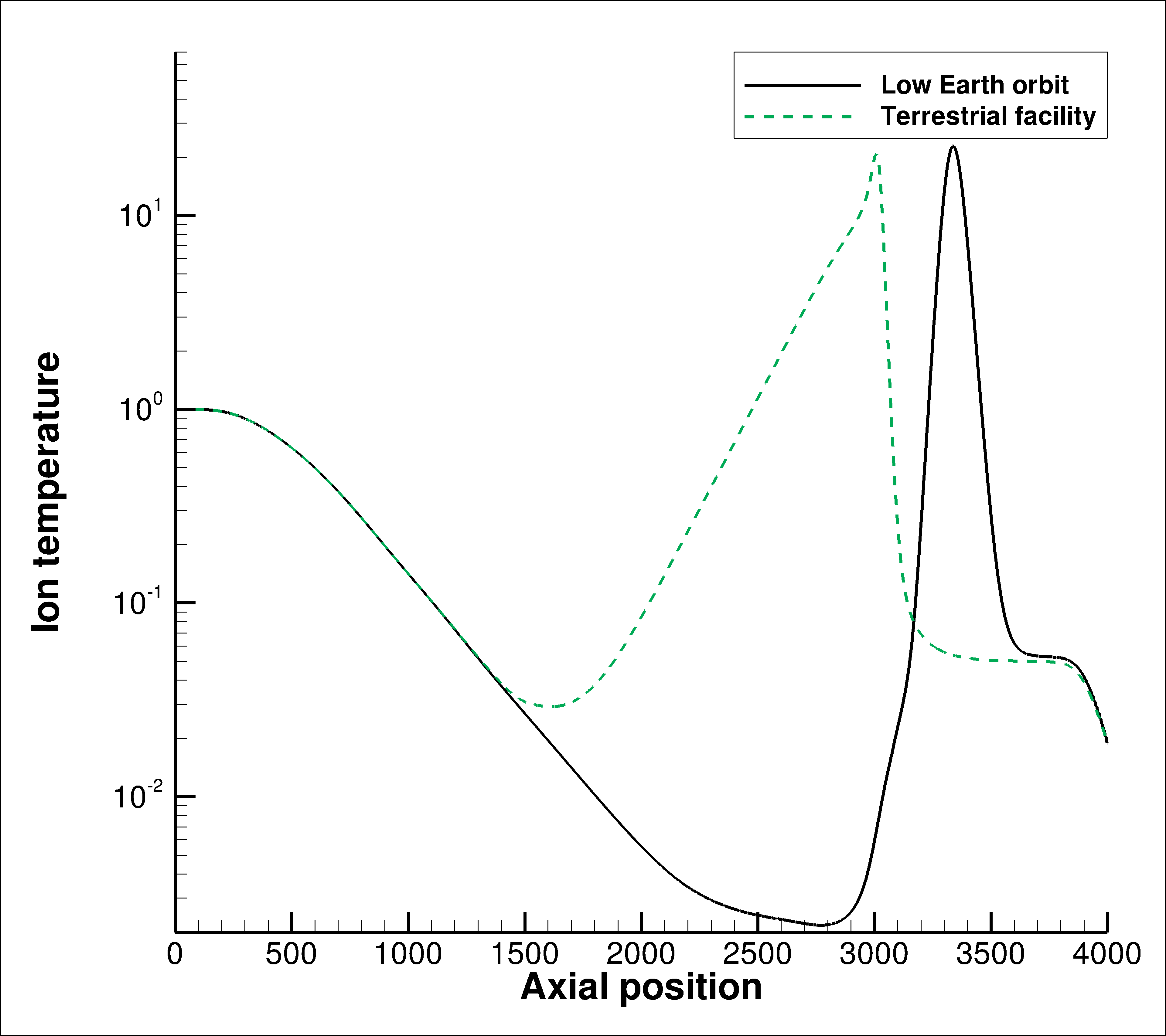}
    }
      \centerline{
    (c)
    \includegraphics[trim=50 50 50 50, clip, width=0.42\linewidth,valign=t]{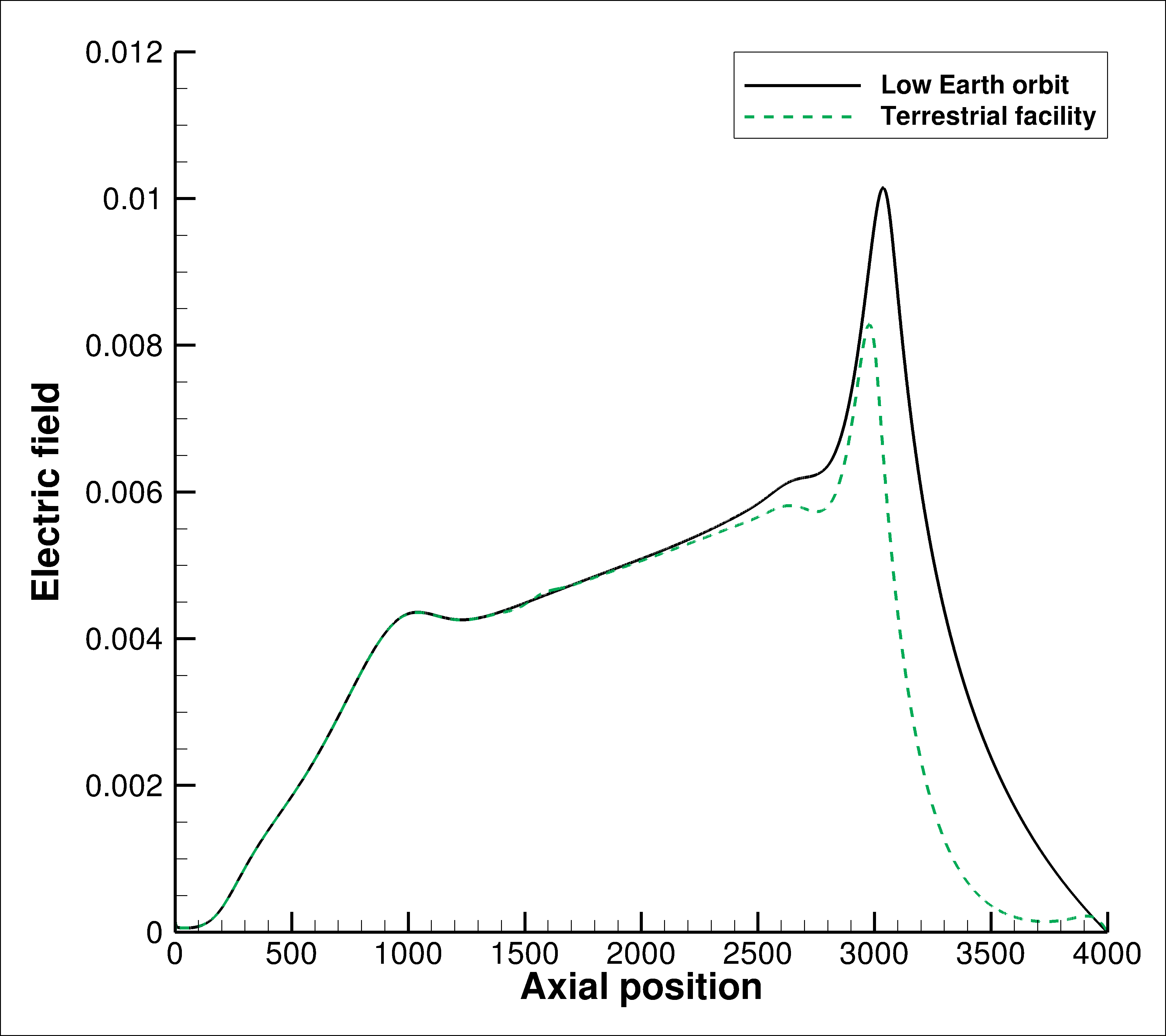}
    \quad
    (d)
    \includegraphics[trim=50 50 50 50, clip, width=0.42\linewidth,valign=t]{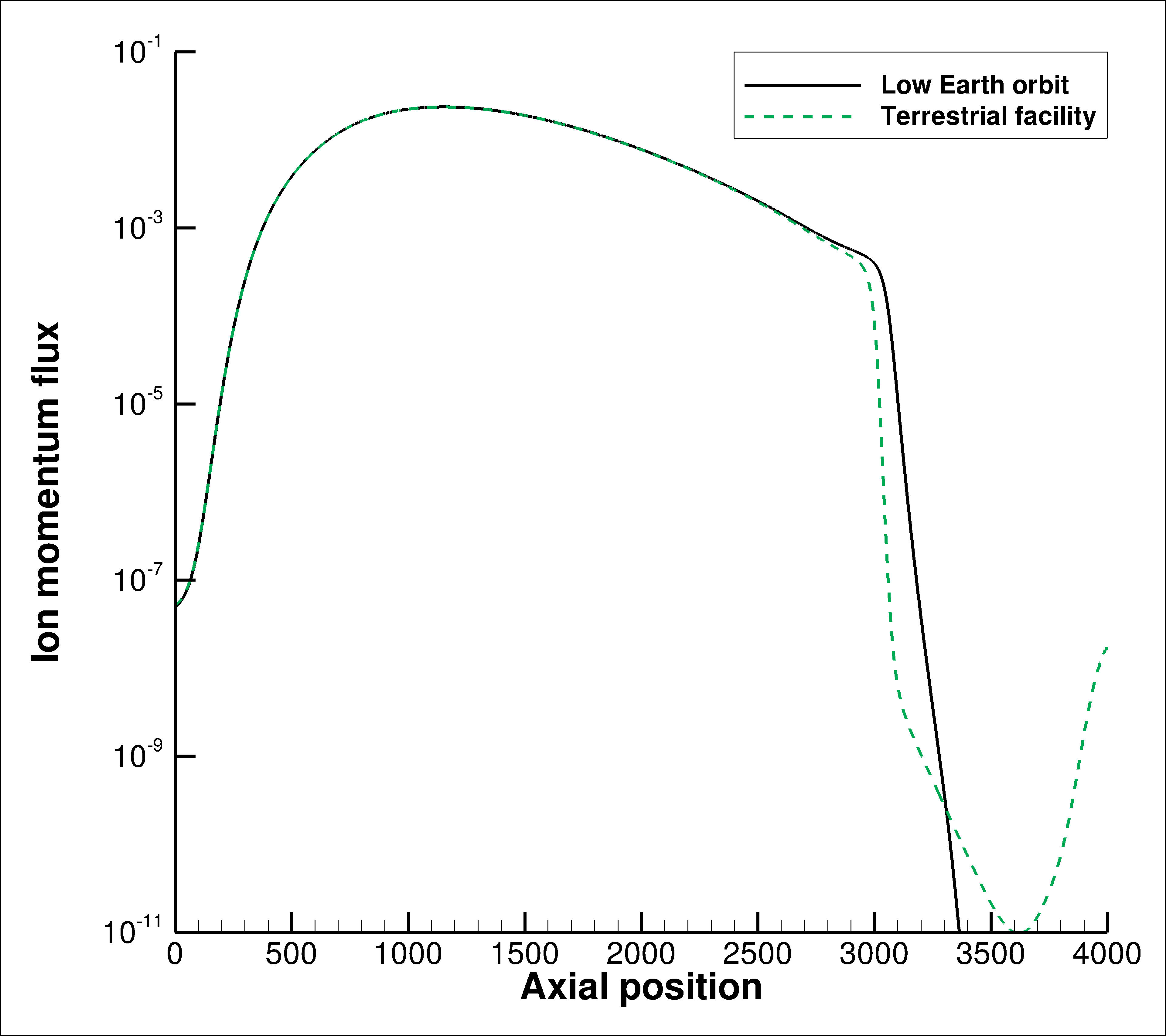}
    }
      \caption{Comparison of ion mean axial velocity (a), ion temperature (b), electric field (c), and ion momentum flux (d) for the baseline simulations with a variation in the ambient pressure, all at $t=288$.}
    \label{fig:len1fac}
    \end{figure}

    An important step in reconciling field and laboratory observations is determining the influence of facility effects, in particular the role of background pressure in thruster performance. Computational solvers can bridge the two through the development of suitable numerical experiments. Here, we consider the direct modification of the ambient pressure in the baseline simulations using
    \begin{equation}
    \frac{n_{i,1}}{n_{i,2}} \sim \frac{n_{e,1}}{n_{e,2}} \sim \frac{n_{n,1}}{n_{n,2}} = 10^4.
    \end{equation}
    Figure~\ref{fig:len1fac} plots a comparison of several macroscopic quantities for the two considered sets of density ratios. As one would expect, more frequent collisions with the background gas retard plume expansion and suppress charge separation, as seen in panels (a) and (c), respectively. In addition, increased collisionality increases the tendency to thermodynamic equilibrium, and the ion ``temperature" does not dip as severely. The momentum flux near the expansion front is also modified with potential impact on the resulting thrust measurements.
   
    \section{Conclusions}\label{sec:conc}

    Laser ablation plasma thrusters are a promising space propulsion concept for lightweight payload delivery. Determination of the lifetime and performance of these thrusters requires a detailed understanding of the dynamics of expansion of the ablated plasma plume. Owing to the transient and inhomogeneous nature of the expansion process, along with the broad range of physical regimes that it spans, grid-based DK solvers are an increasingly relevant method to characterize these flows due to their elimination of statistical noise. This work discusses justification of the employment of a kinetic solver for plume expansion and explores avenues to reduce the cost of DK solvers associated with their higher dimensionality, drawing on grid-point requirements derived in previous work~\citep{chan2022grid}. Thruster-relevant metrics yielded by the DK solver are also discussed for a range of domain lengths from the expansion front scale to the thruster scale.
    
    Employment of a DK solver to a 1D1V plume expansion model problem reveals that the underlying particle distribution functions, in particular the ion distribution, are in a highly nonequilibrium state. Processes contributing to nonequilibrium include the ambipolar acceleration of ions by electrons, the corresponding retardation of electrons by fast ions, and the preference for forward-moving particles at the expansion front, all of which contribute to non-Maxwellian features in the particle distribution. A fully kinetic code is necessary to capture these features and their downstream physical implications in detail. A detailed analysis of the moments of the ion distribution function further reveals that thermalization does not have sufficient time to occur in the region preceding the expansion front, bearing in mind that the considered baseline simulations are about a mean free path and mean free time in length and duration, respectively. The baseline simulation plots in this work may be revealing of the internal structure of the expansion front of a quasineutral plasma plume or ball. A detailed analysis of the energetics of the system also sheds light on the intricate interplay between bulk and random kinetic energy of the ions and electrons, as well as the electrostatic potential energy associated with the corresponding electric field, all of which vary significantly with downstream distance. In particular, the ion random kinetic energy is representative of the ion momentum flux, which is crucial to predicting thrust. Thruster performance can thus be sensitive to the degree of thermalization at various plume locations.
    
    The highly inhomogeneous nature of the problem suggested by the aforementioned moments and energies, together with the momentum flux, lends itself to the question of whether the cost of a DK simulation may be reduced via a nonuniform grid. Analysis of the local Debye length and charge density of the plume, bearing in mind the high degree of thermodynamic nonequilibrium in the process, suggests that the grid size in the considered domain may be relaxed by up to two orders of magnitude for the baseline simulation length without significantly compromising on accuracy. Nonuniform grid simulations relaxing the grid size by an order of magnitude are able to achieve satisfactory convergence with respect to their corresponding uniform grid simulations. The viability of a nonuniform grid enables the simulation of longer domain lengths approaching typical thruster dimensions, and simulations of up to 100 times the baseline domain length and simulation duration were performed with cost savings of up to 20 times. Velocity coarsening has the potential to add a further factor of 2 to 4 in cost savings. While rapid thermalization is absent in a wide region behind the expansion front even with an extended domain length and simulation duration, the varying degree of collisionality across the domain can still impede a fully self-similar description of the complete solution, especially in the highly collisional upstream region where coherent particle advancement may be enhanced.
    
    Apart from direct computation of the momentum flux, a DK solver can also be valuable in determining the influence of facility effects. An increased ambient pressure is shown to decrease the peak electric field and ion mean axial velocity. The ion momentum flux in the vicinity of the expansion front is also modified by increased collisions with the background gas. Putting these observations together, the DK method can shed light on the sensitivity of thruster performance to operating conditions without being susceptible to statistical noise in the various metrics.
    
    The simulations of this work suggest that thruster-scale DK simulations can be viable with the deployment of a nonuniform grid. The general feasibility of a nonuniform DK solver increases its viability of application to a broad range of plasma problems of interest, with the potential to yield crucial physical insights into nonequilibrium processes and their associated energetics that elude traditional analysis methods and numerical solvers.

%

    \section*{Acknowledgments}
    
    The authors would like to acknowledge A. Vazsonyi, A. Raisanen, R. Chaudhry, M. Petrusky, C. Cui, and P. Moin for discussions helpful to this work, as well as computational resources from the Blanca Research Computing condo computing service of the University of Colorado. The work is supported by the Air Force Office of Scientific Research, Grant \#FA9550-21-1-0045.

    \bibliography{references}

\end{document}